%% file: main.tex
\documentclass[preprint,5p,10pt]{elsarticle}

\usepackage[ruled,linesnumbered,vlined]{algorithm2e}
\usepackage{algorithmic}
\usepackage[caption=false,font=footnotesize]{subfig}
\usepackage{url}
\usepackage[table]{xcolor}
\usepackage{nameref}
\usepackage{paralist}
\usepackage{multicol}
\usepackage{multirow}
\usepackage{graphicx}
\usepackage[fleqn]{amsmath}
\usepackage{amsfonts}
\usepackage{tcolorbox}
\usepackage{soul}
\usepackage{flushend}
\usepackage{capt-of}
\usepackage{lmodern}
\usepackage{booktabs}

\def\code#1{\texttt{#1}}

\SetCommentSty{mycommfont}
\let\oldnl\nl% Store \nl in \oldnl
\newcommand{\nonl}{\renewcommand{\nl}{\let\nl\oldnl}}% Remove line number for one line

\journal{Computer Networks}

\begin{document}

\begin{frontmatter}

\title{+Tour: Recommending Personalized Itineraries for Smart Tourism}

\author[label1]{João Paulo Esper} \ead{joaopauloesper@dcc.ufmg.br}
\author[label2]{Luciano de S. Fraga}
\author[label3]{Aline C. Viana}
\author[label2]{Kleber Vieira Cardoso}
\author[label2]{Sand Luz Correa}

\cortext[cor1]{Corresponding author.}

\address[label1]{Computer Science Department (DCC), Federal University of Minas Gerais (UFMG),\\Belo Horizonte, Brazil - 31270-901}
\address[label2]{Institute of Informatics (INF), Federal University of Goiás (UFG), Goiânia, Brazil - 74690-900}
\address[label3]{Inria Saclay Île-De-France, 1 Rue Honoré d'Estienne d'Orves, Palaiseau, France - 91120}

\begin{abstract}
Next-generation touristic services will rely on the advanced mobile networks' high bandwidth and low latency and the Multi-access Edge Computing (MEC) paradigm to provide fully immersive mobile experiences. As an integral part of travel planning systems, recommendation algorithms devise personalized tour itineraries for individual users considering the popularity of a city's Points of Interest (POIs) as well as the tourist preferences and constraints. However, in the context of next-generation touristic services, recommendation algorithms should also consider the applications (e.g., social network, mobile video streaming, mobile augmented reality) the tourist will consume in the POIs and the quality in which the MEC infrastructure will deliver such applications. In this paper, we address the joint problem of recommending personalized tour itineraries for tourists and efficiently allocating MEC resources for advanced touristic applications. We formulate an optimization problem that maximizes the itinerary of individual tourists while optimizing the resource allocation at the network edge. We then propose an exact algorithm that quickly solves the problem optimally, considering instances of realistic size. Using a real-world location-based photo-sharing database, we conduct and present an exploratory analysis to understand preferences and users’ visiting patterns. Using this understanding, we propose a methodology to identify user interest in applications. Finally, we evaluate our algorithm using this dataset. Results show that our algorithm outperforms a modified version of a state-of-the-art solution for personalized tour itinerary recommendation, demonstrating gains up to 11\% for resource allocation efficiency and 40\% for user experience. In addition, our algorithm performs similarly to the modified state-of-the-art solution regarding traditional itinerary recommendation metrics.

%Personalized Tour Itinerary Recommendation (PTIR) algorithms have been successfully explored to enhance the user quality of experience. However, the literature lacks works that tackle the joint problem of designing PTIR algorithms and allocating resources at the network edge to fulfill new preferences that will be available with the prevalence of advanced mobile networks and mobile augmented reality applications. In this work, we address this problem and propose an efficient algorithm to solve it. We evaluate our solution using a realistic methodology based on data analysis on a real-world dataset. Results demonstrate that our proposal produces fast and high-quality solutions, showing gains up to 11\% for resource allocation efficiency and 40\% for user experience when compared with a modified state-of-the-art PTIR algorithm.
\end{abstract}

\begin{keyword}

Travel itinerary recommendation \sep Next-generation touristic services \sep Multi-access edge computing \sep Advanced mobile networks.


\end{keyword}

\end{frontmatter}

\input{Sections/01-Introduction}
\input{Sections/02-Related-Work}
\input{Sections/03-System-Model}
\input{Sections/04-Problem-Formulation}
\input{Sections/05-Solution}
\input{Sections/06-Data-Characterization}
\input{Sections/07-Experimental-Evaluation}
\vspace{-0.35cm}
\input{Sections/08-Conclusions-And-Future-Work}

\vspace{-0.35cm}

\section*{Acknowledgment}

The authors thank Dante Campos for his contributions to the optimization model. This work was supported by CAPES, the Brasil 6G project under Grant 01245.020548/2021-07, and LaMCAD/UFG.

\vspace{-0.35cm}

\bibliographystyle{elsarticle-num} 
\bibliography{ref}

\end{document}

%% file: Sections/01-Introduction.tex
\section{Introduction}\label{introduction}

Tourism plays a fundamental role in the economy of many countries. Currently, the massive size of the tourism industry requires Information and Communication Technology (ICT) to make services more intelligent, enhance user (tourist) experience, and improve user engagement. Thus, the last decades have witnessed an explosion of digital services to build Smart Tourism~\cite{eu-capital:23}{,} where users actively participate in their trips. Similarly, the prevalence of mobile devices and advancements in wireless communication allowed the emergence of location-based social networks (LBSNs)~\cite{silva-urban:19}, constituting a rich source of spatial-temporal datasets describing social network users' mobility and interest information. These datasets have been commonly leveraged in the last few years by itinerary recommendation systems aiming to offer more personalized services~\cite{lim-tour:18, shini-extensive:20, yochum-linked:20,halder-survey:24}. 

Personalized Tour Itinerary Recommendation (PTIR) systems are another important component of Smart Tourism. Given the set of Point-of-Interests (POIs) of a city, the time availability for the tour visit, and the user's preferences, the mission of PTIR systems is to identify a tour itinerary (i.e., a sequence of ordered POIs) that maximizes the user's experience while adhering to her time availability constraint. Previously proposed PTIR systems assume that trip experience depends on POI popularity~\cite{choudhury-automatic:10}, user's interest in the POI category (e.g., Shopping, Entertainment, etc.)~\cite{brilhante-where:13, yu-personalized:16, lim-personalized:18}, or other user's preferences mined from LBSNs~\cite{chen-persolnalized:20, chen-trip:23}.

With the deployment of 5G networks worldwide and the future Beyond 5G extension, a new trend in Smart Tourism is Mobile Augmented Reality (MAR) technologies, which allow users to project computer-generated augmentations on top of real-world images using mobile devices (e.g., smartphones or wearables). This task, however, involves processing-hungry computer vision and rendering algorithms, which are hard to fulfill by resource-constrained mobile devices. Thus, realizing MAR requires a compromise between mobility, battery life, and performance. One way to achieve this compromise is to exploit the processing capabilities of Multi-access Edge Computing (MEC) and offload the MAR computation to the wireless network edge~\cite{chen-empirical:17,zhou-5g:24}. Edge computation offloading results in enhanced user experience and smaller footprints of the bandwidth-demanding media traffic to the wireless network~\cite{siriwardhana-survey:21}. Consequently, as advanced mobile networks become prevalent and MAR applications use more sophisticated video formats and resolutions, the user experience when visiting a POI will also depend on the applications (e.g., social network, Mobile Video Streaming (MVS), MAR) she will consume, and the quality that the ICT infrastructure will deliver such applications.

\begin{figure}[t]
    \centering
    \includegraphics[width=0.50\textwidth]{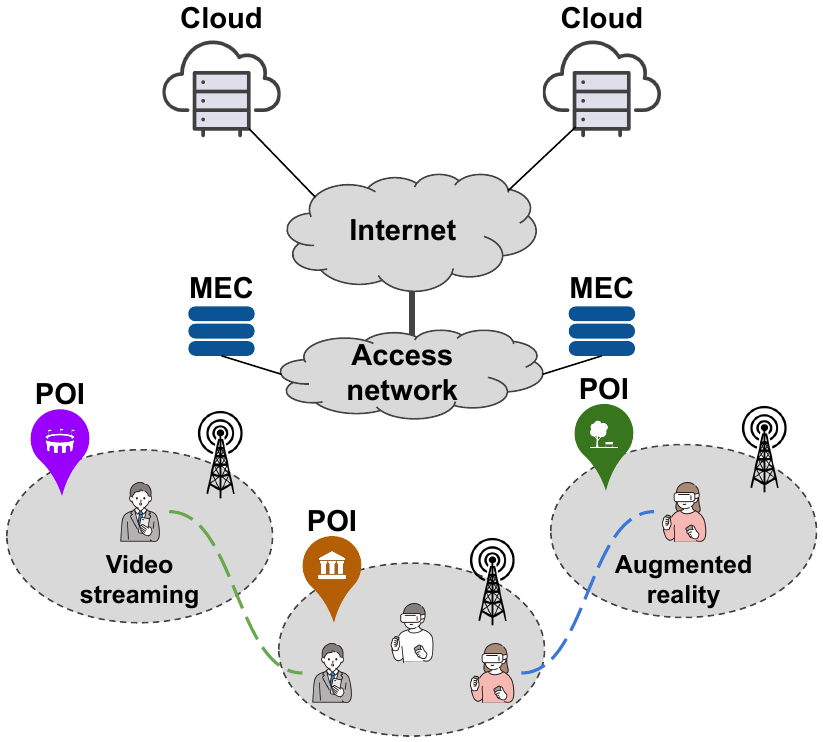}
    \caption{User spots with {advanced mobile networks} ICT infrastructure.}
    \label{fig:ICT-Infrastructure}
\end{figure}

In the above context, neglecting to account for the edge resources when recommending a tour itinerary can create a frustrating experience for users once they seek an advanced, combined physical and virtual visit experience. Thus, the coupling between MEC resource allocation and PTIR systems is the {primary} motivation for this work. Consider, for example, a Metropolitan Tourism Centre (MTC) that provides smart tourism services. The MTC leases network and computing resources from a network operator as a network slice. The network slice comprises mobile wireless coverage in each POI, MEC server capabilities near the POIs, and remote cloud computing resources accessed through the Internet, as illustrated in Figure~\ref{fig:ICT-Infrastructure}. The MTC also offers a PTIR system to help users build their tour itineraries according to their preferences and the applications they consume. However, instead of issuing recommendations for itineraries that rank as the top most relevant according to individual preferences, {the PTIR system proposes an itinerary to individual users that is still attractive according to their preferences but, at the same time, optimizes the MEC resource allocation when running the selected applications.} Although such recommendations can slightly affect the recommended POIs, they aim at higher MEC resource allocation efficiency and, thus, a better global (multi-users) physical and virtual experience.

This work advances the state-of-the-art {recommendation of} personalized tour itineraries by considering new preferences {that will be} available with the prevalence of advanced mobile networks and MAR applications. In this context, we jointly approach the design of PTIR algorithms and the resource allocation at the network edge, making the following contributions:

\begin{itemize}

\item We formulate an optimization problem for recommending personalized tour itineraries for users considering their preferences, the applications they will consume during the tour, and the corresponding resource allocation required at the network edge. The objective is to find a single itinerary for each user so that the set of chosen itineraries maximizes the sum of the profits perceived by all the users, prioritizing the MEC resource allocation while satisfying the users and the infrastructure constraints. This paper significantly extends our previous paper~\cite{fonseca-personlized:19}, which, to the best of our knowledge,
was the first work to formulate {and evaluate} this problem (cf. Section~\ref{sec:related-work}). 

\item We devise the next-generation of Touristic services (+Tour), an algorithm that efficiently solves the above problem using {Dynamic Programming and Mixed Integer Linear Programming~\cite{hillier2015introduction}}. 

\item We then infer the potential interest of users in types of applications or services. For this, we automatically extract POI popularity and user preferences on POIs from a real-world location-based photo-sharing application (Flickr). The resulting dataset describes users' visits in 13 tourist cities on four continents. We then conduct and present an exploratory analysis of this dataset to understand preferences and users' visiting patterns. We then propose a methodology to identify user interest in applications based on such investigations.

{\item We introduce two new metrics, namely, \textbf{Allocation Efficiency} (\textbf{AE}) and \textbf{User Experience} (\textbf{UE}) to assess the performance of the algorithms on allocating resources in the network edge and on the overall user experience provided by the recommended itinerary.}

{\item We provide a dataset based on real-world POI visitation history from 13 cities spread across 4 continents. This dataset includes 20461 valid tour sequences generated by 8407 users who visited 401 POIs divided across 20 unique categories.}

\item Using the collected dataset, we evaluate the effectiveness of +Tour in three different resource utilization scenarios. In addition, we compare the results obtained by +Tour with those produced by a modified version of PersTour~\cite{lim-personalized:18}, a state-of-the-art solution for recommending personalized tour itineraries. +Tour outperforms the modified version of PersTour in all scenarios, showing gains up to 11\% for resource allocation efficiency and 40\% for user experience. In addition, +Tour performs similarly to {a modified version of} PersTour in traditional metrics for evaluating PTIR systems. We also show that +Tour can optimally solve the problem for instances of realistic size {(250 users per instance in an environment with the scale outlined in Table~\ref{tab:dataset})} in a reasonable amount of time {(Figure~\ref{fig:time})}.

\item We make the +Tour source code and the processed dataset publicly available.

\item {Finally, we highlight that our proposal induces the users to cooperate by strategically recommending each tourist itinerary using information from a group of tourists that may compete for resources. An MTC adopting +Tour would benefit from this induced cooperative behavior, which could improve the users’ experience and, consequently, retain current clients and attract new ones. Even in other contexts, cooperative behavior is generally desired. For example, the authors of \cite{rehman2022ITS} show how a transport company can improve resilience and security by using a cooperative intelligent transportation system. In \cite{fang2023COVID}, the authors investigate the impact of cooperative behavior between social organizations during the COVID-19 pandemic outbreak in Shanghai and find consistent advantages of cooperative multi-organization management. While it is well-known that cooperation among regional departments plays an important role in facilitating the equitable disposal of construction waste across regions, through modeling and simulation, the authors of \cite{mengdi2024recycling} are able to identify cross-regional cooperation strategies for recycling enterprises, enhancing the efficiency of construction waste treatment and fostering the coordinated development of urban areas.}

\end{itemize}

The rest of this paper is organized as follows. Section~\ref{sec:related-work} discusses the related work. Section~\ref{sec:system-model} introduces the system model. The problem formulation is formalized in Section~\ref{sec:problem-formulation}, while +Tour is detailed {and analyzed} in Section~\ref{sec:solution}. Data characterization is presented in Section~\ref{sec:data-characterization}, and the experimental evaluation is detailed in Section~\ref{sec:evaluation}. Section~\ref{sec:conclusions} concludes the paper and outlines future work.

%% file: Sections/02-Related-Work.tex
\section{Related Work}\label{sec:related-work}

\begin{figure*}[t]
\centering
\includegraphics[width=0.8\textwidth]{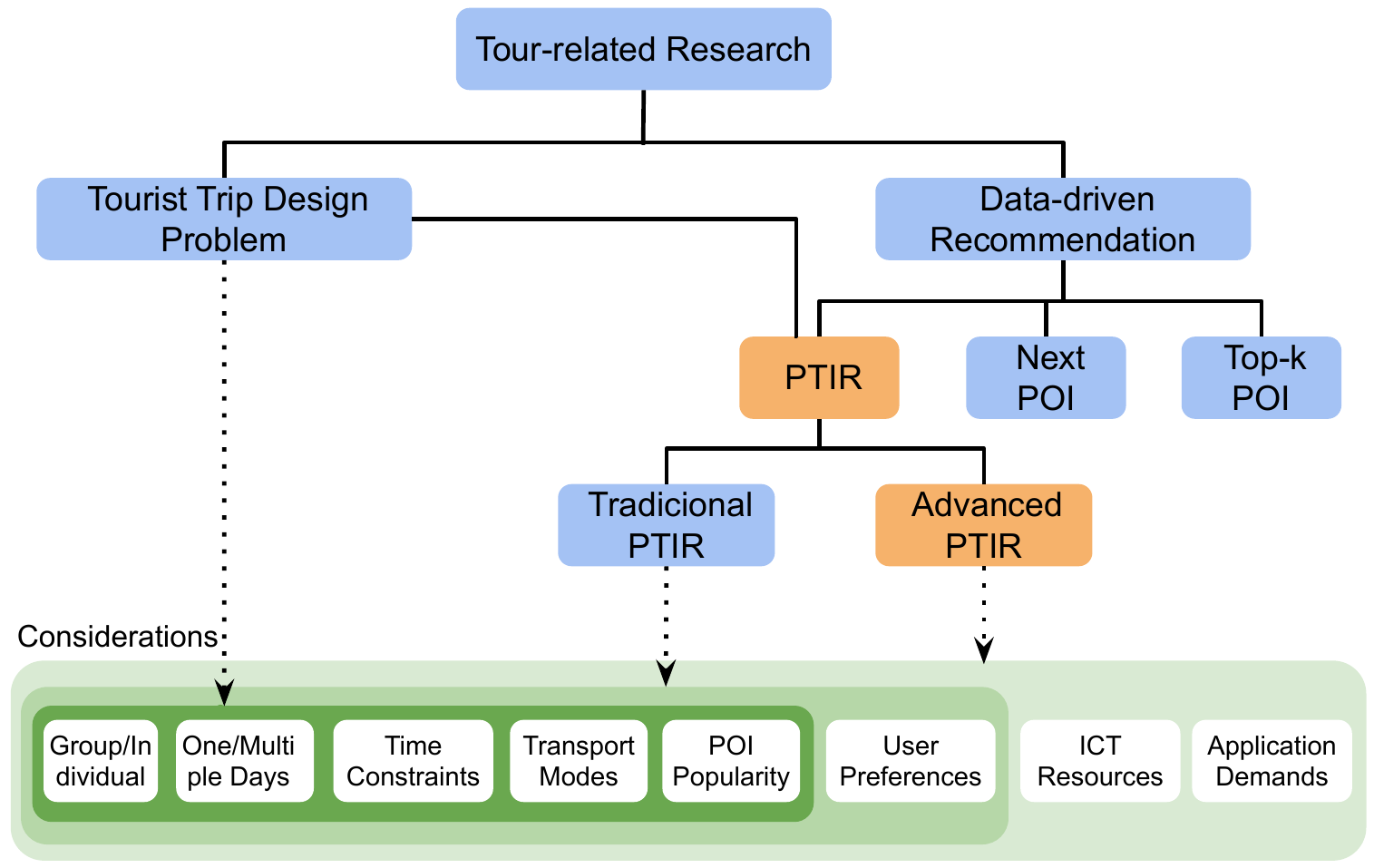}
\caption{{Tree of classification for tour-related research. Adapted from~\cite{lim-tour:18}. Yellow boxes indicate the topic of interest.}}
\label{fig:classification}
\end{figure*}

{As illustrated in Figure~\ref{fig:classification}, the literature related to tour itinerary generation is broadly divided into two categories: Tourist Trip Design Problem (TTDP) and Data-driven Recommendation~\cite{lim-tour:18}. Next, we discuss relevant works in both categories.}

{\subsection{TTDP}
Indeed, the problem of recommending personalized tour itineraries has its roots in the Operational Research community as the Tourist Trip Design Problem. The authors in~\cite{vansteenwegen-mobile:07} defined the generic TTDP as an extension of the Orienteering Problem (OP)~\cite{vansteenwegen-orienteering:11}, where the objective is to schedule an optimal path (i.e., the path that maximizes the collected profit) for the user, given the set of candidate POIs, the travel time among them (e.g., identified by the transport mode), the profit of each POI (e.g., identified by the POI popularity), the visiting duration at a POI, starting/ending points for the tour, and the user time budget. The TTDP can be further specialized depending on the number of users involved (i.e., a single traveler or group of tourists) and the number of routes to be generated (i.e., one-day or multiple-day tours). A comprehensive survey on using OP to model multiple variants of the TTDP was presented in~\cite{gavalas-survey:14}. An important characteristic of works on TTDP is the lack of individual preferences associated with individual users. Consequently, every user providing the same starting/ending points and time budget as input will be recommended for the same tour itinerary.}

{\subsection{Data-driven Recommendation} 
With the popularity of smartphones and LBSNs, several works have focused on data-driven approaches to better model user preferences when recommending POIs or itineraries. The primary goal of these works is to exploit the tourism content available on the Web to enable a deeper understanding of user preferences and to issue personalized recommendations. As illustrated in Figure~\ref{fig:classification}, the literature on data-driven recommendation can be divided into three groups: next POI recommendation, top-k POI recommendation, and personalized tour itinerary recommendation (PTIR). Works on next POI recommendation~\cite{chen-comprehensive:17,liao-multi-context:18,fan-deep:18,sassi-location:19} aims to identify the next location a tourist will likely visit based on her previous trajectories, while studies on top-k location~\cite{logesh:19,zhou-adversarial:19,rahmani-joint:20} recommend multiple POIs as part of a ranked list. Different from those works but closely related, works on PTIR recommend multiple POIs as a connected path, taking the same considerations as the TTDP but with the additional challenge of planning an itinerary that appeals to the user's interest. Next, we detail existing works on PTIR and differentiate them from our proposal.}

{\subsubsection{Existing PTIR solutions.}} 
{Aiming to recommend personalized tour itineraries for a single tourist, the authors in~\cite{choudhury-automatic:10} are one of the first to combine an optimization solution for an OP variant with the mining of users' past trajectories based on geo-tagged photos extracted from Flickr\footnote{https://www.flickr.com/}. Their approach maximizes POI popularity while keeping the user's total time budget. To achieve this goal, transit time between POI and POI visiting time duration are computed as the median and the seventy-fifth percentile of all users, respectively.}

{The work of~\cite{choudhury-automatic:10}  was further refined in~\cite{brilhante-where:13,yu-personalized:16,lim-personalized:18} by assigning categories to POIs and using them to determine the user interests. Particularly, in~\cite{brilhante-where:13} and~\cite{yu-personalized:16}, the user interest is computed based on the frequency the user visits a POI of a particular category. This approach is usually referenced in the literature as frequency-based user interest. In~\cite{lim-personalized:18}, the authors presented PersTour, which advanced the state-of-the-art by introducing the concept of time-based user interest. In this concept, the user’s level of interest in a POI category is based on her time spent at such POIs relative to the average user. PersTour also innovated by personalizing POI visit duration using this time-based user interest. The authors showed that time-based user interest and personalized visit durations reflect users' real-life tours more accurately than frequency-based tourist interest solutions and average visit duration. Thus, PersTour has been recognized as the state-of-the-art one-day, single-traveler PTIR exact solution by many works (e.g.,~\cite{zhang-encoder:24,halder-survey:24}).}  

{Different from~\cite{choudhury-automatic:10,brilhante-where:13,yu-personalized:16,lim-personalized:18} that focus on recommending a one-day tour, in~\cite{friggstad:18} and~\cite{zhong-optimization:23}, the focus is on the multiple-day case. Particularly, the authors in~\cite{friggstad:18} extended the OP and proposed a PTIR algorithm for the multiple-day case where the objective is to optimize for the value of the worst tour. User interest is mined from a private Google dataset containing historical visits to tourist attractions; the duration of a POI visit is set to the median time spent at the POI among all visits; and the transportation mode allows multiple options, using the quickest available one. The authors in~\cite{zhong-optimization:23} addressed the challenge of recommending personalized multi-day tours by considering the time windows of each POI and the different transportation modes to visit them. The proposed problem is modeled as a multi-day TTDP with time windows using data from Chongqing, a popular city in Southwest China.} 

{The PTIR problem was also studied in the context of a group of tourists. In this setting, the objective is to recommend a one-day (\cite{lim-towards:16,yin-group:19}) or multiple-day(\cite{kargar-socially:21}) tour itinerary considering the group members’ heterogeneous preferences. Next, we discuss the novelty of our work.}

{\subsubsection{+Tour positioning.}} 
{Our work focuses on an optimization approach for recommending personalized itineraries for an individual user considering a one-day tour. Hence, it is conceptually closer to the investigations performed in~\cite{choudhury-automatic:10,brilhante-where:13, yu-personalized:16,lim-personalized:18}. However, +Tour differs distinctly from existing PTIR works by considering not only POI popularity and POI category as user preferences but also new applications that will be available with advanced mobile networks. In particular, we consider the user's physical and virtual visit experience. As illustrated in Figure~\ref{fig:classification}, this distinct view requires additional considerations on ICT resource allocation and service application demands. In addition, as ICT resources are shared among tourists during their tours, to issue personalized itineraries for each user, our PTIR solution must take into account a multi-user perspective. Thus, we propose a new formulation to the PTIR problem that captures the joint problem of recommending personalized tour itineraries for multiple individual users while efficiently allocating MEC resources to enable advanced user service applications. Table~\ref{tab:related-work} summarizes the main characteristics of the existing PTIR works for a single traveler and how our work differs from them.}

\begin{table*}[t]
\centering

\footnotesize
\caption{Comparison of our work with relevant literature on PTIR for a single traveler. Time constraints and POI popularity are omitted from the table since they are considered in all PTIR works.}
\label{tab:related-work}
\begin{tabular}{ccccccc}\toprule
& &\multicolumn{5}{c}{\textbf{Considerations}} \\
\cmidrule(lr){3-7}
\textbf{Paper} & \textbf{Dataset} &
\textbf{Routing} & 
\textbf{Transport} & \textbf{User} & 
\textbf{ICT} &  \textbf{Application} 
\\
             &               &
             & 
\textbf{Mode} & \textbf{Preferences} & 
\textbf{Resources}  &  \textbf{Demands} \\
\midrule
\cite{choudhury-automatic:10} & Flickr &
one-day & 
- & POI popularity & 
- & - \\ \hline
\cite{brilhante-where:13} & Flickr & 
one-day & 
walking & POI category (frequency-based) & 
- & - \\ \hline
\cite{yu-personalized:16} & Jie Pang city & 
one-day & 
walking & POI category (frequency-based) & 
- & - \\ \hline
\cite{lim-personalized:18} & Flickr & 
one-day & 
walking & POI category (time-based) & 
- & - \\ \hline
\cite{friggstad:18} & Google's & 
multiple-day & 
multiple modes & POI popularity & 
- & - \\ \hline
\cite{zhong-optimization:23} & Chongqing city & 
multiple-day & 
multiple modes & POI popularity & 
- & - \\ \hline
our work & Flickr & 
one-day & 
walking & POI category (time-based) & 
$\checkmark$ & $\checkmark$ \\
  & & & & and applications & & \\\bottomrule
%\cite{friggstad:18} & x & x & x & x & x & x & x & x  \\ \hline
%\cite{zhong-optimization:23} & x & x & x & x & x & x & x & x  \\ \hline
\end{tabular}
\end{table*}

{In~\cite{fonseca-personlized:19}, we introduced n5GTour, a preliminary version of this work presented in a conference paper. The present work extends~\cite{fonseca-personlized:19} in many ways. First, as with many existing PTIR solutions, n5GTour assumes that the recommended itinerary contains a minimum of three POIs. +Tour extends n5GTour by relaxing this constraint, recommending itineraries with one or more POIs. As we will show in Section~\ref{sec:data-characterization}, this relaxation is reasonable since a significant number of itineraries in the real world include less than three POIs, and the literature usually neglects the visiting patterns of these itineraries. Second, different from~\cite{fonseca-personlized:19}, we present a complete description of +Tour and analyze its computational complexity. Third, we challenge our solution with a more prominent and representative dataset of 13 cities distributed among 4 continents. In contrast, in~\cite{fonseca-personlized:19}, only 4 cities in the same continent were considered. Fourth, we add a data characterization to extract the user visiting patterns of the new dataset. Supported by our data characterization, we propose a more realistic methodology to infer the applications more likely to be used by tourists based on their visit durations. In~\cite{fonseca-personlized:19}, the applications used by tourists are randomly chosen. Finally, we comprehensively evaluate our solution considering multiple cities and different resource utilization scenarios.}

%% file: Sections/03-System-Model.tex
\section{System Model}\label{sec:system-model}

\begin{figure}[t]
    \centering
    \includegraphics[width=0.5\textwidth]{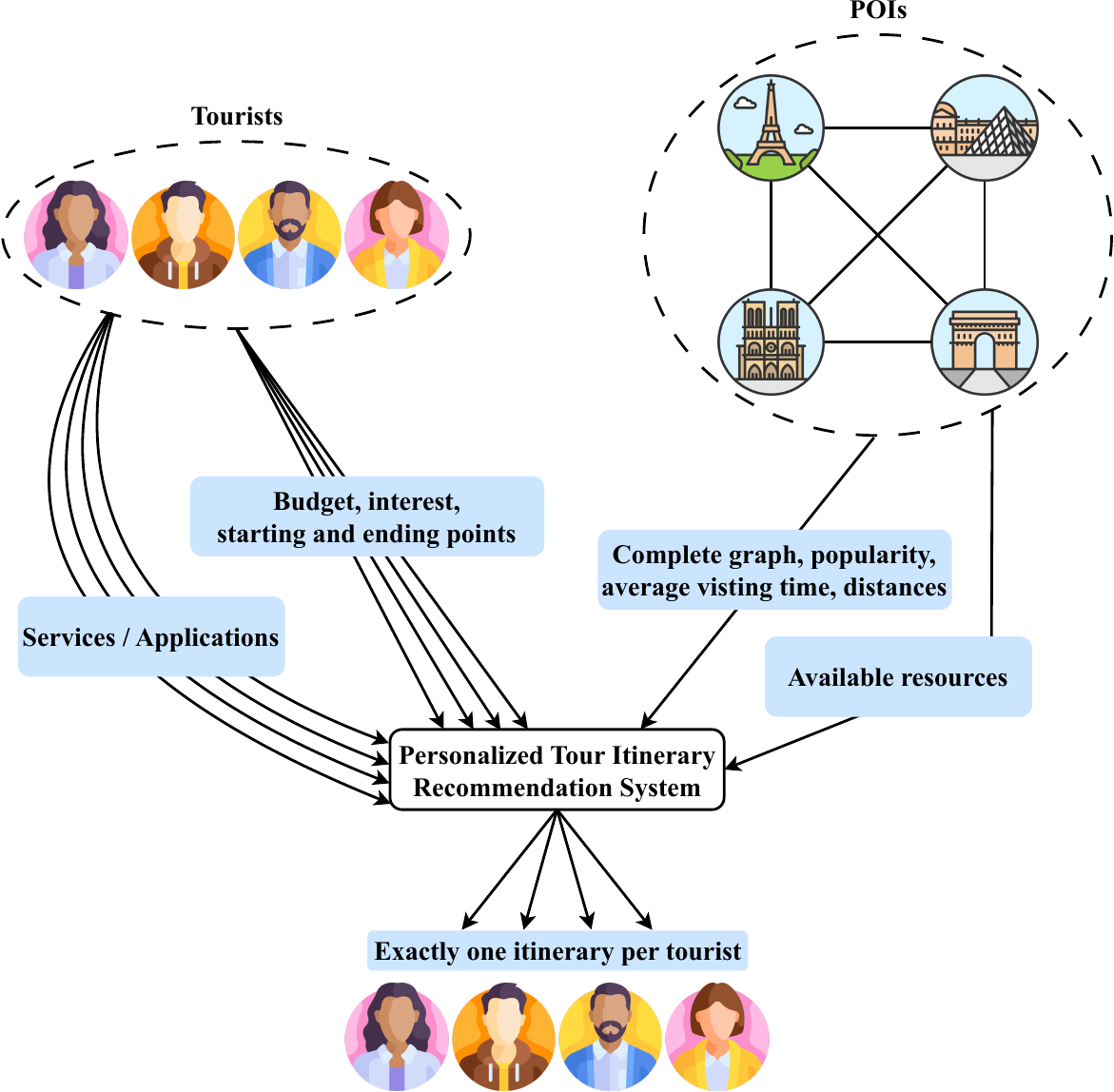}
    \caption{{System model.}}
    \label{fig:System-Model}
\end{figure}

Our tour recommendation problem takes place in a specific city with a set of POIs, the set of applications offered by the MTC to be consumed during the tour, the ICT infrastructure to support the tourist virtual experience, and a set of tourists. In the next subsections, we describe the system model{, which is depicted by Figure~\ref{fig:System-Model},} in detail. Table~\ref{tab:notation} summarizes the notations and definitions used in this work.

\begin{table}[htb]
\centering
\footnotesize

{\caption[skip=0pt]{Notations and definitions used in the system model, problem formulation, and solution.}}
\label{tab:notation}
\begin{tabular}{rlp{60mm}}
    \hline
    & \textbf{Symbol} & \textbf{Description}  \\
    \hline
    {\multirow{10}{*}{\rotatebox[origin=r]{90}{Sets}}}
    & $\mathcal{V}$ & Set of POIs (in a city) \\
    & $\mathcal{E}$ & Set of edges connecting vertices\\
    & $\mathcal{C}$ & Set of all POI categories \\
    & $\mathcal{U}$ & Set of users (tourists) \\
    & $\mathcal{U}_d$ & Set of users whose tours are scheduled to the day $d$\\
    & $\mathcal{M}$ & Set of MEC hosts \\
    & $\mathcal{A}$ & Set of applications offered by the MTC \\
    & $\mathcal{A}_{u}$ & Set of applications user $u$ is willing to use \\
    & $\mathcal{S}$ & Set of all travel histories \\
    & $S_u$ & Travel history of user $u$ \\
    \hline
    {\multirow{6}{*}{\rotatebox[origin=r]{90}{Items}}}
    & $n$ & Number of POIs in a city \\
    & $v_0$ & Virtual POI where every itinerary begins \\
    & $v_{n+1}$ & Virtual POI where every itinerary ends \\
    & $v_i$ & A POI in $\mathcal{V}$\\
    & $u$ & A user in $\mathcal{U}$ \\
   & $\tau$ & Index of the time slot where MEC infrastructure resources are allocated \\
    \hline
    {\multirow{8}{*}{\rotatebox[origin=r]{90}{POI/MEC input}}}
    & $c_{i, j}$ & Travel time between POI $v_{i}$ to POI $v_{j}$ \\
    & $pop(v_{i})$ & Popularity of POI $v_i$ \\
    & $dur(v_{i})$ & Expected visiting time of POI $v_i$ \\
    & $cat(v_{i})$ & Category of POI $v_i$ \\
    & $lat(v_{i})$ & Latitude of POI $v_i$ \\
    & $long(v_{i})$ & Longitude of POI $v_i$ \\
    & $\lambda({v_{i}})$ & Total network resource available at POI $v_{i}$ \\
    & ${\psi}_{m}$ & Total computing resource available at MEC $m$ \\
    \hline
    {\multirow{7}{*}{\rotatebox[origin=r]{90}{Application input}}}
    & $\lambda_{a}^{min}$ & Minimum network demand, in bps, of application $a \in \mathcal{A}$ \\
    & $\lambda_{a}^{max}$ & Maximum network demand, in bps, of application $a \in \mathcal{A}$ \\
    & $\psi_{a}^{min}$ & Minimum computing demand, in reference cores, of application $a \in \mathcal{A}$ \\
    & $\psi_{a}^{max}$ & Maximum computing demand, in reference cores, of application $a \in \mathcal{A}$ \\
    \hline
    {\multirow{9}{*}{\rotatebox[origin=r]{90}{User input}}}
    & $int_u(c)$ & Interest of user $u$ in the POI category $c$ \\
    & $b_u$ & Time budget of user $u$ to complete the tour\\
    & $d_u$ & Date when the tour of user $u$ will take place \\
    & $\lambda_{u}^{min}$ & Minimum network demand of user $u$ during a tour \\
    & $\lambda_{u}^{max}$ & Maximum network demand of user $u$ during a tour \\
    & $\psi_{u}^{min}$ & Minimum computing demand of user $u$ during a tour \\ 
    & $\psi_{u}^{max}$ & Maximum computing demand of user $u$ during a tour \\
\hline
\end{tabular} 
\end{table}

\subsection{Main elements of the model}
\label{sec:model-1}

\subsubsection{\textbf{POIs}}
For a city with $n$ POIs, we consider a complete non-oriented graph $G = (\mathcal{V},\mathcal{E})$, with $\mathcal{V} = \{v_{0}, v_{n+1}\} \cup \{v_{1}, \dots, v_{n}\}$ being the set of vertices representing the POIs and $\mathcal{E} = \{(v_{i},v_{j}) \mid v_{i}, v_{j} \in \mathcal{V}\}$ is the set of edges connecting the nodes. $v_{0} \in \mathcal{V}$ and $v_{n+1} \in \mathcal{V}$ are virtual POIs representing, respectively, the starting and ending location of the users' tours. Each edge $(v_{i},v_{j}) \in \mathcal{E}$ is associated with a cost $c_{i,j}$ representing the travel time between vertex $v_{i}$ to vertex $v_{j}$, employing a given mode of transportation. Since $v_{0}$ and $v_{n+1}$ are virtual POIs, we model $c_{0,j}=c_{i,n+1}=0$ $\forall \ i=0,\dots,n$ and $\forall \ j=1,\dots,n+1$. 

Each POI $v_{i} \in \mathcal{V} \setminus \{v_{0}, v_{n+1}\}$ is characterized by the following attributes: the popularity of the POI, denoted by $pop(v_{i}) \in \mathbb{Z}$; the expected time one should spend in the POI to enjoy what it has to offer (i.e., expected POI visiting time), denoted by $dur(v_{i}) \in \mathbb{R}$; the category representing its nature, denoted by $cat(v_{i})$; and the location expressed in terms of latitude and longitude, denoted by $lat(v_{i})$ and $long(v_{i})$ respectively. We denote by $\mathcal{C} = \{1, \dots, C$\} {as} the set of all POI categories so that $cat(v_{i}) \in \mathcal{C}, \ \forall \ v_{i} \in \mathcal{V} \setminus \{v_{0}, v_{n+1}\}$.

\subsubsection{{\textbf{Service Applications}}}

The MTC offers a set $\mathcal{A} = \{1, \dots, A\}$ of applications to enrich the touristic experience in every POI. Each application $a \in \mathcal{A}$ has specific requirements in terms of network and computing resources, expressed as:

\begin{itemize}
    \item $\lambda^{min}_{a}$ and $\lambda^{max}_{a}$: the minimum and maximum network demand (i.e., demanded traffic volume) measured in bps, respectively;
    \item $\psi^{min}_{a}$ and $\psi^{max}_{a}$: the minimum and maximum computing demand (i.e., demanded processing load) measured in reference core (RC)\footnote{Unit of measure that represents the processing capacity of a reference CPU core.}, respectively.
\end{itemize}

\subsubsection{{\textbf{Advanced mobile networks ICT Infrastructure}}}

In each POI $v_{i} \in \mathcal{V} \setminus \{v_{0}, v_{n+1}\}$, the network resources are provided through a set of {wireless base stations, e.g., gNBs (gNodeBs) in a 5G network}. To keep the model generic, we decide to represent network resources simply by the capacity, i.e., not considering the wireless channel aspects. The whole model's complexity and the time scale of the problem also suggest that this {simplification} is a reasonable approach. We denote by $\lambda({v_{i}})$ the total network resource available at POI $v_{i} \in \mathcal{V} \setminus \{v_{0}, v_{n+1}\}$. 

The computing resources are provided by MEC hosts that are reachable from the access network, as shown in Figure~\ref{fig:ICT-Infrastructure}. We assume that a POI $v_{i} \in \mathcal{V} \setminus \{v_{0}, v_{n+1}\}$ can be served by any MEC host. Let $\mathcal{M} = \{1, \dots, M\}$ denote the set of MEC hosts. We represent by ${\psi}_{m}$ the total computing capacity available at MEC host $m \in \mathcal{M}$.

Applications should run on MEC hosts whenever possible since this results in shorter response times and improved virtual experience. When no resource is available on the edge, applications run in the remote cloud with degraded performance regarding communication delay. {In traditional offload literature, the user device can also run the applications completely. We assume this is not the case in our context because the MTC does not want to deal with issues such as regular updates in the software of the user devices, hardware incompatibility, and security breaches, among others. These issues can be minimized or solved by keeping the client software minimalist while the rest of the application runs remotely on the edge or in the cloud. We argue that this approach is reasonable for an enterprise scenario such as the one of the MTC.} 

\subsubsection{\textbf{Tourists}}
We consider a set $\mathcal{U} = \{1, \dots, U\}$ of users (tourists) where each user $u \in \mathcal{U}$ is described by the following attributes: the user's interest in each POI category $c \in \mathcal{C}$, denoted by $int_{u}(c) \in \mathbb{R}$; the preferred day for the tour, denoted by $d_u$; and the time budget to complete the tour, i.e., the maximum available time for visits,  represented by $b_u \in \mathbb{R}$. The user is also characterized by the set $\mathcal{A}_{u} \in 2^{\mathcal{A}}$, representing the applications she is willing to use during the tour, where $2^{\mathcal{A}}$ is the power set of $\mathcal{A}$. The set $\mathcal{A}_{u}$ determines the minimum and maximum demand for network and computing resources required to run the user applications during the tour. We represent this demand as described in the following:
\begin{itemize}
    \item $\lambda_{u}^{min} = \sum\limits_{a \in \mathcal{A}_{u}} \lambda^{min}_{a}$: the minimum network demand, in bps, required by user $u$ during a tour;
    \item $\lambda_{u}^{max} = \sum\limits_{a \in \mathcal{A}_{u}} \lambda^{max}_{a}$: the maximum network demand, in bps, required by user $u$ during a tour;
    \item $\psi_{u}^{min} = \sum\limits_{a \in \mathcal{A}_{u}} \psi^{min}_{a}$: the minimum computing demand, in reference core (RC), required by user $u$ during a tour; 
    \item $\psi_{u}^{max} = \sum\limits_{a \in \mathcal{A}_{u}} \psi^{max}_{a}$: the maximum computing demand, in reference core (RC), required by user $u$ during a tour.
\end{itemize}

Finally, by modeling the starting and ending location of all users' tour itineraries as virtual POIs, we allow an itinerary to involve one or more real-world POIs.

%Previous work on PTIR algorithms~\cite{lim-personalized:18, fonseca-personlized:19} model the starting location and the ending location of the user's tour as two different POIs in the city. As a consequence of such modeling, a tour itinerary must contain at least two POI visits. However, this is not inline with many real-world visiting scenarios, such as parks or museums, where visits last longer and, after visiting such places, the tourists may want to return to their accommodations. We observed this behaviour in our datasets. Although representative, these visiting scenarios with one POI visit are often neglected by previous works. To avoid this limitation, in this work, we model the starting and ending location of all users' tour itineraries as, respectively, the virtual POIs $v_{0} \in \mathcal{V}$ and $v_{n+1} \in \mathcal{V}$, allowing an itinerary to be a path involving one or more real-world POIs.

\subsection{Visiting-related {elements of the Model}}
\label{sec:model-2}
Similar to~\cite{choudhury-automatic:10,brilhante-where:13,lim-personalized:18}, we define POI popularity, expected POI visiting time, and user's interest based on users' past travel histories extracted from LBSNs. Given the user $u \in \mathcal{U}$ who has visited $r$ POIs in a city, we define her travel history as an ordered sequence $S_u= ((v_1, t_{v_1}^a, t_{v_1}^d), \dots, (v_r, t_{v_r}^a, t_{v_r}^d))$, where each triple $(v_l, t_{v_l}^a, t_{v_l}^d),$ $\ l=1,\dots, r$, represents a visit at POI $v_l$, the arrival time $t_{v_l}^a$ at POI $v_l$, and the departure time $t_{v_l}^d$ from POI $v_l$, $v_{l} \in \mathcal{V} \setminus \{v_{0}, v_{n+1}\}$. The user's visit duration at a POI $v_{l} \in \mathcal{V} \setminus \{v_{0}, v_{n+1}\}$ is computed as $t_{v_l}^d - t_{v_l}^a$.

Given a set $\mathcal{S}$ of all travel histories in a city, i.e., $\mathcal{S}=\bigcup_{u \in \mathcal{U}} S_u$, the popularity of a POI $v_{i} \in \mathcal{V} \setminus \{v_{0}, v_{n+1}\}$ is defined based on the number of times $v_i$ has been visited:%, i.e.: 
\begin{equation}
\label{eq:pop}
pop(v_{i}) = \sum\limits_{S_u \in \mathcal{S}}\sum\limits_{v_{x} \in S_u} \delta(v_{x},v_{i}),
\end{equation}
where $\delta(v_{x},v_{i})=$~1 if $v_{x}=v_{i}$ or $\delta(v_{x},v_{i})=$~0 otherwise.

The expected (average) visiting time of a POI $v_i \in \mathcal{V} \setminus \{v_{0}, v_{n+1}\}$ is defined as follows:
\begin{equation}
\label{eq:dur}
dur(v_{i}) = \frac{\sum\limits_{S_u \in \mathcal{S}} \sum\limits_{v_{x} \in S_u} (t_{v_x}^d - t_{v_x}^a)\delta(v_{x},v_{i})}{pop(v_{i})},
\end{equation}

Similar to~\cite{lim-personalized:18}, we assume a time-based user interest approach where the user's interest in a certain category $c$ is based on the time she spent at a POI of that category relative to the expected visiting time at that POI. The intuition is to determine the user's interest level in a certain category by computing the time she spent at POIs of that category compared to other users. Thus, we define the interest level of a user $u \in \mathcal{U}$ in a category $c \in \mathcal{C}$ as:
\begin{equation}
\label{eq:intu}
     int_u(c) = \sum\limits_{v_{x} \in S_u} \frac{(t_{v_x}^d - t_{v_x}^a)}{dur(v_{x})} \gamma(cat(v_x),c), \\
\end{equation}
where $\gamma(cat(v_{x}),c)=$~1 if $cat(v_x)=c$, or $\gamma(cat(v_{x}),c)=$~0 if $cat(v_x) \neq c$.

\subsection{{Cost element associated with a POI visit of the Model}}
\label{sec:model-3}

The capability of computing the user's interest based on the time-based approach using past travel histories renders an effective solution to personalize the recommended time for a tourist to spend at a certain POI. Formally, given a POI $v_{i} \in \mathcal{V} \setminus \{v_{0}, v_{n+1}\}$ of category $cat(v_i) \in \mathcal{C}$, we can recommend user $u$ to spend the following time at POI $v_{i}$: $int_u(cat(v_i))dur(v_{i})$. 
This time is based on the user's interest level in category $cat(v_i)$ multiplied by the average time spent at POI $v_i$. The rationale for this equation is that the greater the user's interest in the category $cat(v_i)$, the more time she will spend on POI $v_i$ compared to the average user. 

We define the profit perceived by user $u$ when physically visiting a POI $v_i \in \mathcal{V} \setminus \{v_{0}, v_{n+1}\}$ as: 
\begin{equation}
\label{eq:prof_vi}
Prof_{u}(v_i) = \alpha int_u(cat(v_i)) + (1 - \alpha)pop(v_{i}),
\end{equation}

Equation (\ref{eq:prof_vi}) depends on the POI popularity and the user's interest in the POI category. It also depends on the $\alpha \in [0,1]$ value, which can be chosen to emphasize the user's interest or the POI popularity. We also define $Prof_{u}(v_0)=Prof_{u}(v_{n+1})=$~0.

Traveling from POI $v_i$ to POI $v_j$ and visiting $v_j$ consumes the user's budget. Thus, we can define a cost function associated with a POI visit as:
\begin{equation}
\label{eq:cost_vi}
\begin{split}
& Cost_{u}(v_i,v_j) = c_{i,j} + int_u(cat(v_j))dur(v_{j}), \\ 
& {i=0,\dots, n}; {\quad j=1,\dots,n}.
\end{split}
\end{equation}

We also define $Cost_{u}(v_i,v_{n+1})=$~0, $\forall \ i=0,\dots, n$.

%% file: Sections/04-Problem-Formulation.tex
\section{Problem Formulation}\label{sec:problem-formulation}

Given the graph $G=(\mathcal{V},\mathcal{E})$ representing the POIs of the city, the set of users willing to visit the city in a specific day ($\mathcal{U}_d=\{u \in \mathcal{U} \mid d_u=d\}$), the users' preferences ($int_{u}(c), \ \forall \ c \in \mathcal{C}, \ \forall \ u \in \mathcal{U}_d$), the users' time constraints ($b_u, \ \forall \ u \in \mathcal{U}_d$), and the applications the users will consume during the tour ($\mathcal{A}_{u} \in 2^{\mathcal{A}}, \ \forall \ u \in \mathcal{U}_d$), our objective is to find a single itinerary $I_{u*}=(v_0,\dots,v_{n+1})$, starting at $v_0$ and ending at $v_{n+1}$, for each user $u \in \mathcal{U}_d$ so that the set of chosen itineraries, denoted by $\mathcal{I_{*}}$, maximizes the sum of the profits perceived by all the users while prioritizing the MEC resource allocation and satisfying the users and the infrastructure constraints. We named such a problem the MEC-aware Personalized Tour Itinerary Recommendation (MEC-PTIR) problem. Indeed, the MEC-PTIR problem adds the availability of ICT resources and the demand for the chosen applications to the classical PTIR problem. Therefore, the MEC-PTIR problem is more complex than the PTIR one. Since the PTIR problem is NP-hard~\cite{gavalas-survey:14}, we formulate the MEC-PTIR problem as a two-stage optimization problem to address the complexity efficiently.

\subsection{MEC-PTIR problem - First stage}
\label{sec:problem-formulation-first-stage}
We define the first stage of the MEC-PTIR problem as a multi-objective orienteering problem and its integer problem formulation. Formally, for each itinerary $I_u=(v_0, \dots, v_{n+1})$, $u \in \mathcal{U}_d$, let a decision variable $x_{i,j} \in \{0, 1\}$ indicates whether POI $v_i \in \mathcal{V}$ and POI $v_j \in \mathcal{V}$ are visited in sequence in $I_u$ or not. 
The objective of the first stage of the MEC-PTIR problem is to obtain, for each $u \in \mathcal{U}_d$, the set $\mathcal{I}_{u*} = \{I_{u*}^{1}, \dots, I_{u*}^{k}\}$ of itineraries such that the returned physical profit is maximized while the travel cost is minimized, i.e.:\\
\\
\textbf{Stage 1 (MEC-PTIR problem):} $\forall u \in \mathcal{U}_d$
\begin{align}
& {\mathrm{maximize}} \ \sum_{i=0}^{n} \sum_{j=1}^{n+1} x_{i,j}Prof_{u}(v_i)\label{c:profit}&\\
& {\mathrm{minimize}} \ \sum_{i=0}^{n} \sum_{j=1}^{n+1} x_{i,j}Cost_{u}(v_i,v_j) \label{c:cost}&
\end{align}
\begin{align}
& \text{subject to:} \nonumber&\\
& 0 < \sum_{i=0}^{n} \sum_{j=1}^{n+1} Cost_{u}(v_i,v_j)x_{i,j} \leq b_u,\label{c:budget}&\\
&\sum_{j=1}^{n+1} x_{0,j} = \sum_{i=0}^{n} x_{i,n+1} = 1,&\label{c:itinerary}\\
&\sum_{i=0}^{n} x_{i,r} = \sum_{j=1}^{n+1} x_{r,j} \leq 1,  \forall \ r = 1, ..., n, &\label{c:conn_no-rep}\\
& 2 \leq pos(v_i) \leq n + 1, \forall \ i = 1, ..., n+1, &\label{c:no_subtour1}\\
& pos(v_i) - pos(v_j) + 1 \leq n(1 - x_{i,j}), \nonumber& \\
& \qquad\qquad\qquad\qquad\qquad \  \forall \ i,j = 1, ..., n+1.\label{c:no_subtour2}
\end{align}

Equation (\ref{c:profit}) aims to maximize the total collected profit of user $u$ while Equation (\ref{c:cost}) minimizes the travel cost adhering to the user's budget (Constraint~(\ref{c:budget})). The other constraints ensure the proper construction of the itinerary. Constraint~(\ref{c:itinerary}) ensures that the path starts at POI $v_0$ and ends at POI $v_{n+1}$; Constraint~(\ref{c:conn_no-rep}) ensures that the path is connected and no POI is visited more than once. Assuming that $pos(v_i)$ is the position of POI $v_i \in \mathcal{V}$ in itinerary $I_u$, Constraints~(\ref{c:no_subtour1}) and (\ref{c:no_subtour2}) ensure that there are no sub-tours.

For a user $u \in \mathcal{U}_d$, the set $\mathcal{I}_{u*}$ contains all Pareto-efficient itineraries that start at vertex $v_0$, end at vertex $v_{n+1}$, and do not violate the user's time constraint. A solution $I_{u*}^{k} \in \mathcal{I}_{u*}$ is called a Pareto-efficient (or non-dominated) if there is no feasible solution $I_{u*}^{k'}$ that dominates $I_{u*}^{k}$, i.e., that is at least equally as good as $I_{u*}^{k}$ concerning all objective functions, and better than $I_{u*}^{k}$ concerning at least one objective function. The set $\mathcal{I}_{u*}$ is called the Pareto front.  

\subsection{MEC-PTIR problem - Second stage}
\label{sec:problem-formulation-second-stage}
The result of the first stage of the MEC-PTIR problem is the set of Pareto fronts $\mathcal{I} = \bigcup_{u \in \mathcal{U}_d} \mathcal{I}_{u*}$, i.e., the set of all non-dominated itineraries that do not violate the user's time constraint for all users in $\mathcal{U}_d$. Each itinerary $I_{u*}^k \in \mathcal{I}_{u*}$, $\forall \ \mathcal{I}_{u*} \subset \mathcal{I}$, has a physical profit given by:
\begin{equation}
\label{eq:profit}
Prof(I_{u*}^k) = \sum_{v_i \in I_{u*}^k \setminus \{v_{0}, v_{n+1}\}} Prof_{u}(v_i).
\end{equation}

Given the set $\mathcal{I}$, the objective of the second stage of the MEC-PTIR is to find a set $\mathcal{I_{*}}$ that maximizes the sum of the profits perceived by the users while prioritizing the resource allocation at the network edge. In addition, $\mathcal{I_{*}}$ must contain exactly one itinerary for each user $u \in \mathcal{U}_d$.
To achieve this goal, we formulate the second stage of the MEC-PTIR as a Mixed Integer Linear Programming problem.

Let the indicator function $\phi(I_{u*}^k , v_i) \in \{0,1\}$ represents POI visit in an itinerary, with $\phi(I_{u*}^k , v_i)=$~1 if itinerary $I_{u*}^k \in \mathcal{I}$ visits POI $v_{i} \in \mathcal{V} \setminus \{v_{0}, v_{n+1}\}$, and $\phi(I^{j}_{u*}, v_i)=$~0 otherwise. 

Assume that every itinerary in $\mathcal{I}$ starts and ends within a period of time $T$ and that the computing and network resources for these itineraries are allocated in discrete time slots of size $\Delta t$. The time slots are indexed by $\tau \in \mathbb{Z}$ such that $ 1 \leq \tau \leq T$. Given an itinerary $I_{u*}^k \in \mathcal{I}_{u*}$, a POI $v_{i} \in \mathcal{V} \setminus \{v_{0}, v_{n+1}\}$, and a time index $\tau$, we define the indicator function $\rho(I_{u*}^k, v_{i}, \tau) \in \{0, 1\}$, with $\rho(I_{u*}^k, v_{i}, \tau)=$~1 if itinerary $I_{u*}^k$ is visiting POI $v_{i}$ during $\tau$, and $\rho(I_{u*}^k, v_{i}, \tau)=$~0 otherwise. 
%This information can be derived from the start time, sequence of POI visits, and POI visit duration for each itinerary $I_{u*}^k \in \mathcal{I}_{u*}$ obtained from the first stage. 

Let the set of decision variables $y(I_{u*}^k) \in \{0, 1\}$ represents itinerary choices so that $y(I_{u*}^k)=$~1 if itinerary $I_{u*}^k \in \mathcal{I}_{u*}$ composes the solution, and $y(I_{u*}^k)=$~0 otherwise. A MEC host should provide computing resources for each POI visit for each user. We define the decision variable $z(I_{u*}^k,v_{i}, m) \in \{0, 1\}$ for representing MEC host association so that $ z(I_{u*}^k,v_{i},m)=$~1 if MEC host $m \in \mathcal{M}$ is responsible for providing computing resources during a visit to POI $v_{i} \in \mathcal{V} \setminus \{v_{0}, v_{n+1}\}$ in itinerary $I_{u*}^k \in \mathcal{I}_{u*}$, and $z(I_{u*}^k,v_{i},m)=$~0 otherwise. 

Let the decision variables $p(I_{u*}^k,v_{i}) \in \mathbb{R}$ and $q(I_{u*}^k,v_{i},m) \in \mathbb{R}$ represent, respectively, the amount of network and the amount of computing resources (at MEC host $m \in \mathcal{M}$) allocated during a visit to POI $v_{i} \in \mathcal{V} \setminus \{v_{0}, v_{n+1}\}$ in itinerary $I_{u*}^k \in \mathcal{I}_{u*}$. Assuming $Norm(value)$ as a generic function that normalizes a value, we define the objective function of the second stage of the MEC-PTIR problem as:\\
\\
\textbf{Stage 2 (MEC-PTIR):}
\begin{equation} \label{eq:stage2_summary}
\begin{split}
& \mathrm{maximize} \sum\limits_{\mathcal{I}_{u*} \subset \mathcal{I}}\sum\limits_{I_{u*}^k \in \mathcal{I}_{u*}}      y(I_{u*}^k)Norm(Prof(I_{u*}^k))\ +&\\
& \sum\limits_{\mathcal{I}_{u*} \subset \mathcal{I}}\sum\limits_{I_{u*}^k \in \mathcal{I}_{u*}}\sum\limits_{v_i \in \mathcal{V} \setminus \{v_{0}, v_{n+1}\}} \frac{Norm(p(I_{u*}^k,v_{i}))}{2|I_{u*}^k \setminus \{v_{0}, v_{n+1}\}|} \ +&\\
& \sum\limits_{\mathcal{I}_{u*} \subset \mathcal{I}}\sum\limits_{I_{u*}^k \in \mathcal{I}_{u*}}\sum\limits_{v_i \in \mathcal{V} \setminus \{v_{0}, v_{n+1}\}}\sum\limits_{m \in \mathcal{M}} \frac{Norm(q(I_{u*}^k,v_{i},m))}{2|I_{u*}^k \setminus \{v_{0}, v_{n+1}\}|}. &
\end{split}
\end{equation}

Equation~(\ref{eq:stage2_summary}) aims to maximize simultaneously two objectives: (i) the aggregated total collected physical profit of all users in $\mathcal{U}_d$; and (ii) the aggregated total collected virtual profit of all users in $\mathcal{U}_d$ given by the sum of the amount of allocated network resources and the amount of allocated computing resources at the network edge. Together, these objectives ensure that the generated set of itineraries maximizes the sum of the physical profits perceived by the users while prioritizing the resource allocation at the network edge. The $Norm(value)$ function ensures that both objectives stay in the same interval of values. The aggregated physical profit and the aggregated virtual profit have equal weights. Thus, Equation (\ref{eq:stage2_summary}) may choose itineraries that do not have the highest individual physical profit but provide the best balance with the resource allocation, resulting in an improved experience for the set of users as a whole. The objective function represented by Equation~(\ref{eq:stage2_summary}) is subject to the following constraints.

\textbf{Itinerary choice constraints} -- For each user $u \in \mathcal{U}_d$, exactly one itinerary $I_{u*}^k \in \mathcal{I}_{u*}$ must be {selected}, i.e.:
\begin{equation}
\label{eq:one_itinerary_1}
\sum\limits_{I_{u*}^k \in \mathcal{I}_{u*}} y(I_{u*}^k)=1, {\quad \forall \ \mathcal{I}_{u*} \subset \mathcal{I}}.
\end{equation}
%\begin{equation}
%\label{eq:one_itinerary_2}
%y(I_{u*}^k) \in \{0, 1\},
%{\quad \forall I_{u*}^k \in \mathcal{I}_{u*}}, {\quad \forall \mathcal{I}_{u*} \subset \mathcal{I}}.
%\end{equation}

If an itinerary $I_{u*}^k \in \mathcal{I}_{u*}$ is {selected} to compose a solution for a user $u \in \mathcal{U}_d$, we also need to allocate one MEC host to each POI visited in the itinerary, i.e.:
\begin{equation}
\label{eq:mec_itinerary_1}
\begin{split}
& \sum\limits_{m \in \mathcal{M}} z(I_{u*}^k,v_{i},m)=y(I_{u*}^k) \phi(I_{u*}^k, v_{i}), \\
& {\forall \ I_{u*}^k \in \mathcal{I}_{u*}}, {\quad \forall \ \mathcal{I}_{u*} \subset \mathcal{I}}, {\quad \forall \ v_{i} \in \mathcal{V} \setminus \{v_{0}, v_{n+1}\}}.
\end{split}
\end{equation}
%\begin{equation}
%\label{eq:mec_itinerary_2}
%\begin{split}
%z(I_{u*}^k,v_{i},m) \in \{0,1\}, {\quad \forall I_{u*}^k \in \mathcal{I}_{u*}}, {\quad \forall \mathcal{I}_{u*} \subset \mathcal{I}}, {\quad \forall v_{i} \in \mathcal{V}}.
%\end{split}
%\end{equation}  

\textbf{Service demand constraints} -- For each user $u \in \mathcal{U}_d$ and {selected} itinerary $I_{u*}^k \in \mathcal{I}_{u*}$, we also need to select at least the minimum of resources required to run the applications {selected} by user $u$, avoiding allocating more than is needed, i.e.:
\begin{equation}
\label{eq:service_demand_net}
\begin{split}
& p(I_{u*}^k,v_{i}) \geq \lambda_{u}^{min}y(I_{u*}^k)\phi(I_{u*}^k, v_{i}) \ \land \\
& p(I_{u*}^k,v_{i}) \leq \lambda_{u}^{max}y(I_{u*}^k)\phi(I_{u*}^k, v_{i}), \\
& {\forall \ I_{u*}^k \in \mathcal{I}_{u*}}, {\quad \forall \ \mathcal{I}_{u*} \subset \mathcal{I}}, {\quad \forall \ v_{i} \in \mathcal{V} \setminus \{v_{0}, v_{n+1}\}}.
\end{split}
\end{equation}  

Similarly, we ensure proper allocation of computing resources at the network edge with the following constraint:
\begin{equation}
\label{eq:service_demand_mec}
\begin{split}
& q(I_{u*}^k,v_{i},m) \geq \psi_{u}^{min}z(I_{u*}^k,v_{i},m)\phi(I_{u*}^k, v_{i}) \ \land \\
& q(I_{u*}^k,v_{i},m) \leq \psi_{u}^{max}z(I_{u*}^k,v_{i},m)\phi(I_{u*}^k, v_{i}), \\
& {\forall I_{u*}^k \in \mathcal{I}_{u*}}, {\ \forall \mathcal{I}_{u*} \subset \mathcal{I}}, {\ \forall v_{i} \in \mathcal{V} \setminus \{v_{0}, v_{n+1}\}}, {\ \forall m \in \mathcal{M}}.
\end{split}
\end{equation}

\textbf{Resource capacity constraints} -- Finally, we assure that the performed allocations do not exceed the amount of available network and computing resources, at any given time slot:
\begin{equation}
\label{eq:bs_capacity}
\begin{split}
& \sum\limits_{I_{u*}^k \in \mathcal{I}_{u*}} \rho(I_{u*}^k, v_{i}, \tau)p(I_{u*}^k,v_{i}){\leq \lambda_{v_{i}},} \\
& \qquad {\forall \ v_{i} \in \mathcal{V} \setminus \{v_{0}, v_{n+1}\},} {\quad 1 \leq \tau \leq T}.
\end{split}
\end{equation}
\begin{equation}
\label{eq:mec_capacity}
\begin{split}
&\sum\limits_{I_{u*}^k \in \mathcal{I}_{u*}} \sum\limits_{v_{i} \in \mathcal{V} \setminus \{v_{0}, v_{n+1}\}} \rho(I_{u*}^k, v_{i}, \tau) q(I_{u*}^k,v_{i},m){\leq {\psi}_{m},}\\
& \qquad\quad {\forall \ m \in \mathcal{M},} {\quad 1 \leq \tau \leq T}.
\end{split}
\end{equation}

We describe our proposed solution in the next section.
%To solve the MEC-PTIR problem, we propose the next generation of Tourist services assisted by modern mobile networks (+Tour) algorithm, which proceeds in two {stages}. In the first {stage}, +Tour solves the first {part} of the MEC-PTIR problem using an approach based on Dynamic Programming for the Shortest Path Problem with Resource Constraints (SPPRC)~\cite{irnich-shortest:05}. In the second {stage}, +Tour solves the second stage of the MEC-PTIR problem. This solution can be obtained straightforwardly from a linear optimization solver {such as Gurobi, IBM ILOG CPLEX, or SCIP~\cite{meindl2012analysis}.}

%% file: Sections/05-Solution.tex
\section{+Tour Algorithm}\label{sec:solution}

To solve the MEC-PTIR problem, we propose the next-generation of Touristic services assisted by modern mobile networks (+Tour) algorithm described in Algorithm~\ref{alg:rawpir}. As described in Section~\ref{sec:problem-formulation-first-stage}, the first stage of the MEC-PTIR problem is a multi-objective orienteering problem with two conflicting objectives: maximizing the total collected profit and minimizing the travel cost. This problem belongs to the class of NP-hard problems~\cite{feillet-tsp:05}. To solve this part of the problem optimally and efficiently, +Tour uses a variant of the Shortest Path Problem with Resource Constraints (SPPRC)~\cite{irnich-shortest:05}, summarized in Algorithm~\ref{alg:rptimt-5g_stage1}. The second stage of the MEC-PTIR problem is a Mixed Integer Linear Programming problem, which can be obtained straightforwardly from a linear optimization solver such as Gurobi, IBM ILOG CPLEX, or SCIP~\cite{meindl2012analysis}. In the following, Section~\ref{sec:alg-first-stage} describes how we use the SPPRC variant to solve the MEC-PTIR problem's first stage, while Section~\ref{sec:complexity-analysis} provides a complexity analysis of +Tour.

\begin{algorithm} [!ht]
\label{alg:rawpir}

\caption{The +Tour Algorithm}
\DontPrintSemicolon
\small
\SetKwInOut{Input}{Input}
\SetKwInOut{Output}{Output}

\Input{$G = (\mathcal{V},\mathcal{E})$; $\mathcal{U}_d$, $\alpha$; and $int_{u}(c)$, $b_u$, $\mathcal{A}_{u} \in 2^{\mathcal{A}}$, $\ \forall \ c \in \mathcal{C}, \ \forall \ u \in \mathcal{U}_d$; and $pop(v_{i}), dur(v_{i}), cat(v_{i})$ $\forall v_i \in \mathcal{V} \setminus \{v_{0}, v_{n+1}\}$}
\Output{$\mathcal{I}_*$}
\BlankLine
\BlankLine
 $\mathcal{I} \gets \emptyset$ \\
\ForAll{$u \in \mathcal{U}_d$}
{
    \tcp{Stage 1}
    $\mathcal{I}_{u*} \gets$ $\textbf{Algorithm 2}$ ($G = (\mathcal{V},\mathcal{E})$; $\alpha$;\\ 
    \nonl ($pop(v_{i}), dur(v_{i}), cat(v_{i})$) $\forall v_i \in \mathcal{V} \setminus \{v_{0}, v_{n+1}\}$; \\
    \nonl (($int_{u}(c)$ $\forall c \in \mathcal{C}$), $b_u$)) \\
    $\mathcal{I} \gets$ $\mathcal{I} \cup \mathcal{I}_{u*}$ \\ 
    
}
\tcp{Stage 2}
Find $\mathcal{I}_*$ by solving Stage 2 (MEC-PTIR) using $\mathcal{I}$ as input \\
\Return{$\mathcal{I}_*$}
\end{algorithm}

\subsection{Solution for the first stage} 
\label{sec:alg-first-stage}
The SPPRC finds the shortest path among all paths that start from a source node, end at a destination node, and satisfy a set of constraints defined over a set of resources. In this context, a resource corresponds to a quantity, for example, time or money, which varies along a path according to resource extension functions. A resource extension function is defined for every edge in the graph and every resource involved in the problem. Each function provides a lower bound on its corresponding resource related to the next vertex, given the value accumulated in the present vertex. The resource constraints are given as intervals, also known as resource windows, which limit the values that can be taken by the resources at every vertex along a path. These constraints are defined for every vertex and every considered resource. If multiple resources are involved in the problem, as in the first stage of MEC-PTIR, the SPPRC becomes very close to a multi-criteria problem~\cite{irnich-shortest:05} since the paths may not be comparable for different resources. In this context, by solving the SPPRC, we obtain a set of optimal solutions, i.e., a Pareto front.

Indeed, due to the constraint represented by Equation~(\ref{c:conn_no-rep}), to solve the first stage of the MEC-PTIR, +Tour employs the Elementary SPPRC (ESPPRC), a variant of the SPPRC algorithm that finds only elementary paths, i.e., paths in which no vertex is visited more than once. We use two types of resources: constrained (i.e., time) and unconstrained (i.e., profit). We use a dynamic programming methodology to map the first stage of MEC-PTIR into the ESPPRC framework.\\ 

\noindent\textbf{ESPPRC procedure.} For a given itinerary or path $P = (v_0, v_1, ..., v_{p-1}, v_p)$, let $res(P) = v_p$ be the resident vertex of $P$ (i.e., the last vertex of $P$) and $(v_0, v_1, ..., v_{p-1})$ be an example of prefix path of $P$. For the sake of efficiency, itineraries (or paths) in dynamic programming algorithms are represented through labels. Thus, associated with every itinerary $P$ there is a label $R$. A label stores multiple information, for example, its resident vertex, its predecessor edge, its predecessor label, and its current vector of resource values. For the first stage of MEC-PTIR, two resources are critical to be stored in the labels:
\begin{itemize}
    \item $Spent(R) \in \mathbb{R}$: constrained resource that represents the amount of time spent along the itinerary $P$. Time is consumed whenever the user visits a POI or moves from one POI to another. As described in Section~\ref{sec:model-3}, for every user $u \in \mathcal{U}_d$, the time consumed moving from POI $v_i$ to POI $v_j$ and visiting $v_j$ is represented by Equation~\ref{eq:cost_vi}.
    \item $Profit(R) \in \mathbb{R}$: unconstrained resource that represents the profit of the itinerary $P$. As described in Section~\ref{sec:model-3}, for every user $u \in \mathcal{U}_d$, each POI $v_i \in P$ has a physical profit given by Equation~\ref{eq:prof_vi}.
\end{itemize}

In Algorithm~\ref{alg:rptimt-5g_stage1}, function \code{getLabel} obtains a label from a path while function \code{getPath} obtains a path from a label. Algorithm~\ref{alg:rptimt-5g_stage1} operates over two main sets: $\mathcal{P}$, the set of useful paths, and $\mathcal{Q}$, the set of unprocessed paths, i.e., paths that have not yet been extended. The useful paths $P \in \mathcal{P}$ have already been processed, and they are Pareto-optimal paths or prefixes of Pareto-optimal paths. The set $\mathcal{P}$ is initially empty (line 1), while the set $\mathcal{Q}$ is initiated with the trivial path $(v_0)$ (line 2). In summary, the main loop (line 3) consists of: \textbf{1)} selecting and removing a path $Q \in \mathcal{Q}$ (line 4), \textbf{2)} making all feasible extensions from $Q$ (lines 5-10), \textbf{3)} adding $Q$ to $\mathcal{P}$ (line 11), and \textbf{4)} identifying and removing dominated paths from $\mathcal{P} \cup \mathcal{Q}$ (lines 12-17). The algorithm obtains the set of non-dominated paths (or itineraries) that end in vertex $v_{n+1}$ (line 18), discarding eventual prefixes contained in $P$. 

\begin{algorithm}
\label{alg:rptimt-5g_stage1}

\caption{Dynamic programming ESPPRC -- MEC-PTIR Stage 1}
\DontPrintSemicolon
\small
\SetKwInOut{Input}{Input}
\SetKwInOut{Output}{Output}
\Input{$G = (\mathcal{V},\mathcal{E})$; $\alpha$; 
    ($pop(v_{i}), dur(v_{i}), cat(v_{i})$) $\forall v_i \in \mathcal{V} \setminus \{v_{0}, v_{n+1}\}$; ($int_{u}(c)$ $\forall c \in \mathcal{C}$, $b_u$)}
\Output{set $\mathcal{I}_{u*}$ for a user $u \in \mathcal{U}_d$}
\BlankLine
\BlankLine
$\mathcal{P} \gets \emptyset$\\
$\mathcal{Q} \gets \{(v_0)\}$\\
\While{$\mathcal{Q} \neq \emptyset$}
{
    Choose a path $Q \in \mathcal{Q}$ and remove $Q$ from $\mathcal{Q}$\\
    \ForAll{$v_j \in \mathcal{V} \mid (v_j \neq res(Q) \; {\bf and} \; v_j \notin Q)$}
    {
        $R \gets$ getLabel($Q$)\\
        $feasible, R' \gets$ \textbf{Algorithm 3} $(R,v_j,\alpha,pop(v_{j}),dur(v_{j}),int_u(cat(v_j)),b_u)$\\
        \If{$feasible =$ true}
        {
            $Q' \gets$ getPath($R'$)\\
            Add the path $Q'$ to $\mathcal{Q}$
        }
    }
    Add the path $Q$ to $\mathcal{P}$\\
    \ForAll{pair of paths $(P_1,P_2) \in \mathcal{P} \cup \mathcal{Q}$ {\bf and} $res(P_1) = res(P_2)$}
    {
        $R_1 \gets$ getLabel($P_1$)\\
        $R_2 \gets$ getLabel($P_2$)\\
        $dominates \gets$ \textbf{Algorithm 4} ($R_1,R_2$)\\
        \If{$dominates =$ true}
        {
            Remove the path $P_2$
        }
    }
}
Filter $\mathcal{P}$ such that $res(P) = v_{n+1}$ $\forall P \in \mathcal{P}$ \\
\Return $\mathcal{P}$
\end{algorithm}
%Filter $\mathcal{P}$, such that $\forall P \in \mathcal{P} \mid res(P) = v_f^{u}$

Let $\Xi$ be the set of all solutions of the ESPPRC, where each element $\mathcal{X} \in \Xi$ is a set of Pareto-optimal paths, such that:
\setcounter{equation}{22}
\begin{equation}
\label{eq:extension_invariant}
	\exists \mathcal{X} \in \Xi \mid \mathcal{X} \subseteq \{ (Q,P) \mid Q \in \mathcal{Q}, P \in \mathcal{Z}(Q) \} \cup \mathcal{P},
\end{equation}

\noindent where $\mathcal{Z}(Q) = \{P \mid (Q,P) \in \mathcal{F}(v_i,v_j)) \cap \mathcal{G}\}$ represents the set of all feasible extensions. The set $\mathcal{F}(v_i,v_j)$ has all resource-feasible paths from vertex $v_i$ to vertex $v_j$. 

The condition described by Equation (\ref{eq:extension_invariant}) is held for the initialization of Algorithm~\ref{alg:rptimt-5g_stage1} (lines 1-2), since $v_i = v_0$ and $v_j = res(P) = \emptyset$. For every vertex $v_j \in \mathcal{V}$ (line 5), the condition described by Equation (\ref{eq:extension_invariant}) is verified (line 8) before adding the extended path $Q'$ to $\mathcal{Q}$ (line 10). This depends on Algorithm~\ref{alg:extension} (called in line 7) described latter in this section. In addition, to be efficient, Algorithm~\ref{alg:rptimt-5g_stage1} must avoid making extensions over dominated prefix paths, i.e., paths not part of the Pareto front. Thus, after processing every path $Q \in \mathcal{Q}$ (lines 4-11), the algorithm takes every pair of paths $(P_1,P_2) \in \mathcal{P} \cup \mathcal{Q}$ where $res(P_1) = res(P_2)$ (line 12), verifies if $P_1$ dominates $P_2$ (line 16) and, if this is true, removes the path $P_2$ (line 17). This procedure depends on Algorithm~\ref{alg:dominance} (called in line 15) described latter in this section.

\noindent\textbf{Extension procedure.} The goal of Algorithm~\ref{alg:extension} is to verify if path $P$ (described by label $R$) can be extended to vertex $v_j$. After identifying the path $P$ associated with label $R$ (line 1) and the resident vertex $v_i = res(P)$ of this path (line 2), the algorithm includes into label $R'$ (associated with the extended path $P'$) the information about the time consumed for traveling from $v_i$ to $v_j$ and for visiting the vertex (or POI) $v_j$ (lines 3-4). The algorithm also includes into label $R'$ the information about the profit of the path $P'$ (line 5). Finally, the algorithm verifies if the constrained resource $Spend(R')$ is inside its resource window (line 6), which can be seen as the interval $[ 0, b_u - Cost_{u}(v_i,v_j) ]$. The algorithm returns \textbf{true}, if the extension is feasible (line 7), or \textbf{false} (line 9), otherwise.

\begin{algorithm}
\label{alg:extension}

\caption{Extension procedure}
\DontPrintSemicolon
\small
\SetKwInOut{Input}{Input}
\SetKwInOut{Output}{Output}

\Input{\textbf{R} (label), \textbf{$v_j$} (vertex for path extension),\\
        $\alpha$, $pop(v_{j})$, $dur(v_{j})$, $int_u(cat(v_j))$, $b_u$}
\Output{\textbf{true} or \textbf{false} to indicate the feasibility or not of the extension, \textbf{R'} (label of the path extension)}
\BlankLine
\BlankLine
$P \gets$ getPath($R$)\\
$v_i \gets res(P)$\\
\If{$v_j=v_{n+1}$}
{
    $Cost_{u}(v_i,v_j) \gets 0$\\
}
\Else
{
$Cost_{u}(v_i,v_j) \gets c_{i,j} + dur(v_{j})int_u(cat(v_j))$\\
}
$Spent(R') \gets Spent(R) + Cost_{u}(v_i,v_j)$\\
$Profit(R') \gets Profit(R) + Prof_u(v_j)$\\

\If{$Spent(R') \leq b_u$}
{
    \Return{(\textbf{true},R')}
}
\Else
{
    \Return{(\textbf{false},R')}
}
\end{algorithm}

\noindent\textbf{Dominance procedure.} In general terms, dominance rules identify non-useful paths by comparing the resource vector $T(Q)$ and the set of all feasible extensions $\mathcal{Z}(Q)$ with the corresponding $T(P)$ and $\mathcal{Z}(P)$, such that $res(P) = res(Q)$. In the context of the first stage of MEC-PTIR, as described by Algorithm~\ref{alg:dominance}, this comparison consists of verifying if label $R_1$ (corresponding to path $P_1$) provides a higher or equal profit and consumes less or equal time than label $R_2$ (that corresponds to path $P_2$) (line 1). The algorithm returns \textbf{true} if the condition is satisfied (line 2) or \textbf{false} (line 4), otherwise. When the condition is true, path $P_2$  must be removed since it was identified as equal to or worse than $P_1$.

\begin{algorithm}
\label{alg:dominance}

\caption{Dominance procedure}
\DontPrintSemicolon
\small
\SetKwInOut{Input}{Input}
\SetKwInOut{Output}{Output}

\Input{$\mathbf{R_1}$ (label), $\mathbf{R_2}$ (label)}
\Output{\textbf{true} if $\mathbf{R_1}$ dominates $\mathbf{R_2}$, \textbf{false} otherwise}
\BlankLine
\BlankLine
\If{$Profit(R_1) \geq Profit(R_2)$ \textbf{and} $Spent(R_1) \leq Spent(R_2)$ }
{
    \Return{\textbf{true}}
}
\Else
{
    \Return{\textbf{false}}
}
\end{algorithm}

\subsection{Complexity Analysis of +Tour Algorithm}
\label{sec:complexity-analysis}
In this Section, we analyze the complexity of +Tour, represented by Algorithm~\ref{alg:rawpir}, which is composed of Algorithm~\ref{alg:rptimt-5g_stage1} ($A_2$), Algorithm~\ref{alg:extension} ($A_3$), Algorithm~\ref{alg:dominance} ($A_4$), and Stage 2 ($S_2$). In the worst case, the complexity of Algorithm~\ref{alg:rawpir}  is $\mathcal{O}\big(|\mathcal{U}_d| \mathcal{O}(A_2)\big) + \mathcal{O}(S_2)$, i.e., the iteration through all users in $\mathcal{U}_d$ (line 2) times the complexity of Algorithm~\ref{alg:rptimt-5g_stage1} (line 3), plus the complexity of Stage 2 (line 5). The complexity of Algorithm~\ref{alg:rptimt-5g_stage1} is $\mathcal{O} \Big(|\mathcal{Q}| \big( |\mathcal{V}|\mathcal{O}(A_3) + \frac{|\mathcal{P} \cup \mathcal{Q}|  (|\mathcal{P} \cup \mathcal{Q}| - 1)}{2}\mathcal{O}(A_4)\big)\Big) + \mathcal{O}(|\mathcal{P}|)$, i.e., the iteration through all paths in $\mathcal{Q}$ (line 3), times the iteration through all POIs in $\mathcal{V}$ (line 5) times the complexity of Algorithm~\ref{alg:extension} (line 7), plus the combination of path pairs in $\mathcal{P} \cup \mathcal{Q}$ (line 12) times the complexity of Algorithm~\ref{alg:dominance} (line 15), plus the iteration through the paths in $\mathcal{P}$ (line 18). In line 12, since $\mathcal{P} \cap \mathcal{Q} = \emptyset$, in the worst case,  $|\mathcal{P}|$, $|\mathcal{Q}|$ and $|\mathcal{P} \cup \mathcal{Q}|$ are all equal to $\sum_{k=1}^n \frac{n!}{(n-k)!}$ where $n = |\mathcal{V}|$ and $k$ represent the number of extensions. Finally, Algorithms~\ref{alg:extension} and~\ref{alg:dominance} are constant, i.e., their complexity is $\mathcal{O}(1)$.

%% file: Sections/06-Data-Characterization.tex
\color{black}
\section{Data Characterization}\label{sec:data-characterization}

We extract data from Flickr\textsuperscript 1, a real-world location-based photo-sharing application, to derive user travel histories and visiting-related statistics. The purpose of using real-world data is three-fold: (i) to show how real data can feed our model of POI information and user preferences; (ii) to characterize real-world visiting patterns; and (iii) to achieve a more realistic evaluation of the +Tour. With this goal in mind, we collect data from thirteen cities across four different continents, namely Asia (New Delhi and Osaka), Europe (Athens, Barcelona, Budapest, Edinburgh, Glasgow, London, Madrid, and Vienna), North America (Toronto), and Oceania (Perth and Melbourne). These cities are important from a tourist point of view and guarantee variety and diversity in our evaluation. Next, we first describe our methodology for data extraction. Then, we present an exploratory analysis of our dataset.

\subsection{Data extraction methodology}
\label{sec:data-1}

We use the Google Places API\footnote{\url{https://developers.google.com/places/web-service/intro}} to obtain the list of POIs of the cities. Each city identifies each POI by ID, name, latitude, and longitude, as well as the category to which it belongs. The cost $c_{i,j}$ of an edge connecting two POIs $v_i$ and $v_j$ in the same list is computed using the Google Matrix Distance API\footnote{\url{https://developers.google.com/maps/documentation/distance-matrix/start}} in the walking mode.

We use the Flickr API\footnote{\url{https://www.flickr.com/services/developer/api/}} to extract geo-tagged photos and derive user travel histories for each city, except for Melbourne whose data is provided by~\cite{MelbourneDataSet}. Each Flickr photo is tagged with the user ID of the photo owner, timestamp, and coordinates (latitude and longitude).

For each city, we combine its photo dataset with its POI list to generate the set $\mathcal{S}$ of travel histories in the city. First, we match each photo to the corresponding POI using the geo-coordinates of the photo and the POI. A photo is mapped to a POI if its geo-coordinates differ by less than 100 meters according to the Haversine formula. If this condition holds for more than one POI, the photo is associated with the nearest POI, so each photo is mapped to a single POI. Then, we construct the travel history $S_u$ for each user, sorting the photos by user ID and timestamp and grouping consecutive POI visits in the same travel sequence if the travel sequence duration does not exceed 8 hours. Then, for each POI $v_i$ in a travel sequence, we take the time of the first and the last photo taken by user $u$ at $v_i$ as the arrival ($t_{v_i}^a$) and departure ($t_{v_i}^d$) time in that POI, respectively. Finally, we discard travel sequences with cycles (repeated POI visits), with POIs with less than five visits, or with only one photo (since they result in a sequence duration of zero seconds). Table~\ref{tab:dataset} details our resulting dataset, which is also publicly available in a GitHub repository\footnote{\url{https://github.com/LABORA-INF-UFG/plusTour}}.

\begin{table*}[hbt]
\centering
\footnotesize
\caption{Dataset summary.}
\label{tab:dataset}
\begin{tabular}{ccccccccc}
\hline
\textbf{City}&\textbf{Country}&\textbf{Continent}&\textbf{\begin{tabular}[c]{@{}c@{}}Photos\end{tabular}}&\textbf{\begin{tabular}[c]{@{}c@{}}Users\end{tabular}}&\textbf{\begin{tabular}[c]{@{}c@{}}Sequences\end{tabular}}&\textbf{\begin{tabular}[c]{@{}c@{}}Valid sequences\end{tabular}}&\textbf{\begin{tabular}[c]{@{}c@{}}POIs\end{tabular}}&\textbf{\begin{tabular}[c]{@{}c@{}}POI categories\end{tabular}}\\\hline
Athens&Greece&Europe&5026&291&516&509&24&5\\
Barcelona&Spain&Europe&13654&782&1654&1634&29&8\\
Budapest&Hungary&Europe&8090&573&2361&1058&36&6\\
Edinburgh&Scotland&Europe&20620&879&5028&2405&25&6\\
Glasgow&Scotland&Europe&8172&367&2227&969&25&7\\
London&England&Europe&48519&2198&5018&4949&30&10\\
Madrid&Spain&Europe&20881&577&1396&1382&30&8\\
Melbourne&Australia&Oceania&15255&626&5106&2096&84&9\\
New Delhi&India&Asia&2961&186&489&278&19&6\\
Osaka&Japan&Asia&5306&279&1115&540&23&4\\
Perth&Australia&Oceania&2503&99&716&307&19&7\\
Toronto&Canada&North America&30267&846&6057&2721&29&6\\
Vienna&Italy&Europe&22432&704&3193&1613&28&8\\\hline
\textbf{13 cities}&\textbf{10 countries}&\textbf{4 continents}&\textbf{203686}&\textbf{8407}&\textbf{34876}&\textbf{20461}&\textbf{401}&\textbf{20 (unique)}\\\hline
\end{tabular}
\end{table*}

\subsection{Data analysis}
\label{sec:data-2}
Using our dataset, we compute the visiting-related statistics, i.e., POI popularity, expected POI visiting time, and user's interests in a given category for each city. Next, we analyze our dataset's most relevant visiting-related statistics and patterns. Due to space limitations, we present results only for the five cities with the most distinguished patterns, namely London, Melbourne, Osaka, Perth, and Toronto. The complete data analysis is available at the GitHub repository\textsuperscript 6.

\begin{figure*}[!ht]
\centering
    \begin{tabular}{@{}ccccc@{}}
        \includegraphics[width=.18\textwidth]{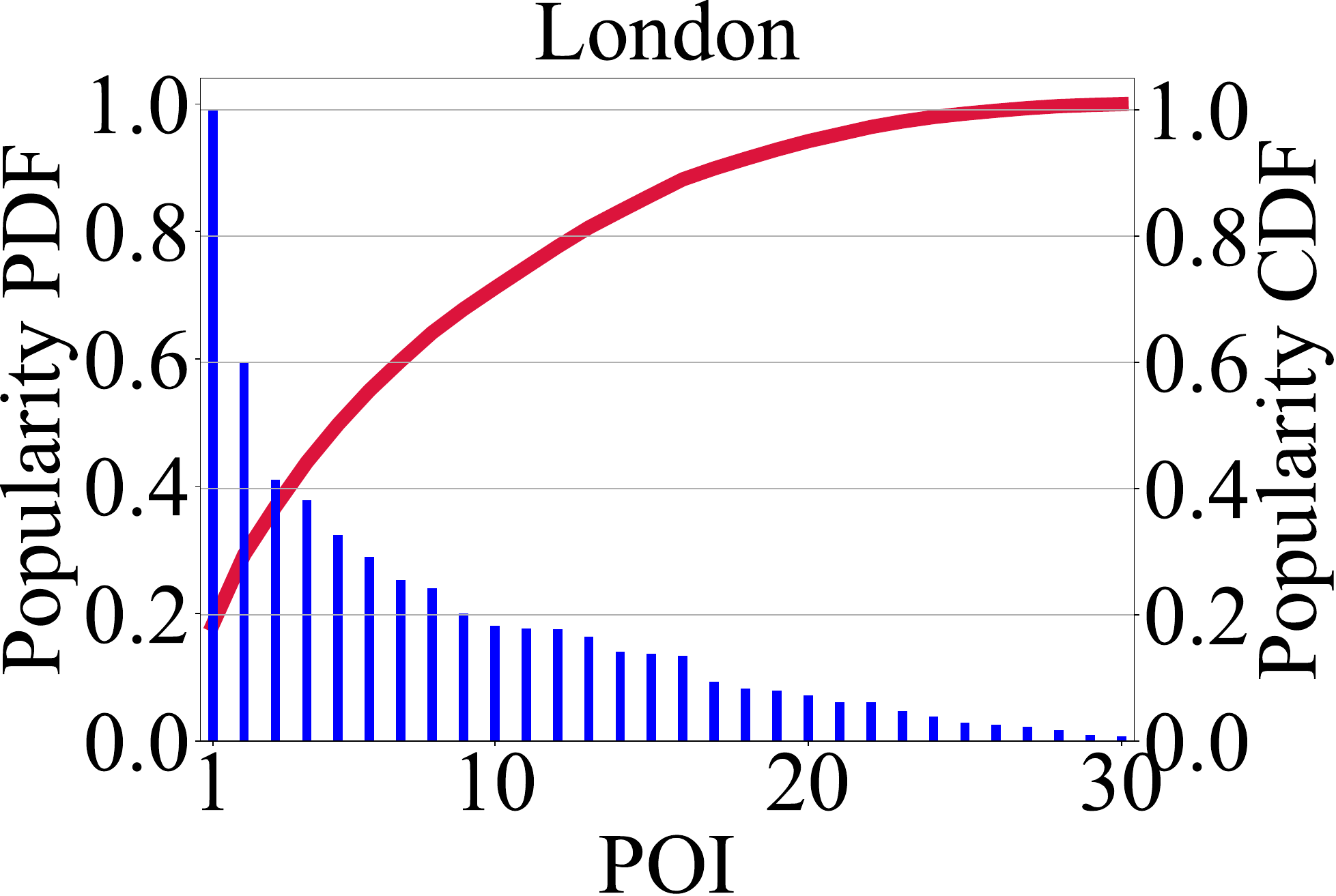} &
        \includegraphics[width=.18\textwidth]{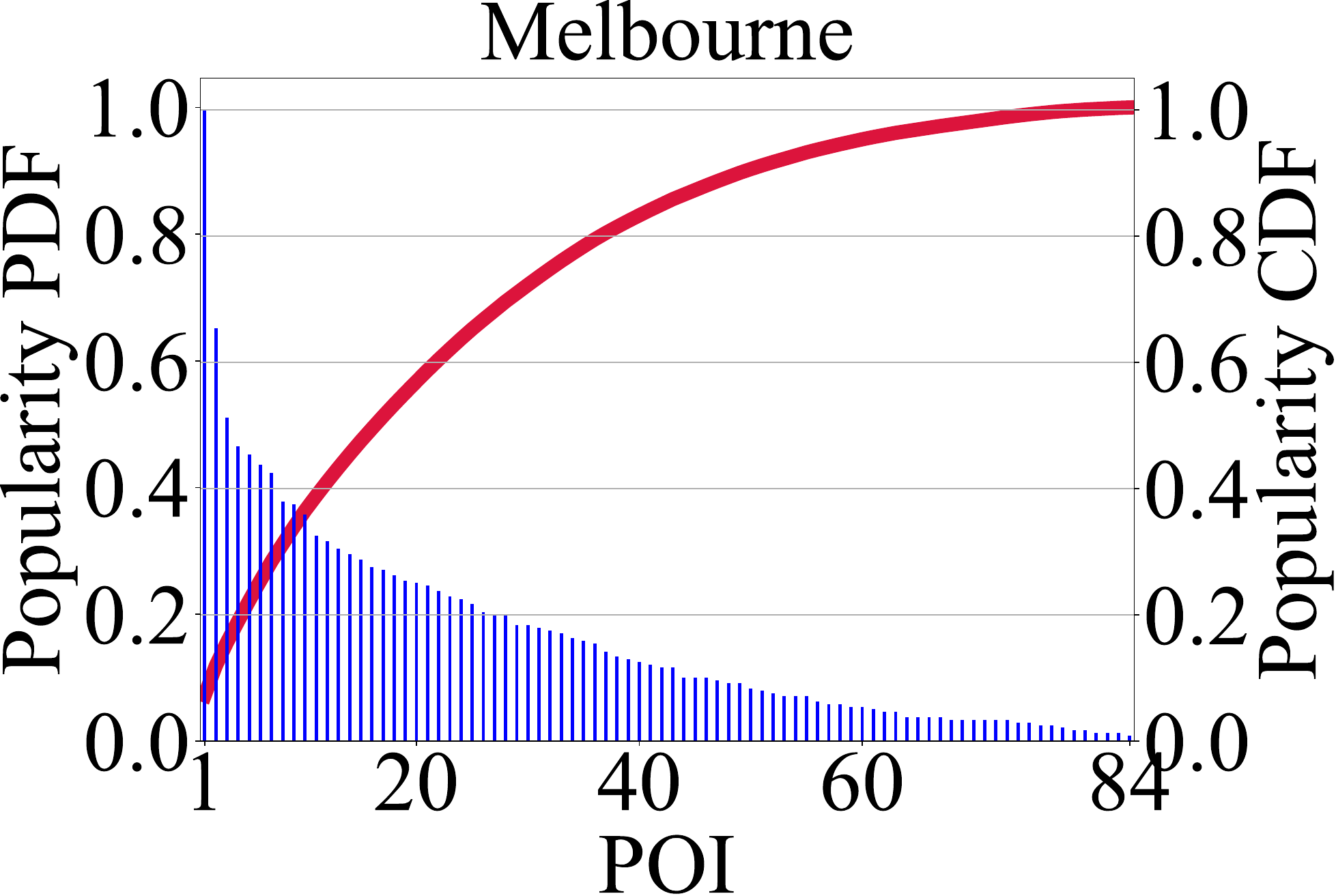} &
        \includegraphics[width=.18\textwidth]{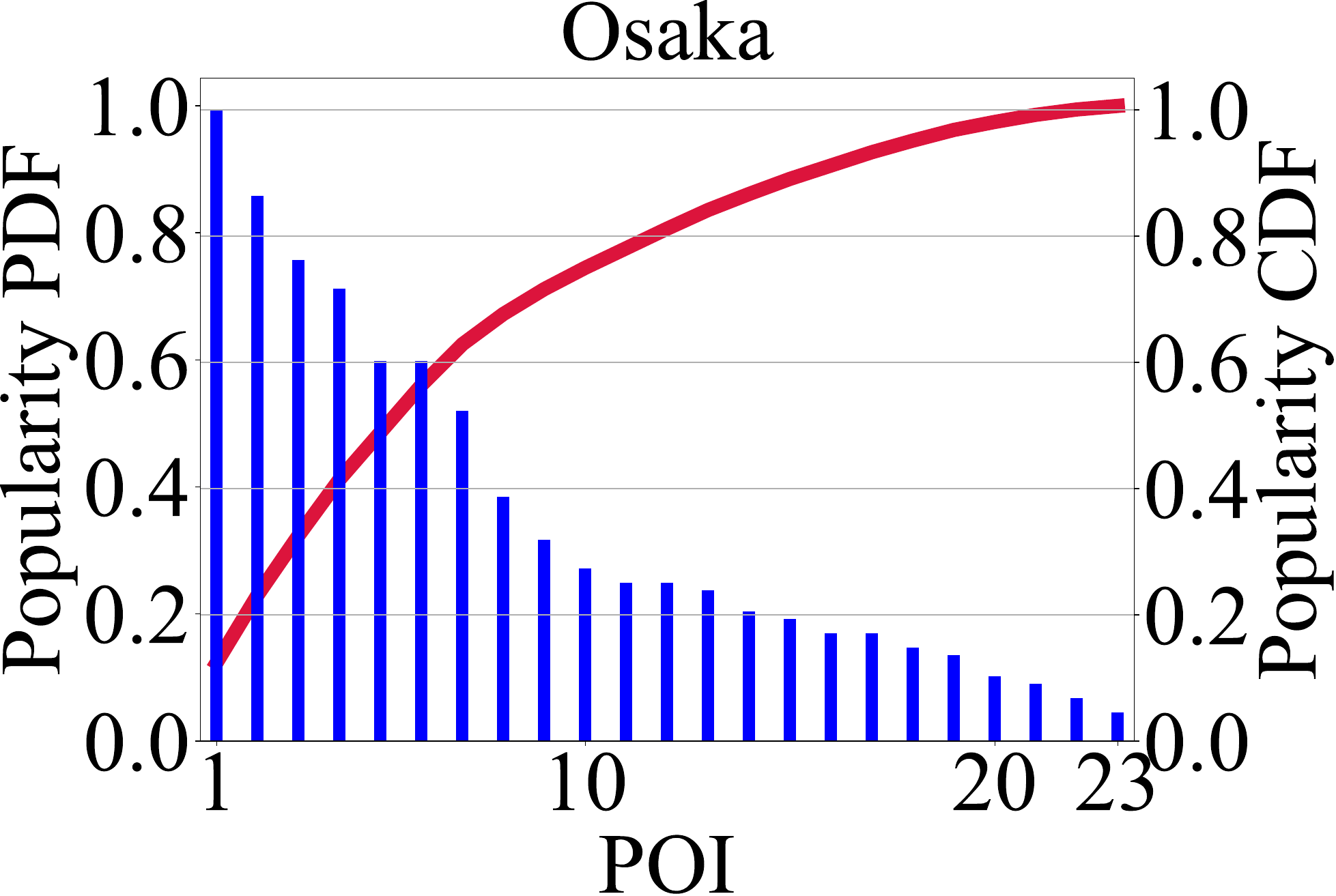} &
        \includegraphics[width=.18\textwidth]{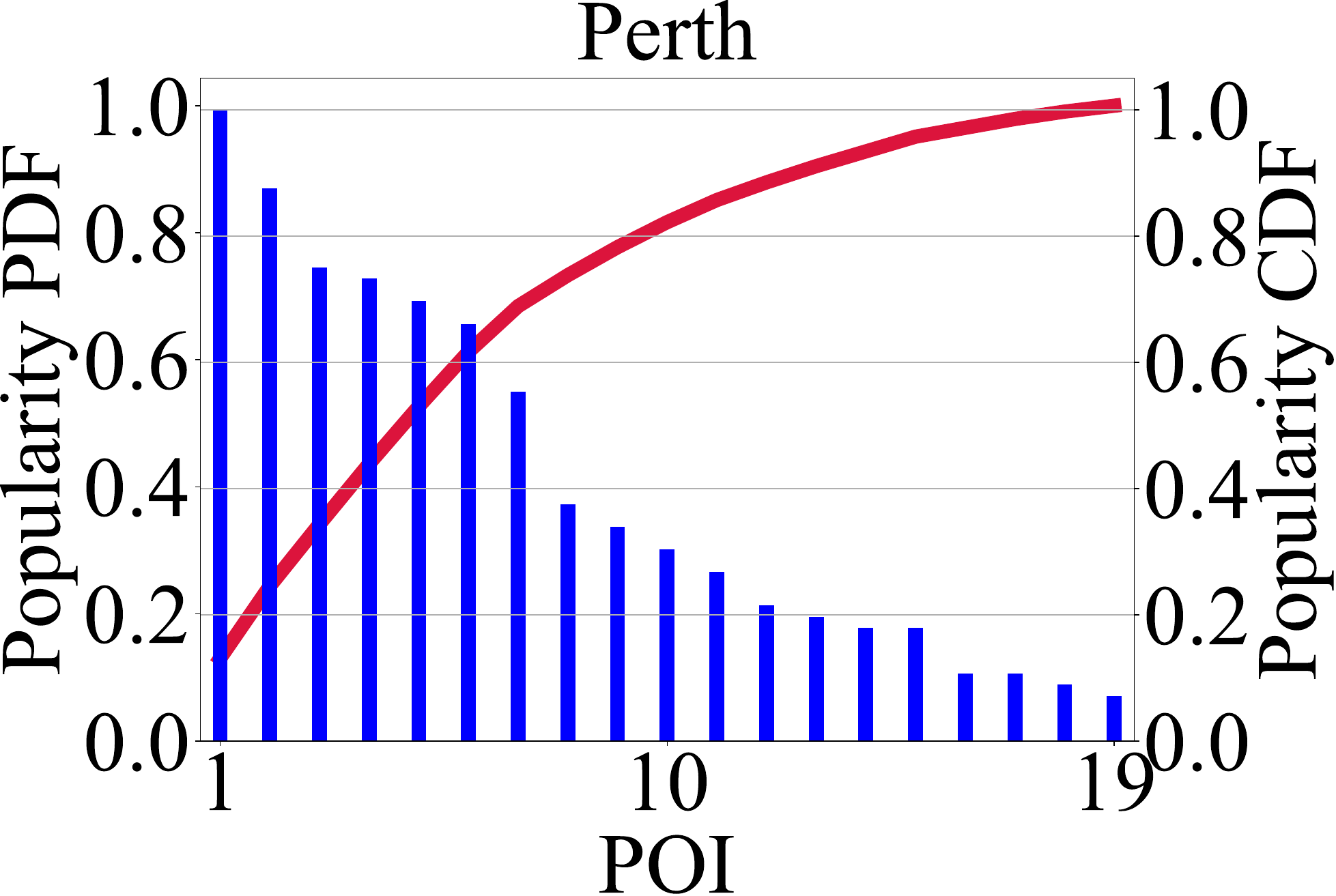} &
        \includegraphics[width=.18\textwidth]{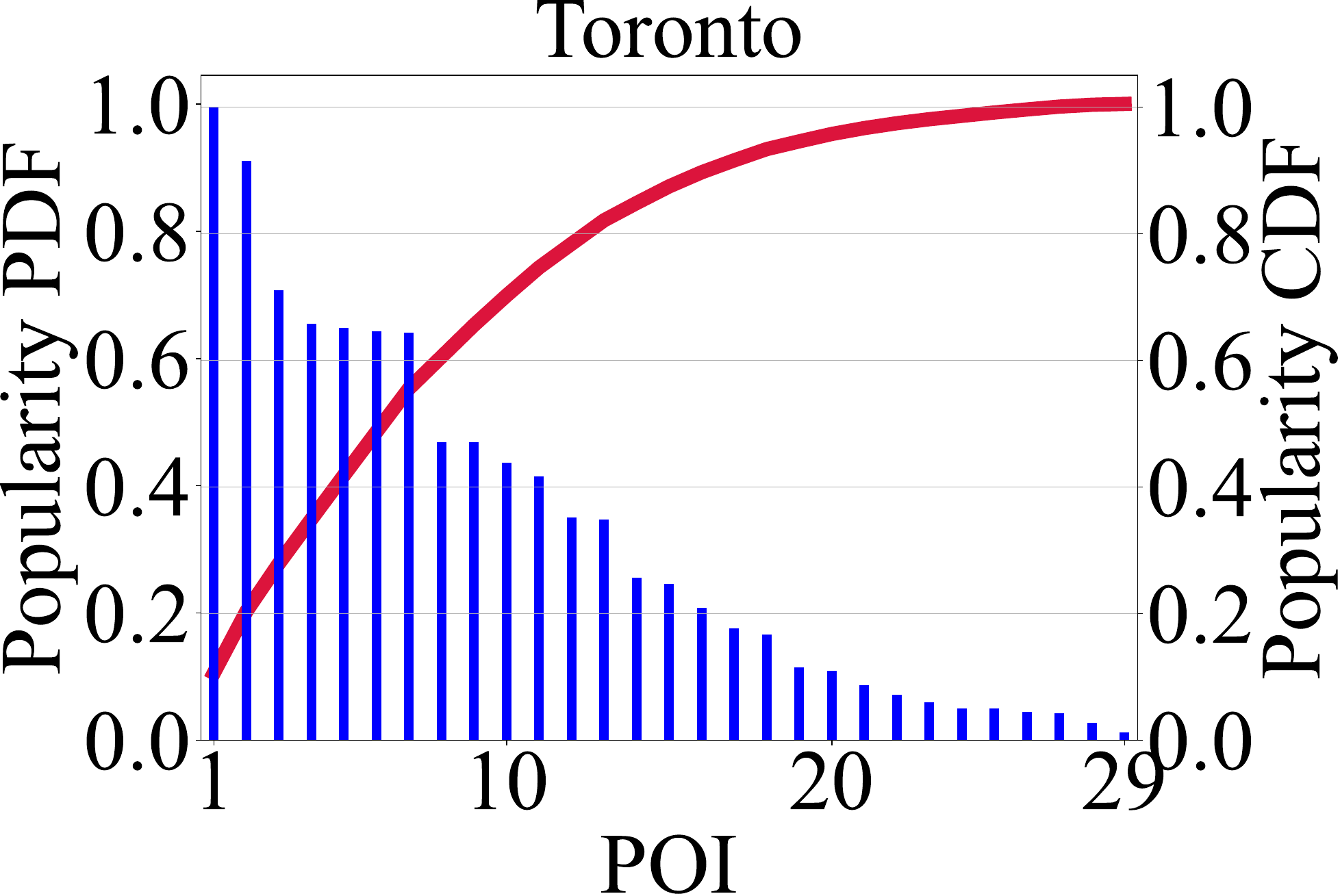}   \\
    \end{tabular}
    \caption{POI popularity Distribution. PDF is presented in blue, and CDF is shown in red. Values have been normalized so that the popularity of the most popular POI equals 1.}
    \label{fig:POIs-Popularity}
\end{figure*}

Figure~\ref{fig:POIs-Popularity} shows the Probability Density Function (PDF) (blue bars) and the Cumulative Distribution Function (CDF) (red line) of the POI Popularity for each analyzed city. POI popularity exhibits similar patterns among most cities, with more than 50\%  of the visits being concentrated in usually 5 POIs, while most POIs are rarely visited. POI popularity is slightly different in London and Melbourne, where a single POI (the Trafalgar Square in London and the City Square in Melbourne) holds the majority of the visits.

\begin{figure*}[!ht]
\centering
    \begin{tabular}{@{}ccccc@{}}
        \includegraphics[width=.18\textwidth]{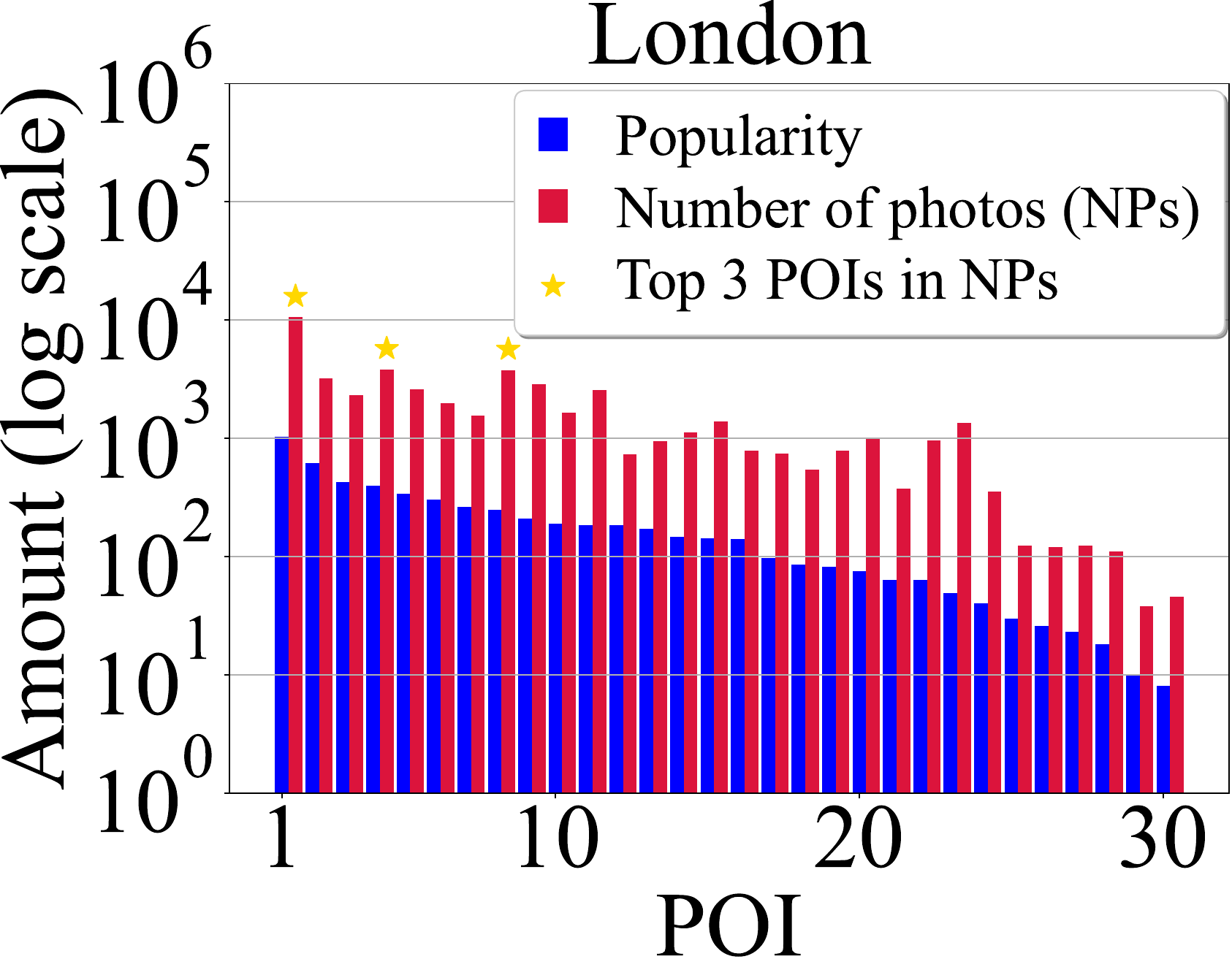} &
        \includegraphics[width=.18\textwidth]{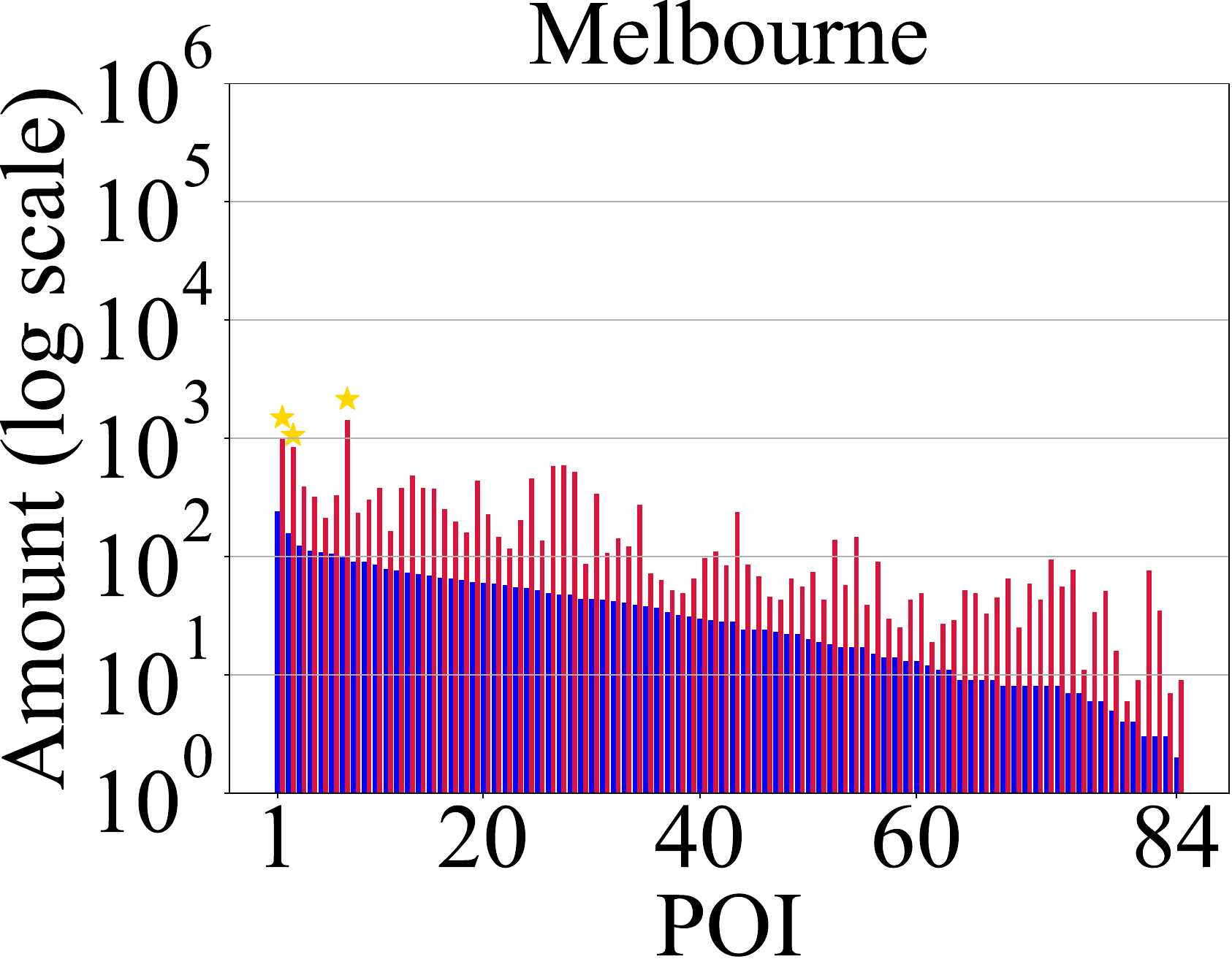} &
        \includegraphics[width=.18\textwidth]{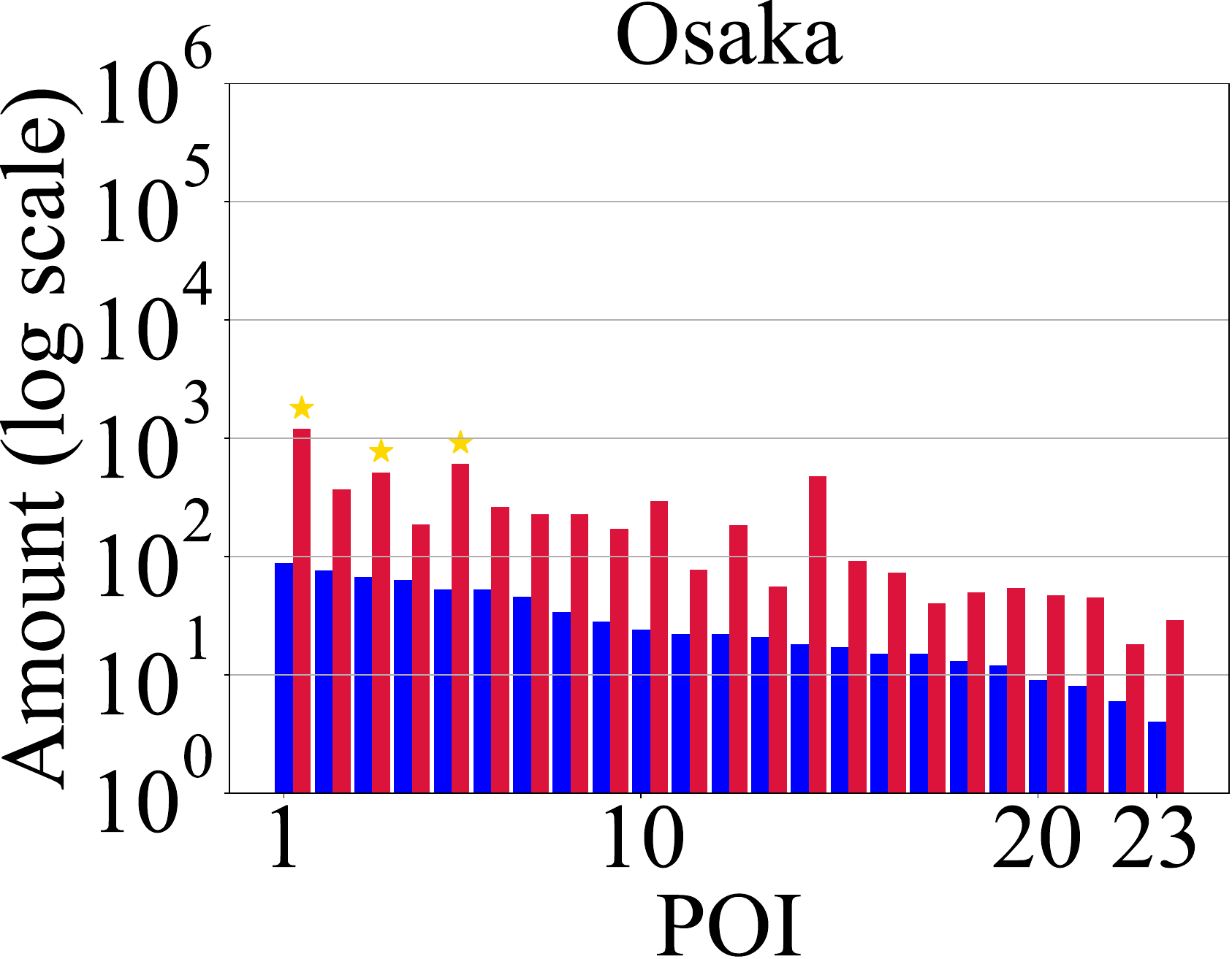} &
        \includegraphics[width=.18\textwidth]{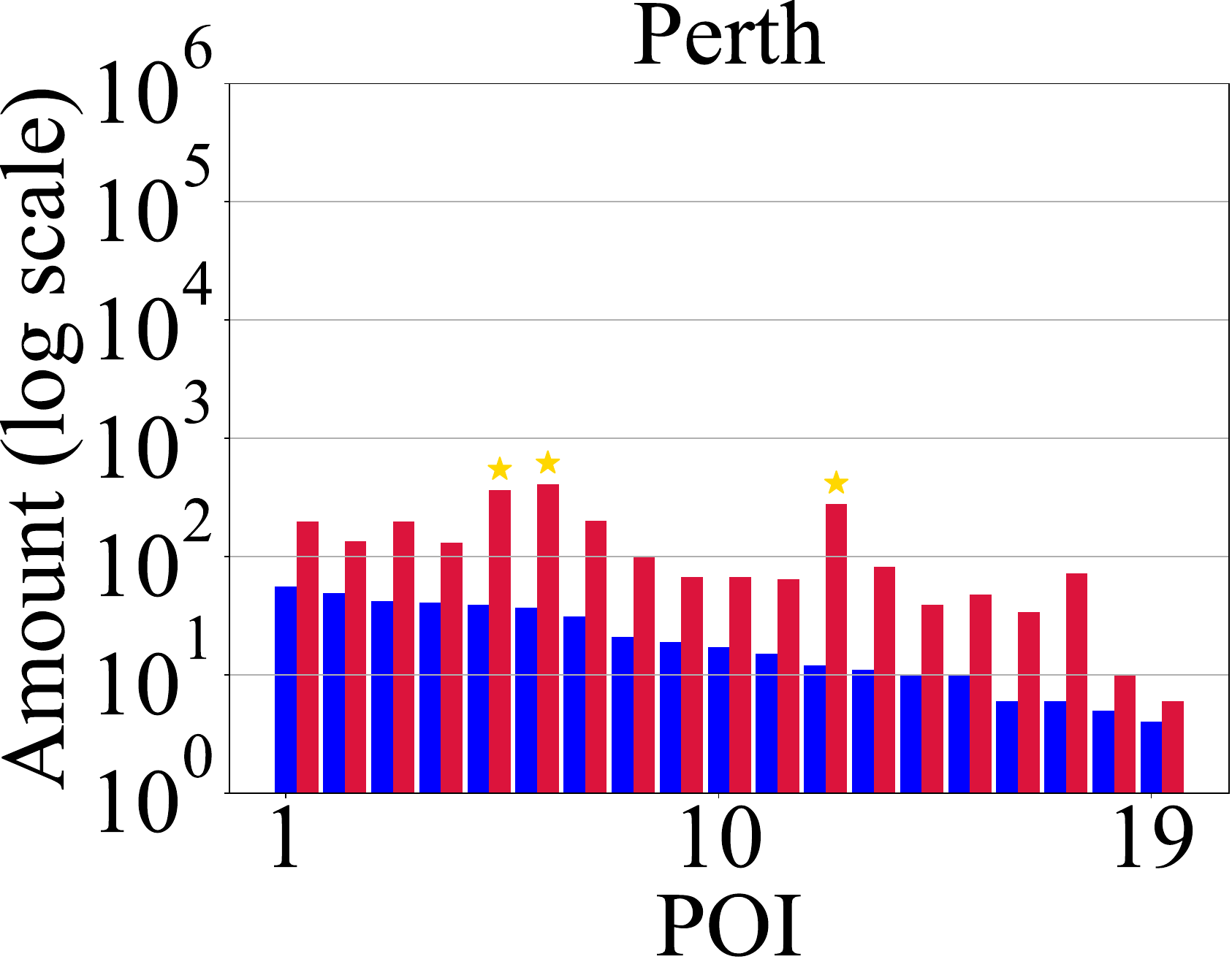} &
        \includegraphics[width=.18\textwidth]{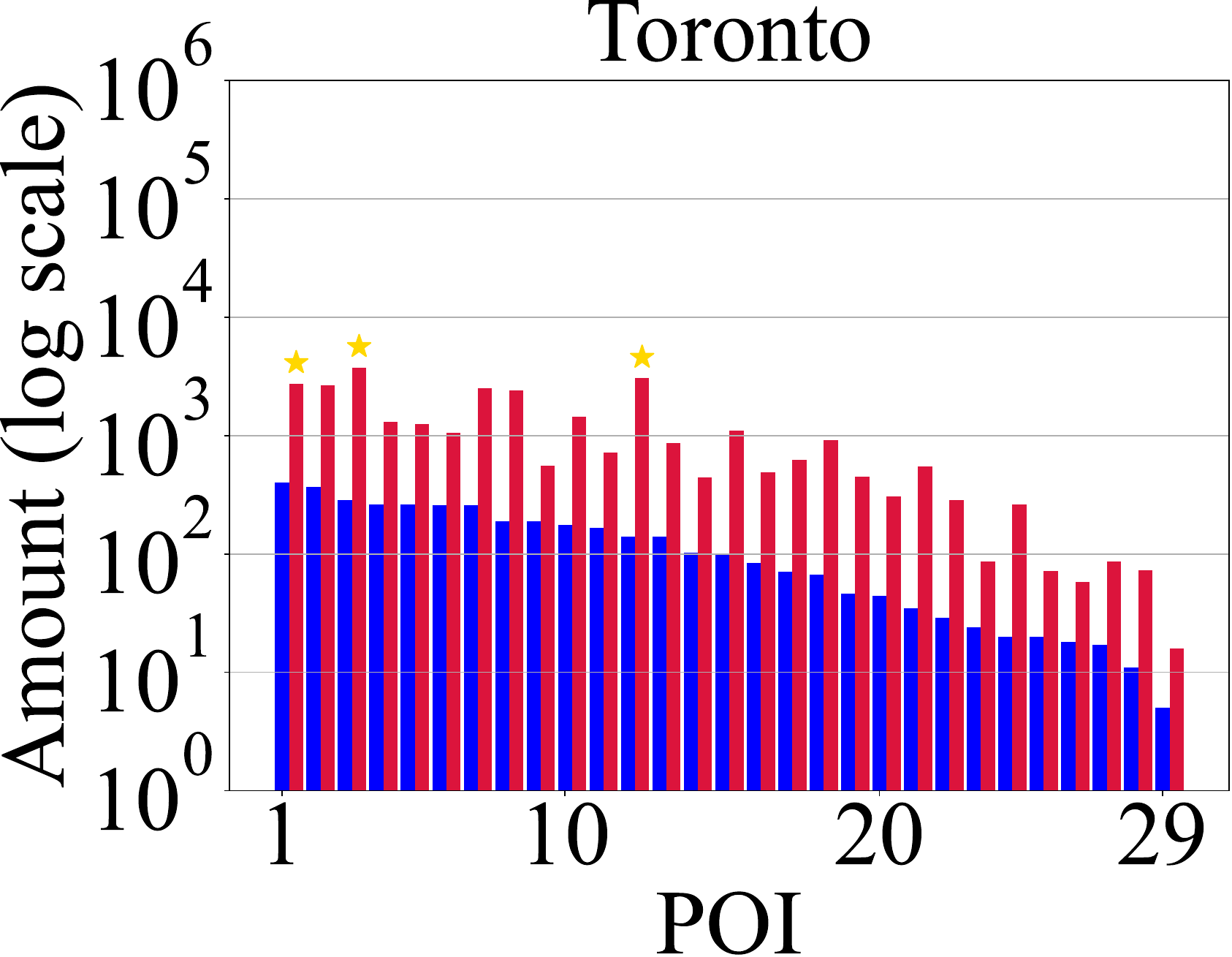}   \\
    \end{tabular}
    \caption{POI Popularity (in blue) and number of photos taken in each POI (in red). POI popularity is given in the number of POI visits. The top 3 POIs in the number of photos are marked with stars.}
    \label{fig:POIs-Popularity-And-Number-Of-Photos}
\end{figure*}

Figure~\ref{fig:POIs-Popularity-And-Number-Of-Photos} shows the relation between POI popularity (blue bars) and the number of photos taken in the POI (red bar). POIs are sorted by popularity, and the top 3 POIs in the number of photos are highlighted with a star. Usually, the top 3 POIs in the number of photos appear among the top 10 POIs in popularity, showing that popularity does not always correlate to the number of photos. Perth presents a greater imbalance between popularity and number of photos per POI.

\begin{figure*}[t]
\centering
    \begin{tabular}{@{}ccccc@{}}
        \includegraphics[width=.18\textwidth]{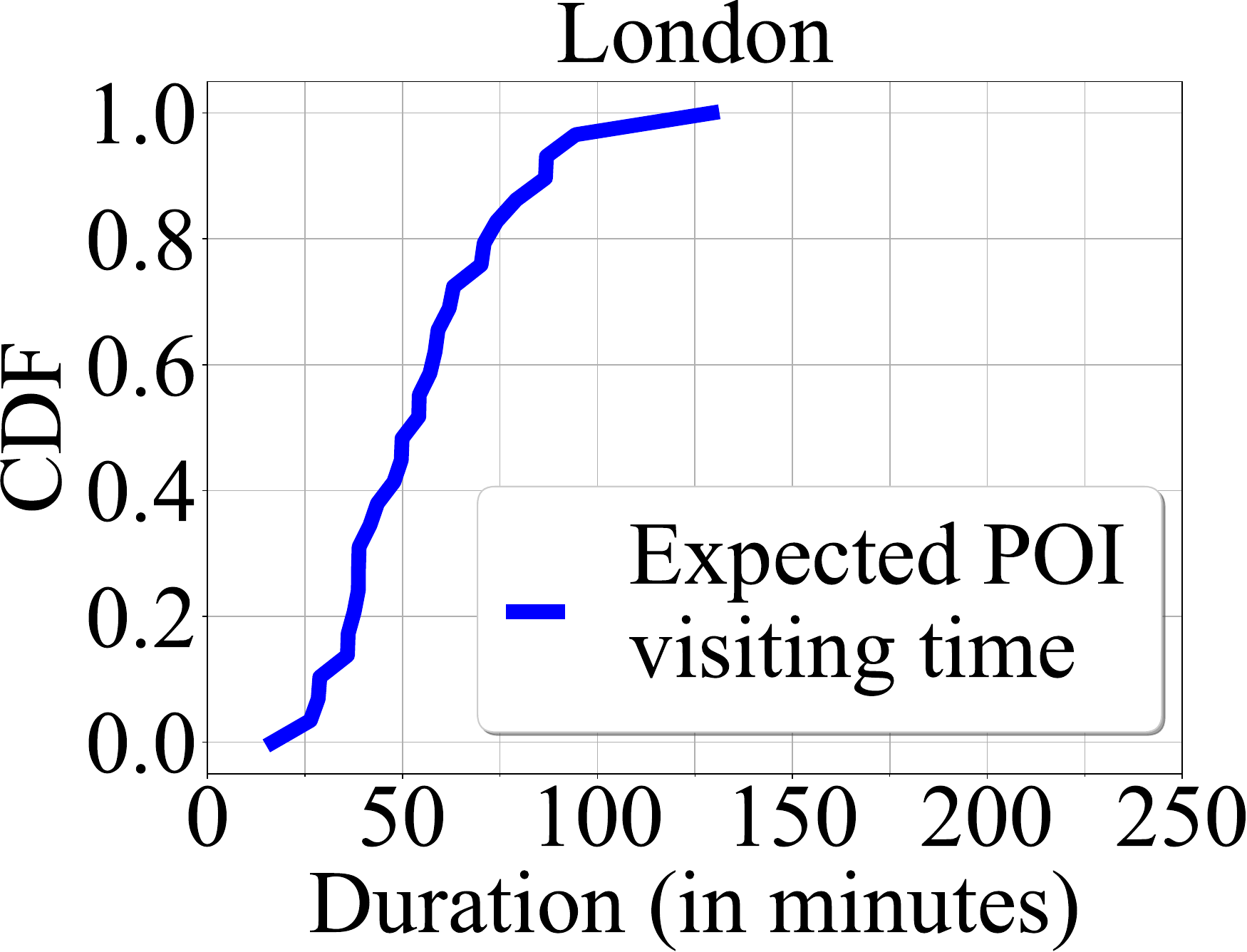} &
        \includegraphics[width=.18\textwidth]{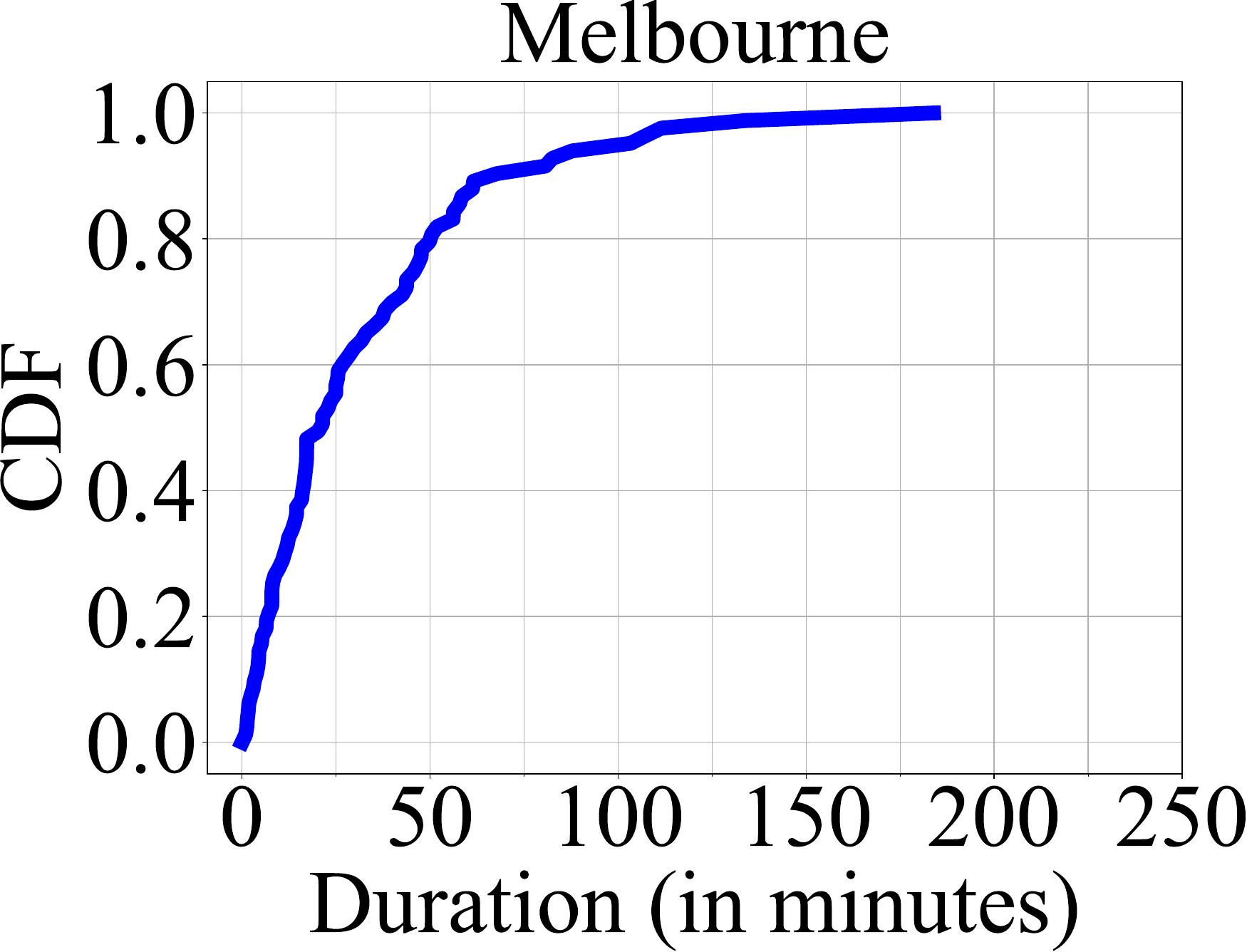} &
        \includegraphics[width=.18\textwidth]{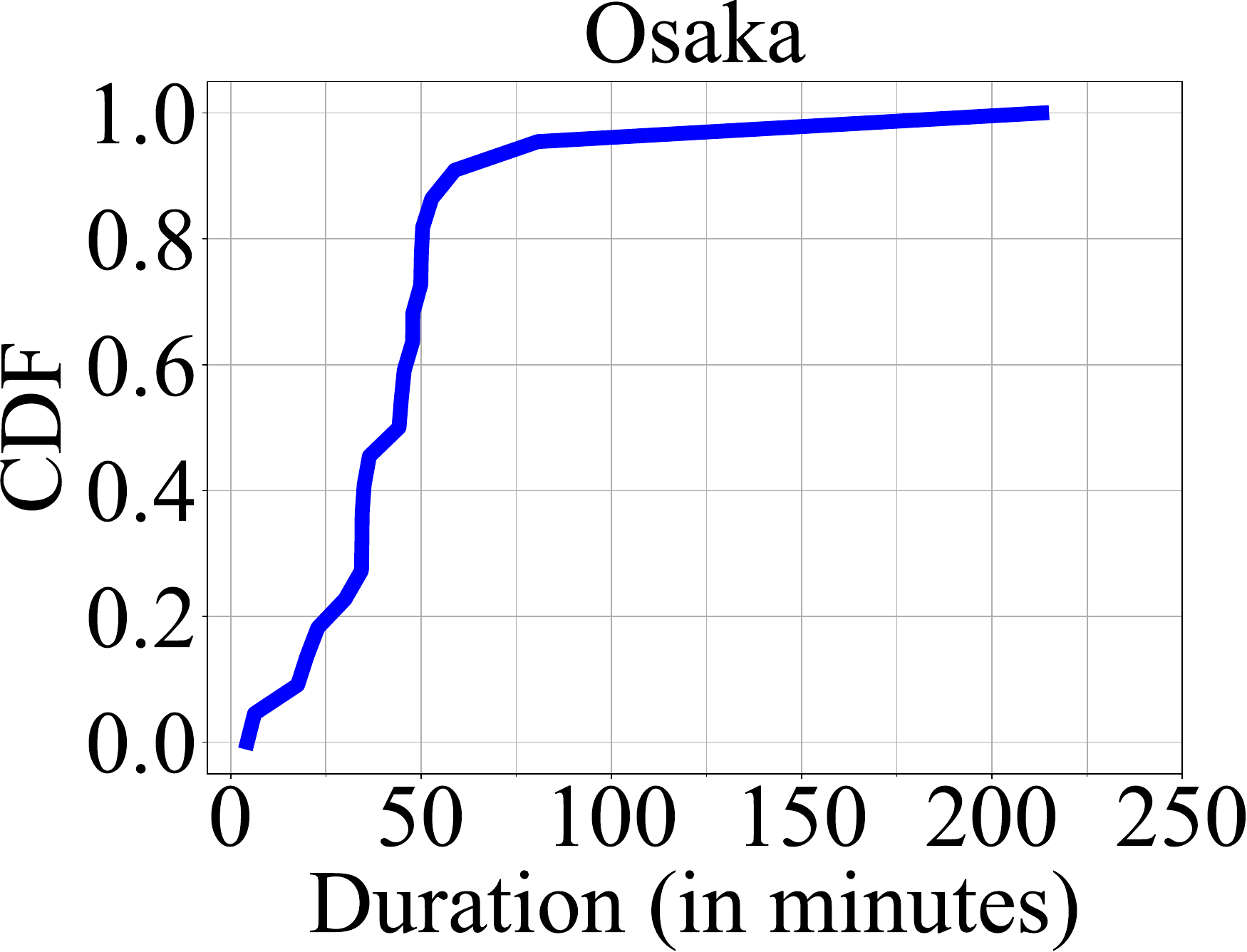} &
        \includegraphics[width=.18\textwidth]{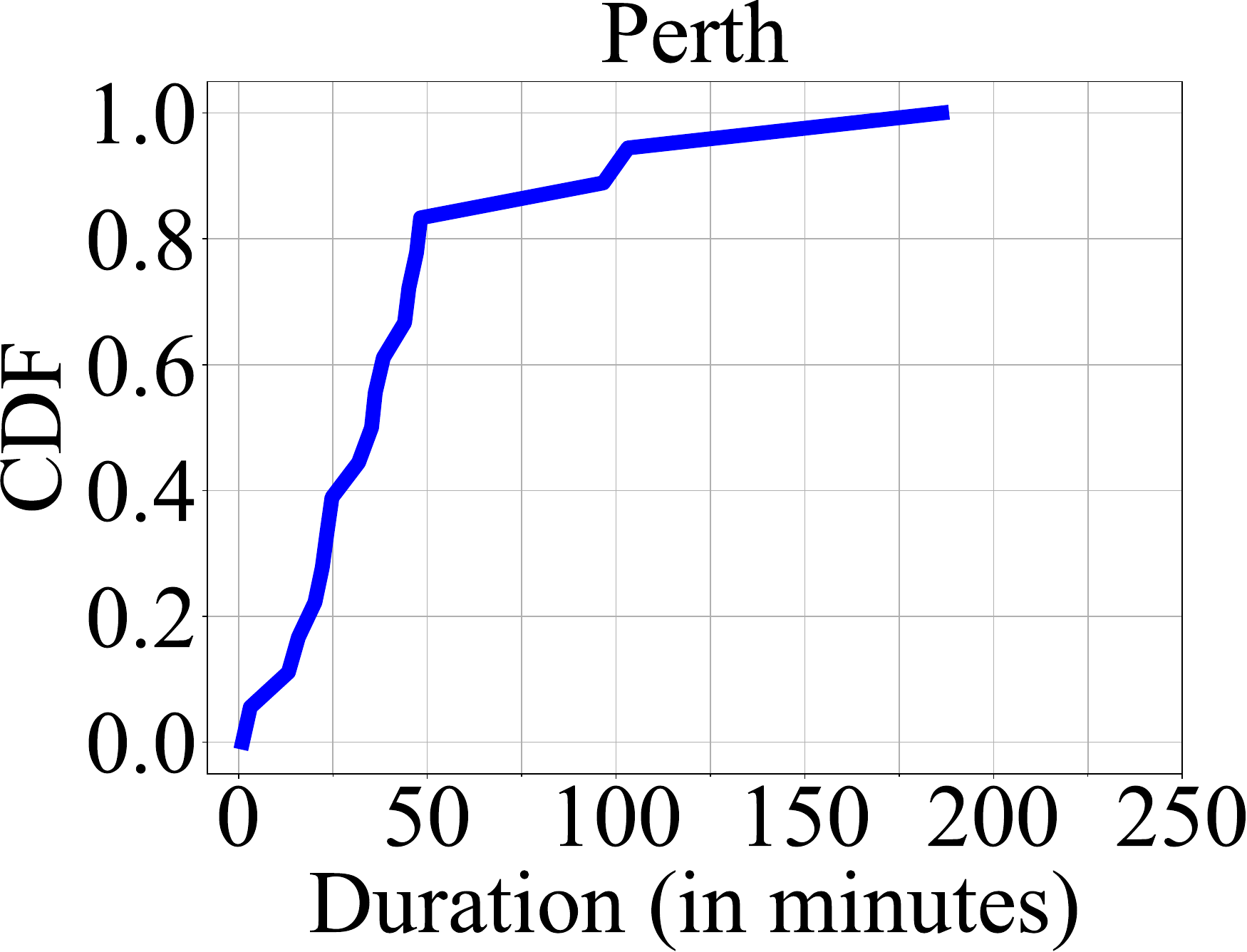} &
        \includegraphics[width=.18\textwidth]{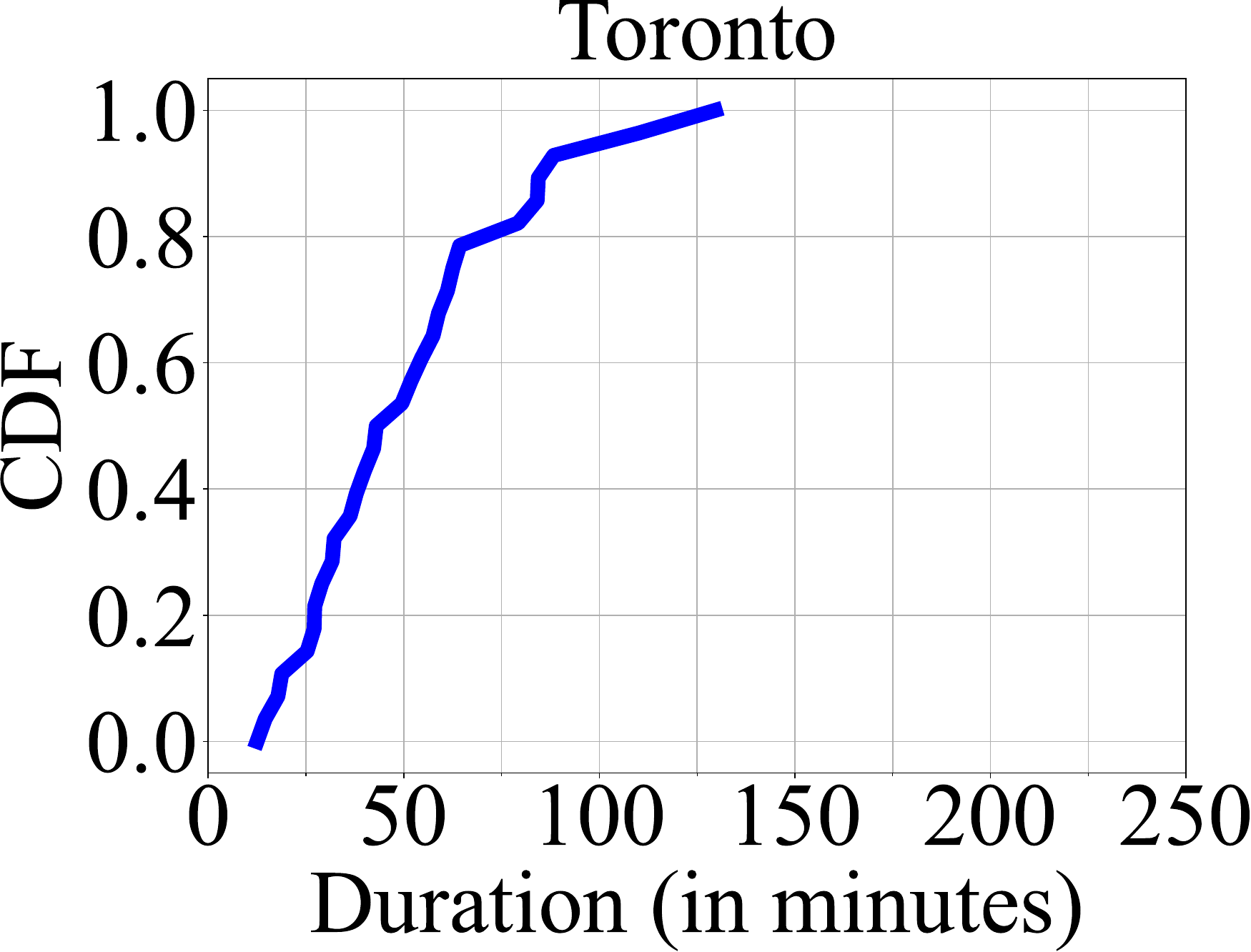}   \\
    \end{tabular}
    \caption{CDF of the expected POI visiting time.}
    \label{fig:POIs-Visiting-Time}
\end{figure*}

\begin{figure}[!ht]
\begin{center}
 	\includegraphics[width=0.65\linewidth]{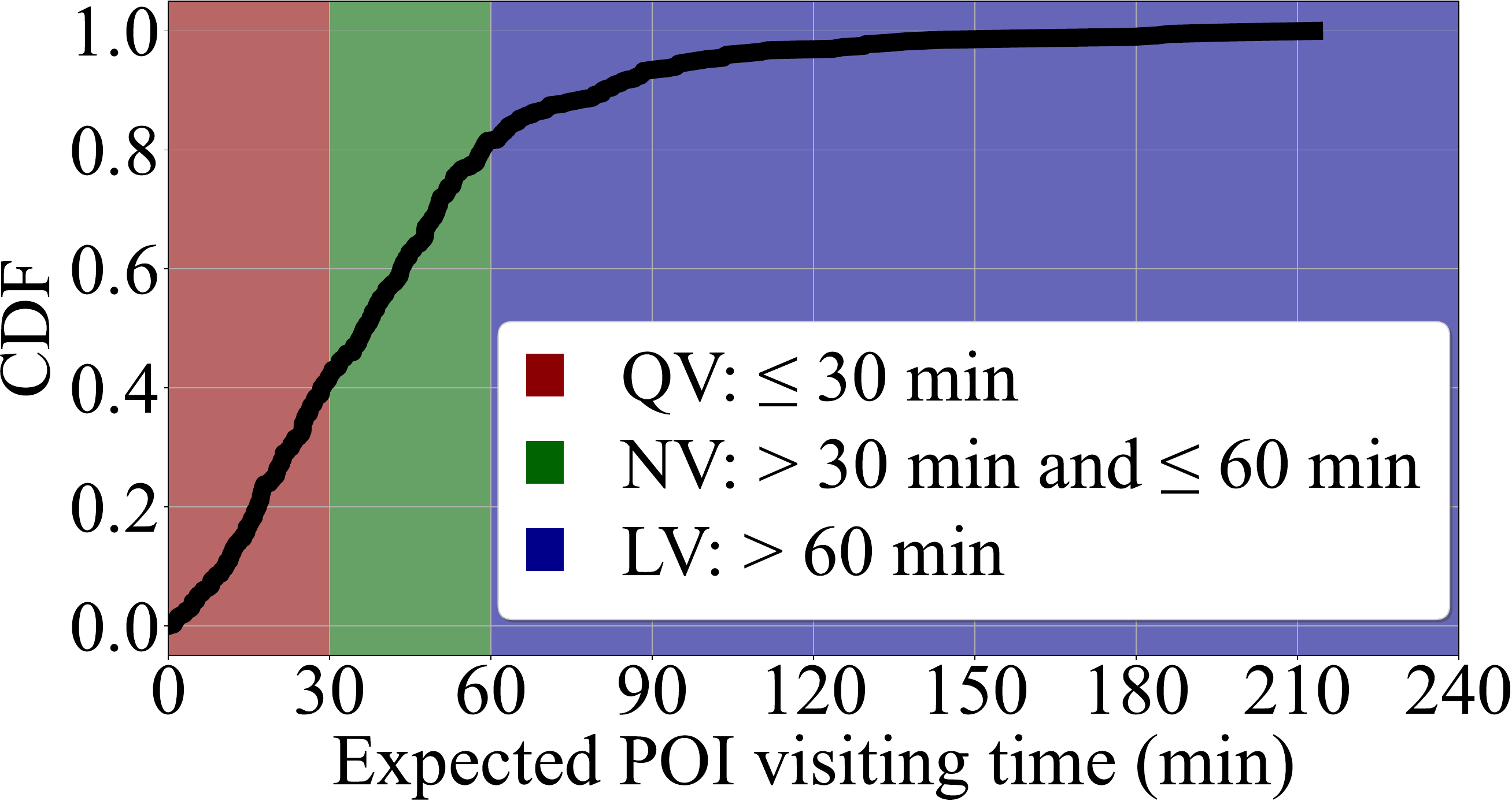}
        \caption{Expected POI visiting time clustering.}
        \label{fig:POIs-Visiting-Time-Clustering}
\end{center}
\end{figure}

Figure~\ref{fig:POIs-Visiting-Time} shows the CDF of the expected (average) POI visiting time. In all analyzed cities, the expected POI visiting time is short, with 50\% lasting less than 50 minutes and 90\% lasting less than 1 hour. Melbourne presents the distribution with the shortest expected POI visiting time, with 50\% lasting less than 30 minutes. Figure~\ref{fig:POIs-Visiting-Time-Clustering} shows a complementary analysis of the CDF of the expected POI visiting time, but here, the CDF is presented considering all POIs in all cities in our dataset. The figure shows three POI groups: \textit{Quick-visited} (QV, $\leq$ 30 minutes, in red), \textit{Normal-visited} (NV, $>$ 30 minutes and $\leq$ 60 minutes, in green), and \textit{Long-visited} (LV, $>$ 60 minutes, in blue).

\begin{figure}[!ht]
\begin{center}
 	\includegraphics[width=0.5\textwidth]{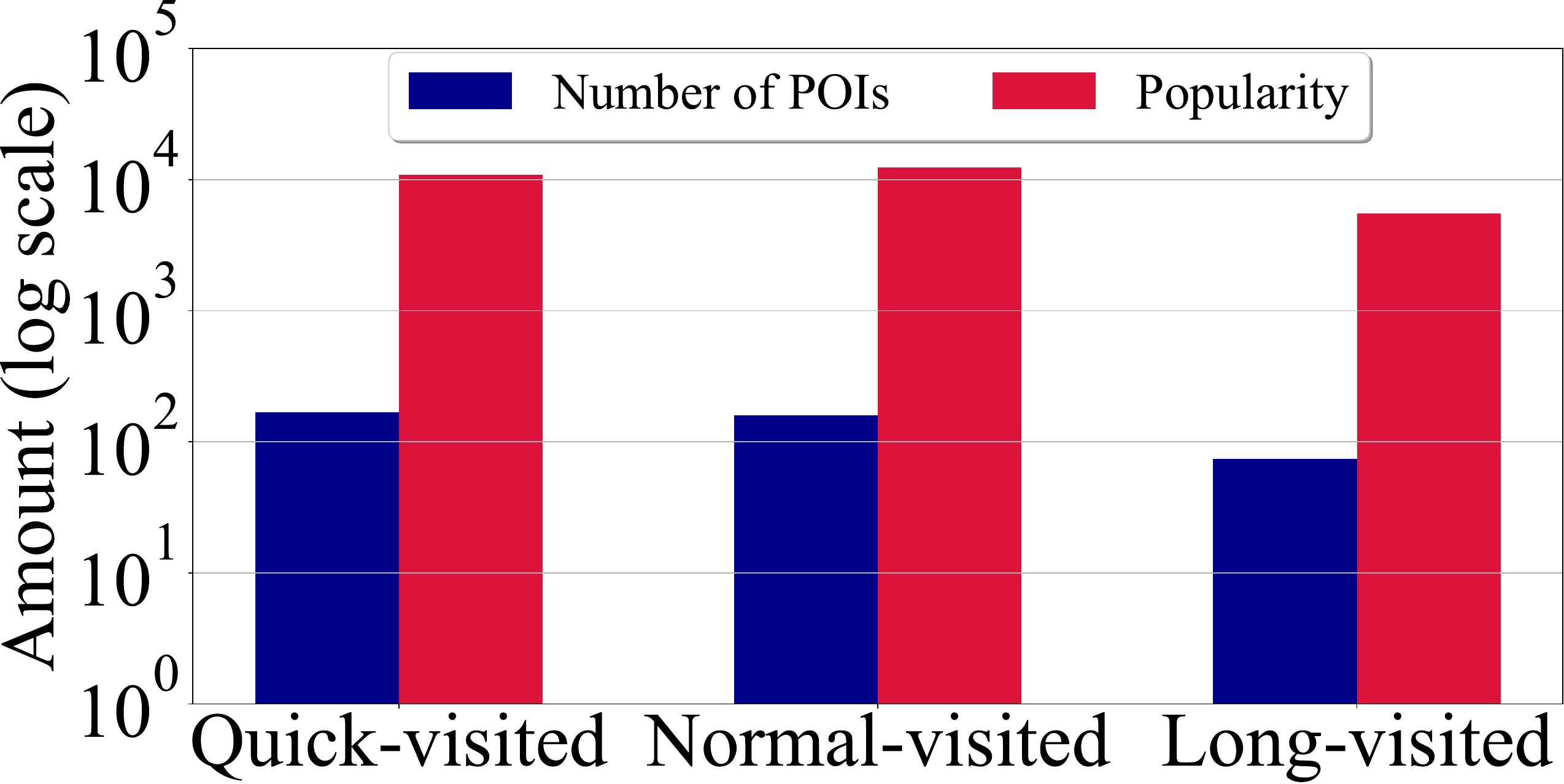}
        \caption{Number of POIs vs. POI Popularity.}
        \label{fig:POIs-Number-Vs-Popularity}
\end{center}
\end{figure}

The result shows that more than 80\% of the POIs are either in the \textit{Quick-visited} group or the \textit{Normal-visited} group, and few POIs go beyond the 1-hour visiting time, confirming the result presented in Figure~\ref{fig:POIs-Visiting-Time}. Figure~\ref{fig:POIs-Number-Vs-Popularity} analyzes the POI popularity (in terms of number of visits) in each group illustrated in Figure~\ref{fig:POIs-Visiting-Time-Clustering}. The \textit{Quick-visited} group contains 168 POIs and a total of 10848 visits; the group \textit{Normal-visited} contains 159 POIs and 12306 visits; and the \textit{Long-visited} group is composed of the remaining 74 POIs with 5516 visits. This result shows that the most popular POIs do not present the highest expected visiting times.

\begin{figure}[!ht]
\begin{center}
 	\includegraphics[width=0.5\textwidth]{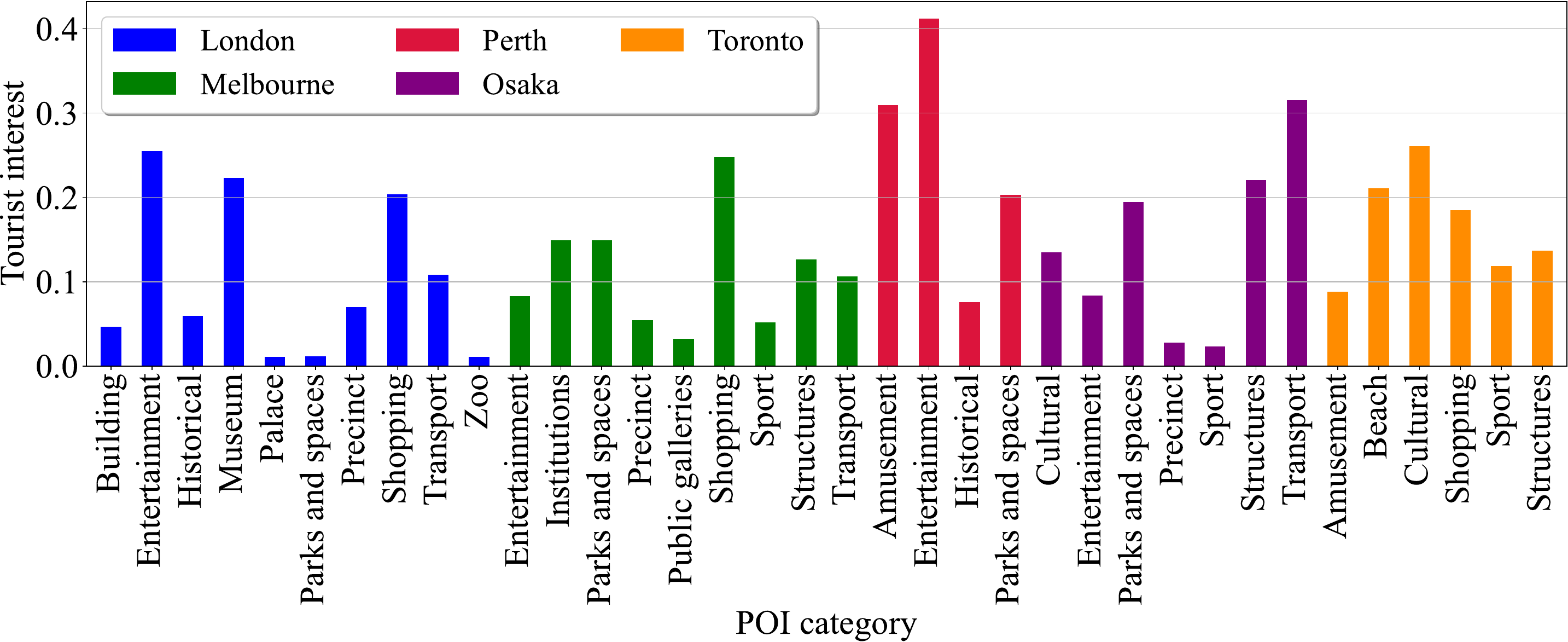}
 	\caption{Distribution of the tourists' interest in the POI categories.}
 	\label{fig:Tourists-Interest}
\end{center}
\end{figure}

The last visiting-related statistic analyzed in our dataset is the distribution of the user's interest in the POI categories, illustrated in Figure~\ref{fig:Tourists-Interest}. The figure shows that the distribution is quite different in the cities. In London and Perth, the user's interest is concentrated in a few categories (e.g., Entertainment, Museum, and Shopping in London and Entertainment and Amusement in Perth). In other cities, it is more evenly distributed. Melbourne is the city where users have visited the most different POI categories.

\begin{figure*}[t]
\centering
    \begin{tabular}{@{}ccccc@{}}
        \includegraphics[width=.18\textwidth]{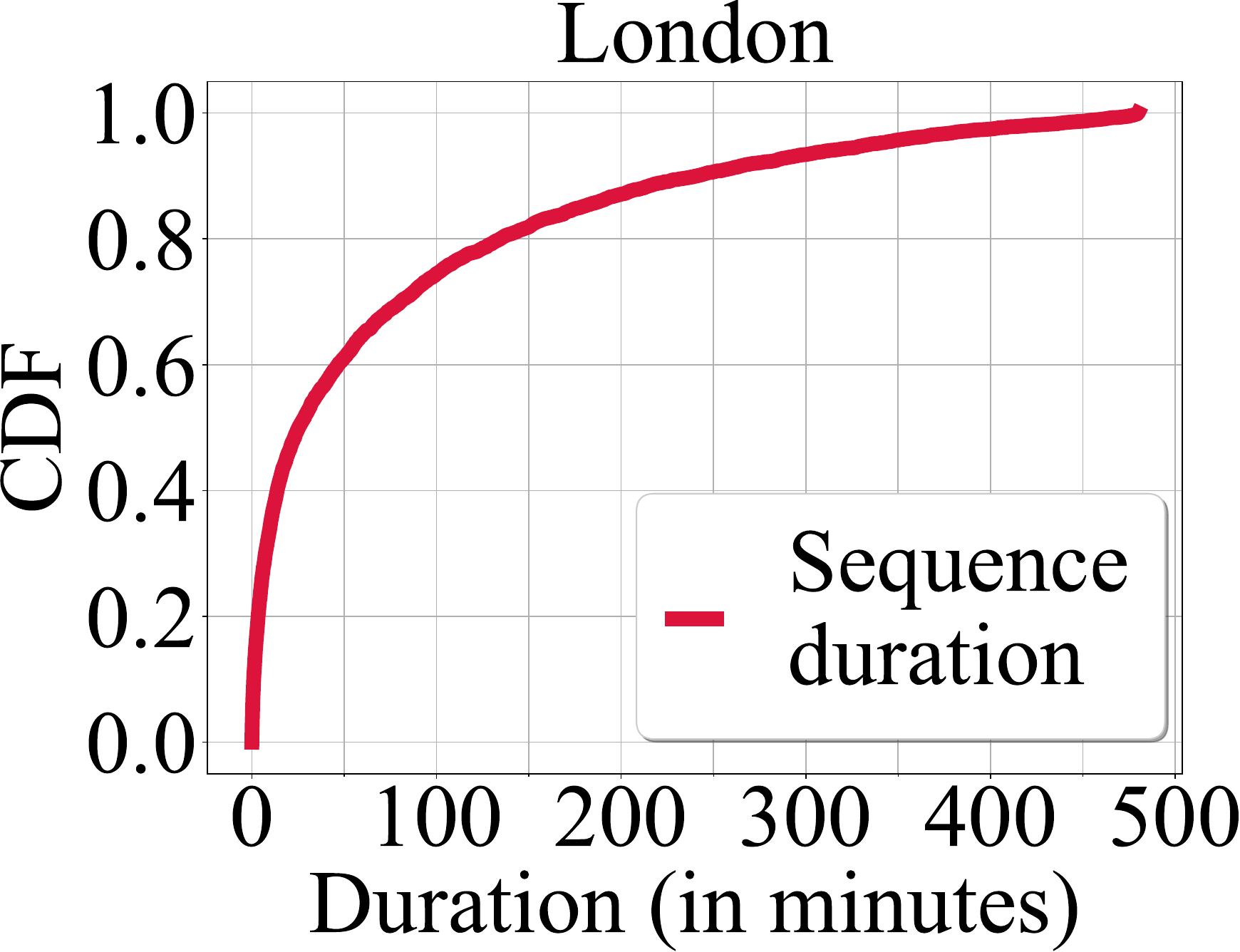} &
        \includegraphics[width=.18\textwidth]{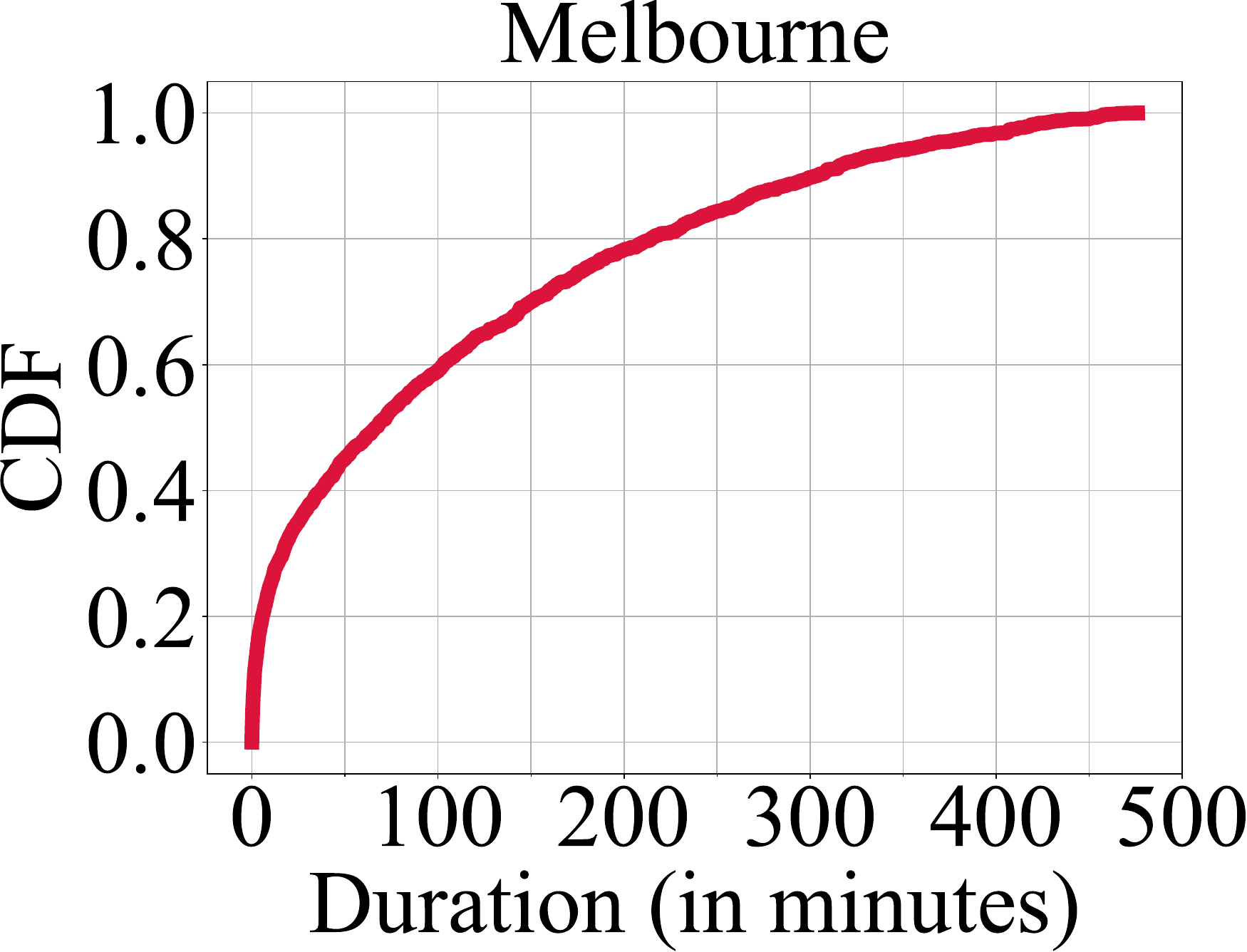} &
        \includegraphics[width=.18\textwidth]{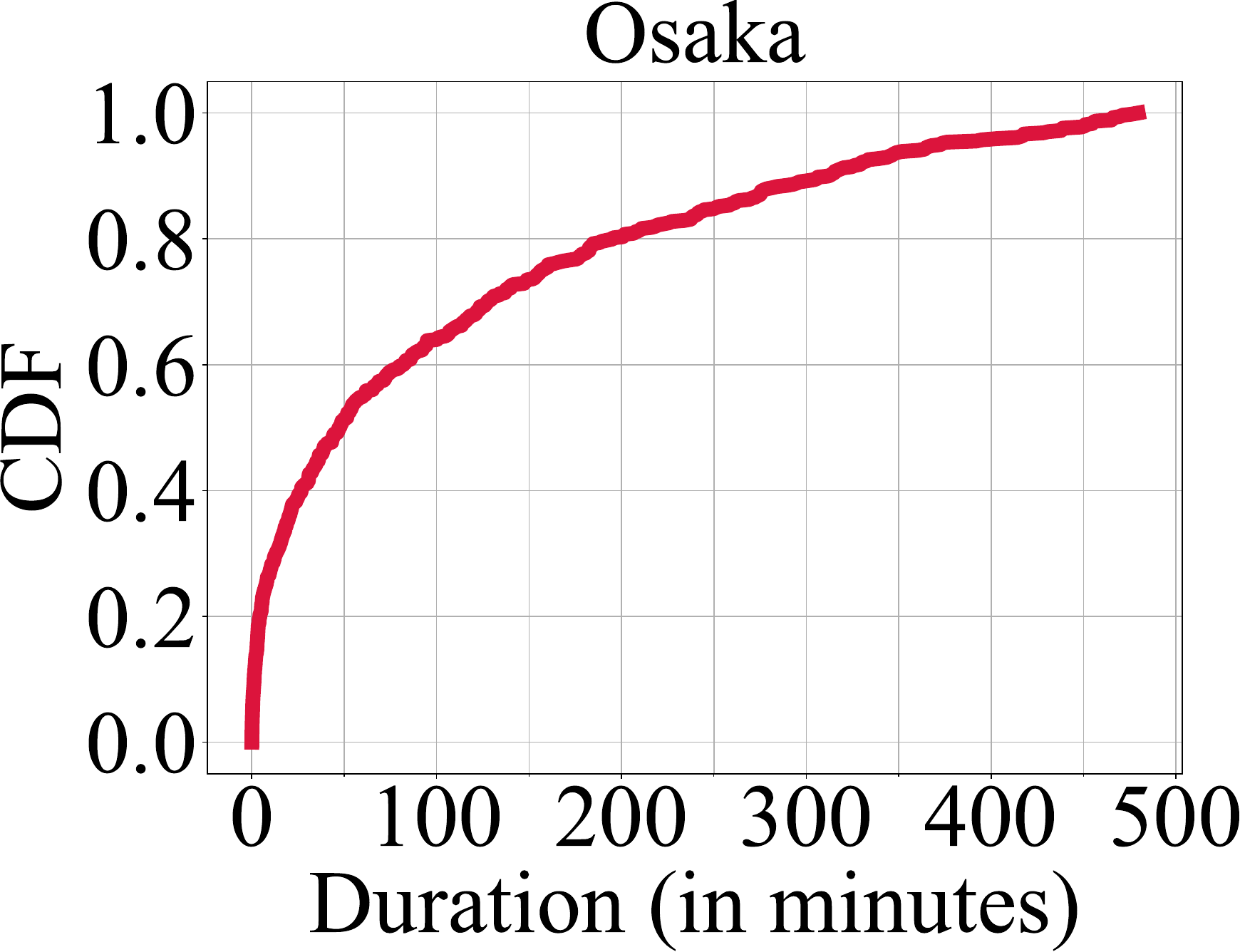} &
        \includegraphics[width=.18\textwidth]{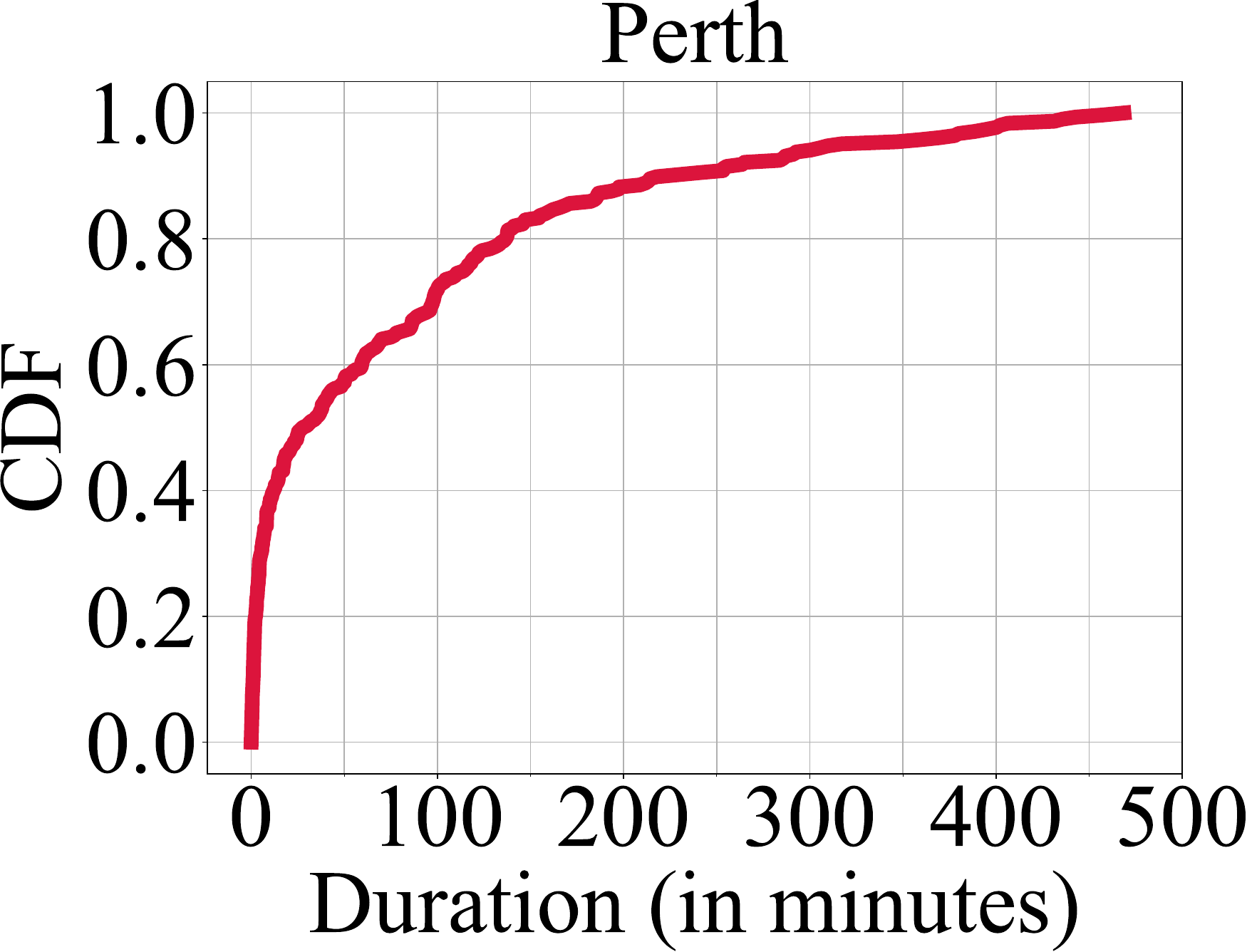} &
        \includegraphics[width=.18\textwidth]{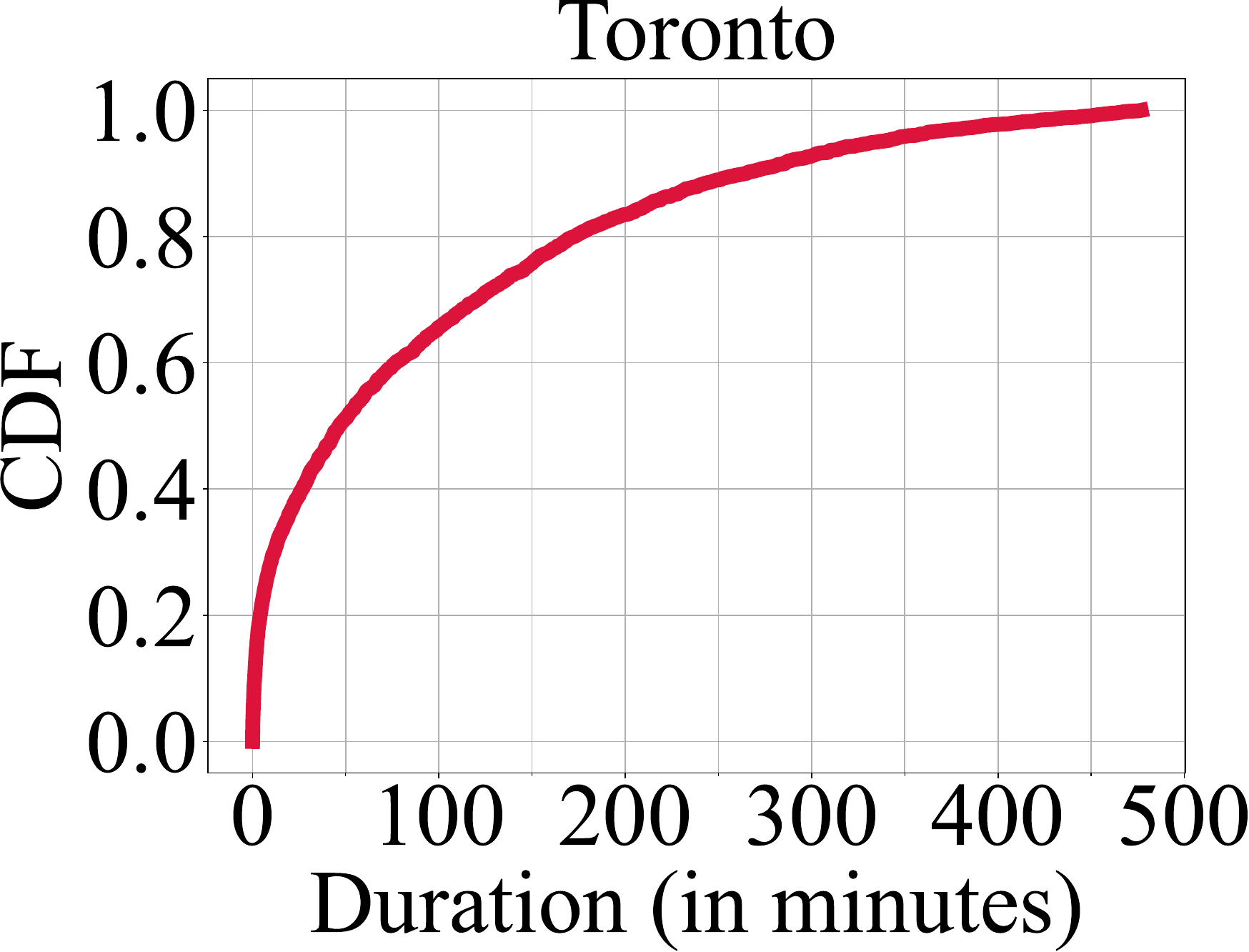}   \\
    \end{tabular}
    \caption{CDF of the travel sequence duration.}
    \label{fig:Sequences-Duration}
\end{figure*}

\begin{figure*}[!ht]
\centering
    \begin{tabular}{@{}ccccc@{}}
        \includegraphics[width=.18\textwidth]{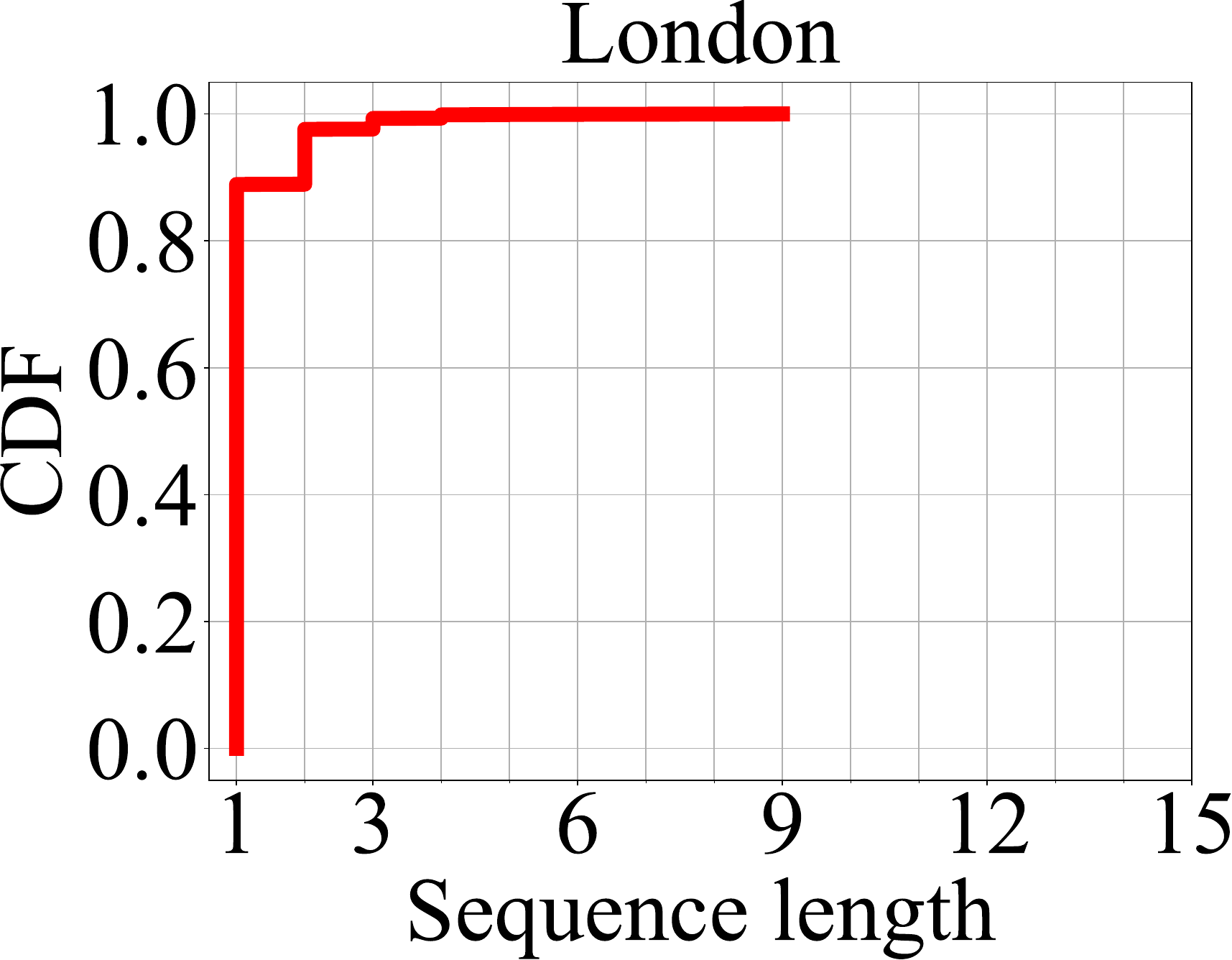} &
        \includegraphics[width=.18\textwidth]{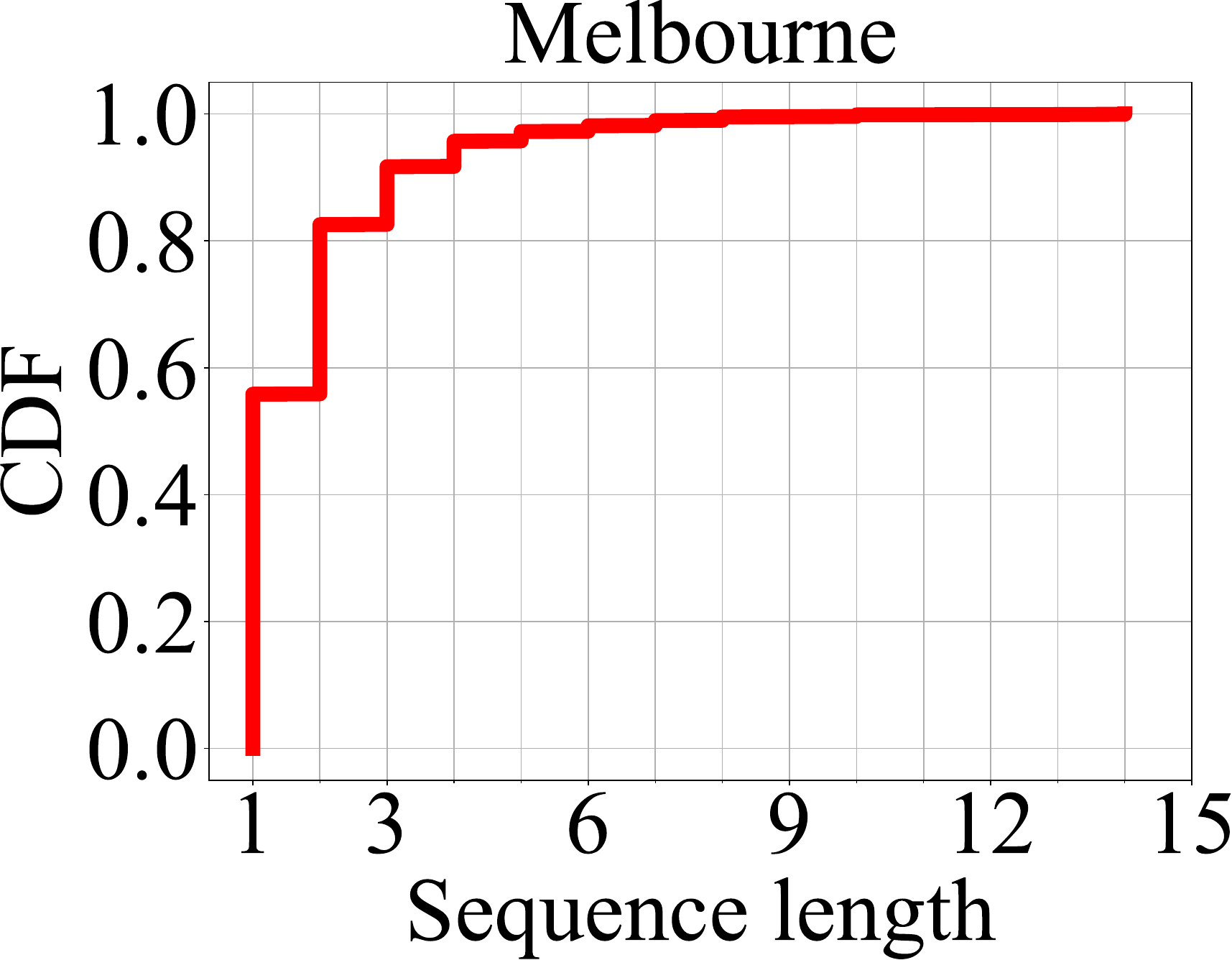} &
        \includegraphics[width=.18\textwidth]{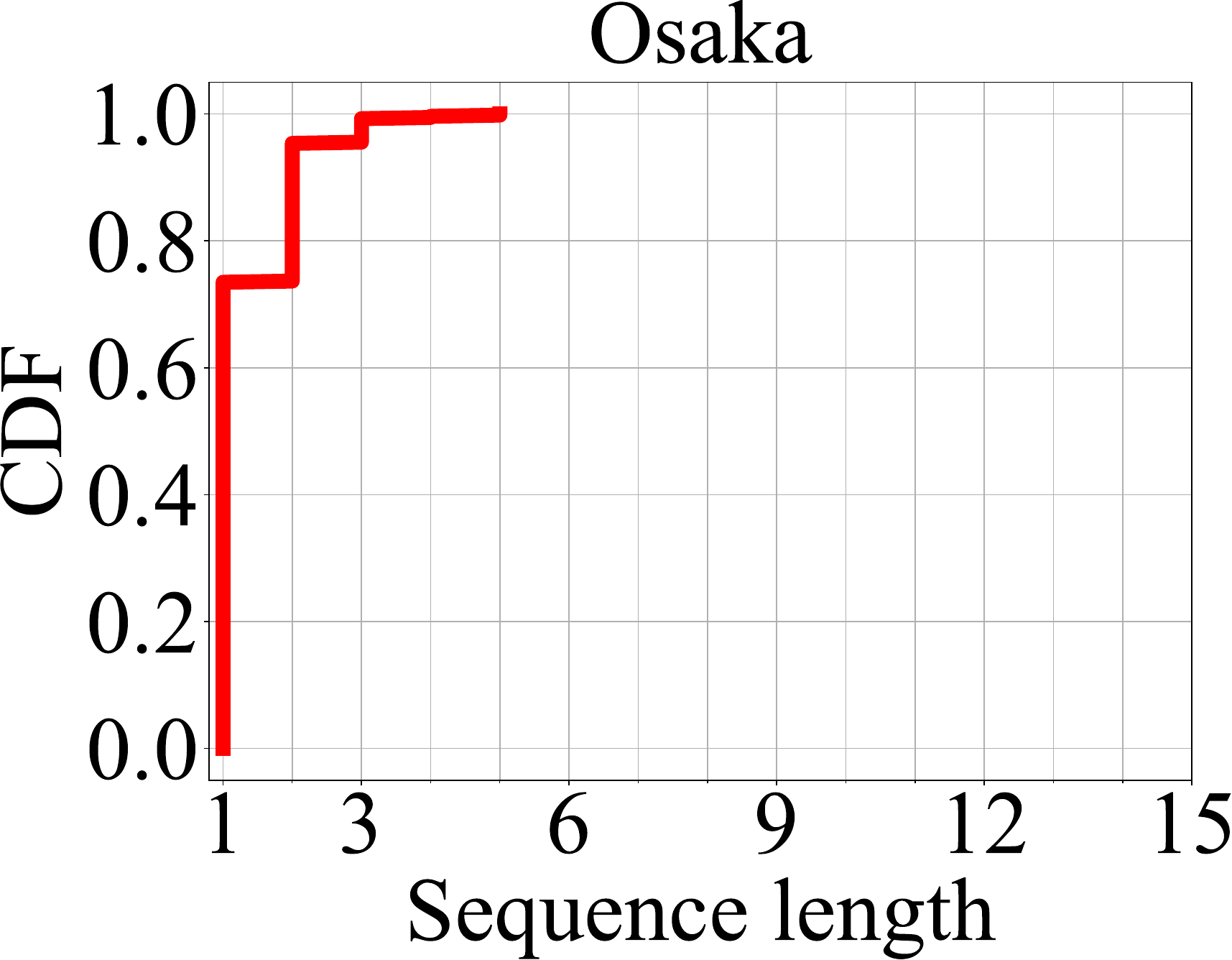} &
        \includegraphics[width=.18\textwidth]{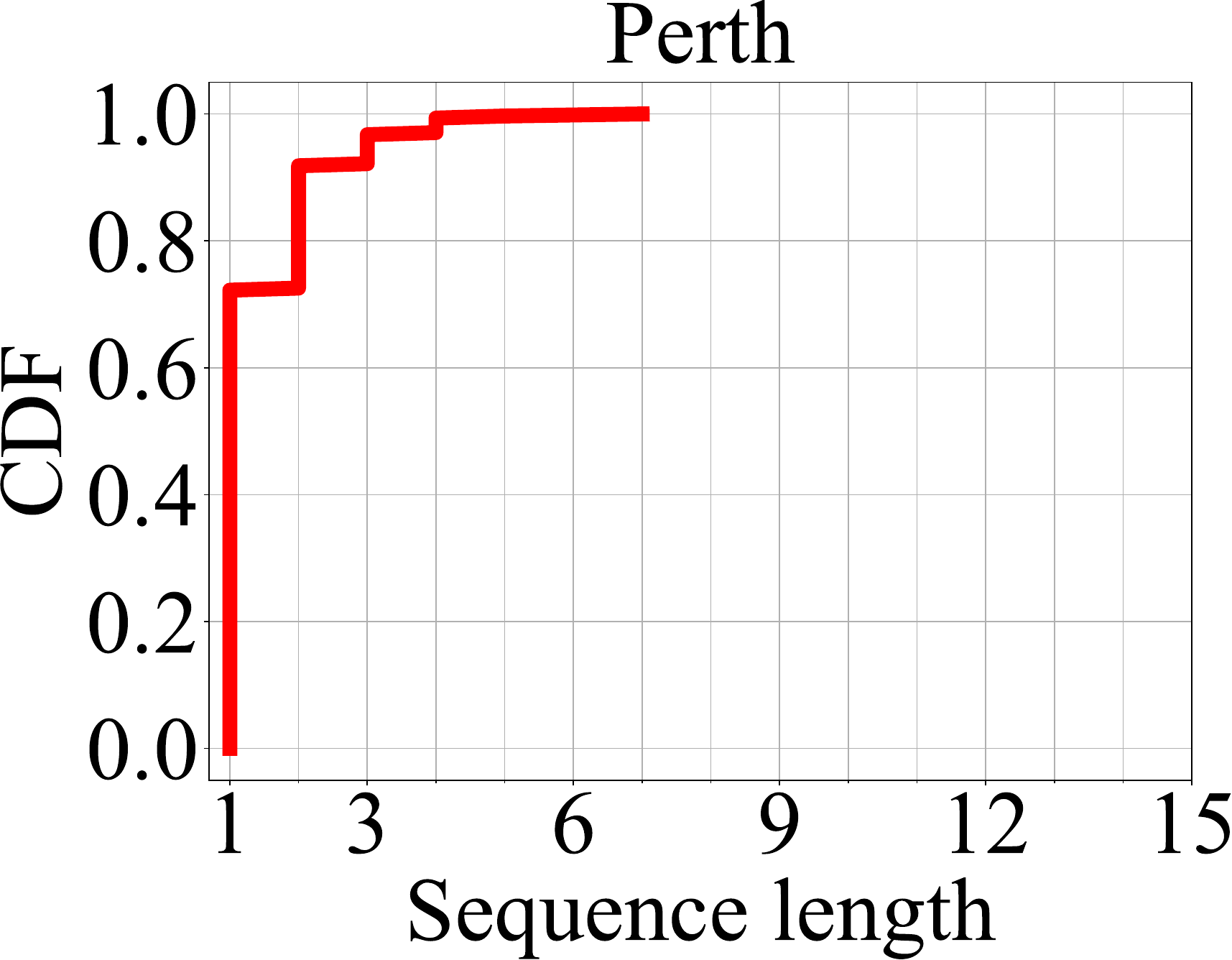} &
        \includegraphics[width=.18\textwidth]{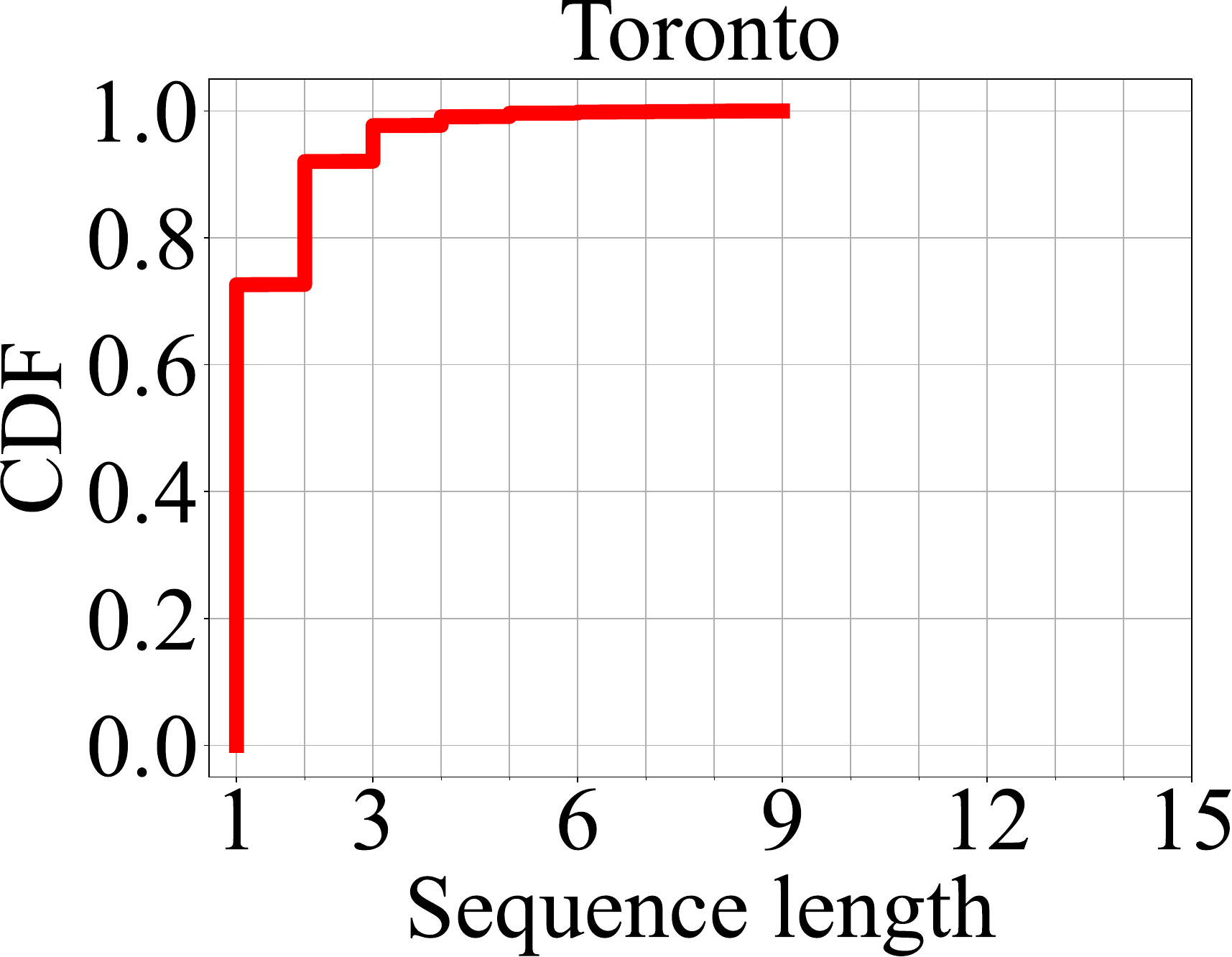}   \\
    \end{tabular}
    \caption{CDF of the travel sequence length.}
   \label{fig:Sequences-Length}
\end{figure*}

To better understand the visiting patterns in our dataset, we analyze the user travel sequences, i.e., $S_u \subset S, \ \forall \ u \in \mathcal{U}$. Figure~\ref{fig:Sequences-Duration} shows the CDF of the travel sequence duration, computed as the time difference between the last photo of the last POI and the first photo of the first POI. The duration of the travel sequences is also short, with 50\% lasting less than 50 minutes in all cities except for Melbourne. Indeed, the travel sequence duration throughout the cities presents a similar behavior, with most sequences having a duration shorter than 2 hours and few travel sequences reaching the 8-hour duration. 

We also analyze the travel sequence length, i.e., the number of POIs composing a travel sequence. Figure~\ref{fig:Sequences-Length} shows the CDF of the travel sequence length. In all analyzed cities, 80\% of the travel sequences are composed of two or fewer POI, evidencing that, in the Flickr dataset, most of the travel sequences are composed of few visits. Indeed, in London, almost 90\% of the travel sequences have a single POI visit. Melbourne stands out for having few sequences with 14 POIs, showing some users with a high exploratory behavior. On the other hand, no tourist in Osaka visited more than 5 POIs in a day.

%% file: Sections/07-Experimental-Evaluation.tex
\section{Experimental Evaluation}\label{sec:evaluation}

In this section, we evaluate the performance of +Tour using our dataset. First, we describe the baseline used for comparison (Section~\ref{sec:eval1}). Then, we describe the evaluation methodology (Section~\ref{sec:eval2}). Finally, we discuss and draw conclusions based on the obtained results (Section~\ref{sec:eval3}).

\subsection{Baseline algorithm}\label{sec:eval1}
Since currently, PersTour~\cite{lim-personalized:18} is the state-of-the-art for PTIR exact solutions targeting the one-day, single-traveler case, we compare +Tour against PersTour. 
%PersTour solves the classical PTIR problem for a single tourist using an Integer Programming Model. 
However, PersTour does not consider MEC resources and service application demands while generating the itinerary. Thus, to use it to solve the MEC-PTIR problem, we adapt the algorithm as follows:

\begin{itemize}
    \item Since PersTour is designed for a single tourist to produce itineraries for every user $u \in \mathcal{U}_d$, we run $|\mathcal{U}_d|$ independent instances of PersTour, i.e., one instance per user.
    \item We then use the single itinerary generated by PersTour for each tourist as input to our implementation that solves the second stage of the MEC-PTIR problem. This approach may lead to infeasible solutions since PersTour generates a single itinerary for each tourist instead of a Pareto-front as the first {stage} of +Tour. When this happens, we solve the problem by relaxing the constraints of minimum allocation for service applications. Consequently, in terms of user experience, the effective performance of PersTour is even worse than the one shown in the following subsections. In the next sections, we refer to this modified version of PersTour as Resource-aware PersTour (RA-PersTour).
\end{itemize}

\subsection{Evaluation methodology}\label{sec:eval2}

In our evaluation, we consider that the MTC offers two applications for tourists: {MAR} and {MVS}, i.e., \\ $\mathcal{A}=\{\textrm{{MAR}}, \textrm{{MVS}}\}$. The requirements of an {MAR} application are taken from~\cite{garcia:18}, while the requirements of a {MVS} application are based on the Netflix stream service website\footnote{https://help.netflix.com/en/node/306}. Table~\ref{tab:appRequirements} shows the network and processing requirements for both applications.

\begin{table*}[!ht]
    \begin{minipage}[!ht]{.45\textwidth}
    \centering
    \footnotesize
    \captionof{table}{Requirements of the applications{, taken from \cite{garcia:18} and Netflix stream service website$^7$.}}
    \label{tab:appRequirements}
    \begin{tabular}{ccccc}
    \hline
    \textbf{Application} & $\boldsymbol{\lambda^{min}_{a}}$ & $\boldsymbol{\lambda^{max}_{a}}$ & $\boldsymbol{\psi^{min}_{a}}$ & $\boldsymbol{\psi^{max}_{a}}$ \\ \hline
    {MAR}             & 1 Mbps     &  10 Mbps & 0.1 RC   & 1 RC \\
    {MVS}               & 1.5 Mbps   &  25 Mbps & 0 (None) & 0 (None) \\
    \hline
    \end{tabular}
    \end{minipage}
\hspace{0.05\textwidth}
    \begin{minipage}[!ht]{.45\textwidth}
    \centering
    \footnotesize
    \captionof{table}{Evaluation scenarios.}
    \label{tab:evalScenarios}
    \begin{tabular}{ccc}
    \hline
    \textbf{Scenario} & $\boldsymbol{\lambda({v_{i}})}$ & $\boldsymbol{{\psi}_{m}}$ \\ \hline
    High network overload     & 75 Mbps   &  37.5 RCs \\
    Medium network overload   & 150 Mbps  &  37.5 RCs \\
    Low network overload      & 300 Mbps  &  37.5 RCs \\
    \hline
    \end{tabular}
    \end{minipage}
\end{table*}

We assume three scenarios for the ICT Infrastructure, as illustrated in Table~\ref{tab:evalScenarios}. Aligned with experimental evidence in~\cite{malandrino:18}, we define the high network overload scenario, where the amount of network resources available at each POI is 75 Mbps. We also define the medium and low network overload scenarios containing double and triple the amount of network resources at each POI. For the computing resources, we consider two MEC hosts ($|\mathcal{M}|= 2$), each one with 37.5 Reference Cores (RCs), i.e., ${\psi}_{m}=$ 37.5 RC, for all scenarios. Based on~\cite{garcia:18}, we assume that one RC is equivalent to an Intel Haswell i7-4770 @3.40GHz processing power.

For each city, we use the set of travel histories ($\mathcal{S}$) to generate a set with 250 users, i.e., for each city, we consider $|\mathcal{U}_d|=$250. We then provide the set of POIs ($G = (\mathcal{V},\mathcal{E})$), the ICT infrastructure scenario (Table~\ref{tab:evalScenarios}), and the set of users ($\mathcal{U}_d$) and their interests (extracted from $\mathcal{S}$) as input to +Tour and RA-PersTour. We evaluate both algorithms using leave-one-out cross-validation~\cite{kohavi:95}, i.e., for each user $u \in \mathcal{U}_d$, we use all travel sequences $S_u$ of $u$ to build her user's interest ($int_{u}(c), \ \forall \ c \in \mathcal{C}$), except the one we want to test. We also use the duration of the testing travel sequence to define, respectively, the user's budget ($b_u$) and the set of applications that she will consume during the tour ($\mathcal{A}_{u}$). The latter is defined based on our analysis presented in Figure~\ref{fig:POIs-Visiting-Time-Clustering} as follows. If the testing travel sequence belongs to the \textit{Quick-visited} group, we randomly assign one of the following options to $\mathcal{A}_{u}$: \{none\} or \{{MVS}\}. Otherwise, we randomly choose one of the following options to $\mathcal{A}_{u}$: \{{MVS}\}, \{{MAR}\}, or \{{MVS}, {MAR}\}. The rationality behind this approach is that users with short budgets will be willing to watch no application or fast {MVS}. Table~\ref{tab:modelParameters} along with Tables~\ref{tab:appRequirements} and~\ref{tab:evalScenarios} summarize the main parameters and values employed in our evaluation.

\begin{table}[htb]
\footnotesize
\centering
\caption{Model parameters.}
\label{tab:modelParameters}
\begin{tabular}{ll}
\hline
\textbf{Parameter} & \textbf{Value} \\ \hline
$|\mathcal{U}_d|$ & 250  \\
$|\mathcal{V}|$ & 19 - 84 \\ 
$|\mathcal{M}|$ & 2 \\ 
$\alpha$ & 0.5  \\
$\Delta{t}$ & 1 minute \\
$T$ & 480 minutes \\
\hline
\end{tabular}
\end{table}

%\begin{figure*}[!ht]
%    \subfloat[All sequences length.]{
%        \begin{tabular}{@{}ccc@{}}
%            \includegraphics[width=.132\textwidth]{Figures/Experimental-Evaluation/Gain/All/High.pdf} &
%            \includegraphics[width=.132\textwidth]{Figures/Experimental-Evaluation/Gain/All/Medium.pdf} &
%            \includegraphics[width=.132\textwidth]{Figures/Experimental-Evaluation/Gain/All/Low.pdf} \\
%        \end{tabular}
%    \label{fig:gain-All}}
%    \hfill
%    \subfloat[Sequences length greater than 3.]{
%        \begin{tabular}{@{}ccc@{}}
%            \includegraphics[width=.14\textwidth]{Figures/Experimental-Evaluation/Gain/G3/High.pdf} &
%            \includegraphics[width=.14\textwidth]{Figures/Experimental-Evaluation/Gain/G3/Medium.pdf} &
%            \includegraphics[width=.14\textwidth]{Figures/Experimental-Evaluation/Gain/G3/Low.pdf} \\
%        \end{tabular}
%    \label{fig:gain-G3}}
%    \caption{+Tour gain compared to RA-PersTour, separated by size of itineraries sequences.}
%    \label{fig:gain}
%\end{figure*}

%############################################################
\begin{table*}[hbt]
\centering

\footnotesize
\caption{Tour gain compared to RA-PersTour.}
\label{tab:gain}
\begin{tabular}{c|ccccc|ccccc|ccccc}
\hline
\multicolumn{16}{c}{\textbf{All sequences lengths}} \\\hline
\multicolumn{1}{c}{} & \multicolumn{5}{c|}{\textbf{High overload}} & \multicolumn{5}{c|}{\textbf{Medium overload}} & \multicolumn{5}{c}{\textbf{Low overload}} \\\hline
 & \textbf{LO} & \textbf{ME} & \textbf{OS} & \textbf{PE} & \textbf{TO} & 
\textbf{LO} & \textbf{ME} & \textbf{OS} & \textbf{PE} & \textbf{TO} & 
\textbf{LO} & \textbf{ME} & \textbf{OS} & \textbf{PE} & \textbf{TO} \\\hline
%------------------------------------------------------------------------------
+T AE & 0.653 & 0.794 & 0.584 & 0.514 & 0.672 & 0.746 & 0.911 & 0.687 & 0.602 & 0.797 & 0.872 & 0.942 & 0.798 & 0.718 & 0.874 \\\hline
RA AE & 0.649 & 0.731 & 0.552 & 0.461 & 0.657 & 0.744 & 0.852 & 0.665 & 0.579 & 0.793 & 0.869 & 0.885 & 0.791 & 0.704 & 0.870 \\\hline
Gain AE &0.67\%&\hspace{-0.2cm}8.64\%&\hspace{-0.2cm}5.67\%&\hspace{-0.2cm}11.47\% &\hspace{-0.2cm} 2.29\% &\hspace{-0.2cm} 0.32\% &\hspace{-0.2cm} 6.95\% &\hspace{-0.2cm} 3.24\% &\hspace{-0.2cm} 3.88\% &\hspace{-0.2cm} 0.49\% &\hspace{-0.2cm} 0.27\% &\hspace{-0.2cm} 6.41\% &\hspace{-0.2cm} 0.80\% &\hspace{-0.2cm} 2.03\% &\hspace{-0.2cm} 0.53\% \\\hline
%------------------------------------------------------------------------------
+T UE & 0.106 & 0.119 & 0.106 & 0.135 & 0.134 & 0.114 & 0.131 & 0.121 & 0.153 & 0.152 & 0.125 & 0.135 & 0.138 & 0.176 & 0.165 \\\hline
RA UE & 0.103 & 0.085 & 0.084 & 0.099 & 0.123 & 0.112 & 0.097 & 0.108 & 0.137 & 0.151 & 0.124 & 0.101 & 0.132 & 0.171 & 0.163 \\\hline
Gain UE &2.25\%&\hspace{-0.2cm}40.6\%&\hspace{-0.2cm}25.7\%&\hspace{-0.2cm}35.6\% &\hspace{-0.2cm} 8.85\% &\hspace{-0.2cm} 1.19\% &\hspace{-0.2cm} 35.5\% &\hspace{-0.2cm} 12.2\% &\hspace{-0.2cm} 11.2\% &\hspace{-0.2cm} 0.84\% &\hspace{-0.2cm} 1.08\% &\hspace{-0.2cm} 33.7\% &\hspace{-0.2cm} 4.42\% &\hspace{-0.2cm} 2.81\% &\hspace{-0.2cm} 0.94\% \\\hline
%------------------------------------------------------------------------------
\multicolumn{16}{c}{\textbf{Sequences greater than 3}} \\\hline
%------------------------------------------------------------------------------
+T AE & 0.904 & 0.863 & 0.843 & 0.840 & 0.845 & 0.937 & 0.952 & 0.917 & 0.901 & 0.932 & 0.941 & 0.990 & 0.952 & 0.949 & 0.982 \\\hline
RA AE & 0.868 & 0.642 & 0.603 & 0.482 & 0.756 & 0.917 & 0.749 & 0.751 & 0.756 & 0.908 & 0.921 & 0.796 & 0.906 & 0.914 & 0.954 \\\hline
Gain AE &4.15\%&\hspace{-0.2cm}34.3\%&\hspace{-0.2cm}39.7\%&\hspace{-0.2cm}74.2\% &\hspace{-0.2cm} 11.8\% &\hspace{-0.2cm} 2.18\% &\hspace{-0.2cm} 27.1\% &\hspace{-0.2cm} 22.2\% &\hspace{-0.2cm} 19.2\% &\hspace{-0.2cm} 2.57\% &\hspace{-0.2cm} 2.17\% &\hspace{-0.2cm} 24.4\% &\hspace{-0.2cm} 5.05\% &\hspace{-0.2cm} 3.77\% &\hspace{-0.2cm} 2.91\% \\\hline
%------------------------------------------------------------------------------
+T UE & 0.538 & 0.355 & 0.419 & 0.640 & 0.565 & 0.555 & 0.379 & 0.472 & 0.726 & 0.625 & 0.560 & 0.389 & 0.480 & 0.789 & 0.654 \\\hline
RA UE & 0.519 & 0.231 & 0.254 & 0.389 & 0.500 & 0.544 & 0.261 & 0.357 & 0.619 & 0.617 & 0.549 & 0.273 & 0.441 & 0.775 & 0.645 \\\hline
Gain UE &3.70\%&\hspace{-0.2cm}53.4\%&\hspace{-0.2cm}65.1\%&\hspace{-0.2cm}64.2\% &\hspace{-0.2cm} 12.9\% &\hspace{-0.2cm} 2.04\% &\hspace{-0.2cm} 45.2\% &\hspace{-0.2cm} 32.2\% &\hspace{-0.2cm} 17.3\% &\hspace{-0.2cm} 1.22\% &\hspace{-0.2cm} 2.02\% &\hspace{-0.2cm} 42.8\% &\hspace{-0.2cm} 8.90\% &\hspace{-0.2cm} 1.80\% &\hspace{-0.2cm} 1.41\% \\\hline
%------------------------------------------------------------------------------
\end{tabular}
\end{table*}
%############################################################

We use \textbf{Recall}, \textbf{Precision}, and \textbf{F-score} to evaluate +Tour and RA-PersTour. We also introduce two new metrics, namely, \textbf{Allocation Efficiency} (\textbf{AE}) and \textbf{User Experience} (\textbf{UE}) to assess the performance of the algorithms on allocating resources in the network edge and on the overall user experience provided by the recommended itinerary. Assuming that $S_{u}$ is the real-world testing travel sequence and $I_{u*}^k$ is the generated itinerary, these metrics are defined as follows.

\begin{itemize}
    \item \textbf{Recall}: the fraction of POIs from the real-world testing travel sequence that also exists in the generated itinerary.
    \item \textbf{Precision}: the fraction of POIs from the generated itinerary that also exists in the real-world testing travel sequence.
    \item \textbf{F-score}: the model accuracy represented by the harmonic mean of \textbf{Recall} and \textbf{Precision}.
    \item \textbf{Allocation Efficiency}: the amount of computing and network resource allocated to the generated itinerary, relative to the maximum amount demanded by the set of chosen applications, i.e.:
    {\setlength{\mathindent}{0cm}
    \begin{align}
    %\begin{split}
        & AE(I_{u*}^k) = \sum\limits_{v_{i} \in \mathcal{V} \setminus \{v_{0}, v_{n+1}\}}\frac{p(I_{u*}^k,v_{i})}{2|I_{u*}^k \setminus \{v_{0}, v_{n+1}\}|\lambda_{u}^{max}} \ + \nonumber \\
        & \sum\limits_{v_{i} \in \mathcal{V} \setminus \{v_{0}, v_{n+1}\}}\sum\limits_{m \in \mathcal{M}}\frac{q(I_{u*}^k,v_{i},m)}{2|I_{u*}^k \setminus \{v_{0}, v_{n+1}\}|\psi_{u}^{max}}.
    %\end{split}
\end{align}}    
    \item \textbf{User Experience}: determines the relation between the allocation efficiency associated with the generated itinerary (virtual experience) and the profit as perceived by the user for this itinerary (physical experience), i.e.:
    \begin{equation}
    UE(I_{u*}^k) = AE(I_{u*}^k)Norm(Prof(I_{u*}^k)).
    \end{equation}
\end{itemize}

\subsection{Results and discussions}\label{sec:eval3}

\begin{figure*}[!ht]
    \begin{tabular}{@{}ccccc@{}}
        \includegraphics[width=.18\textwidth]{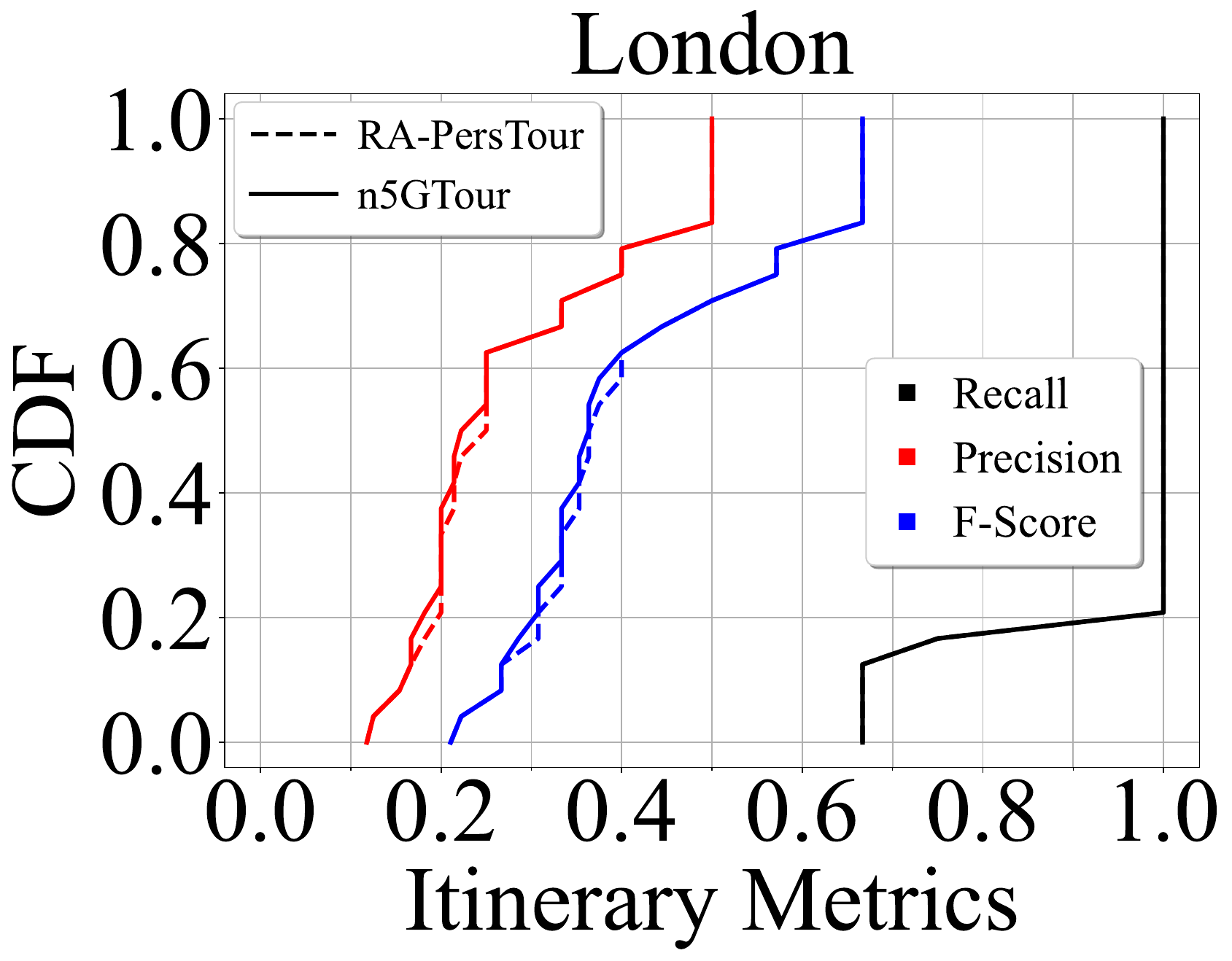} &
        \includegraphics[width=.18\textwidth]{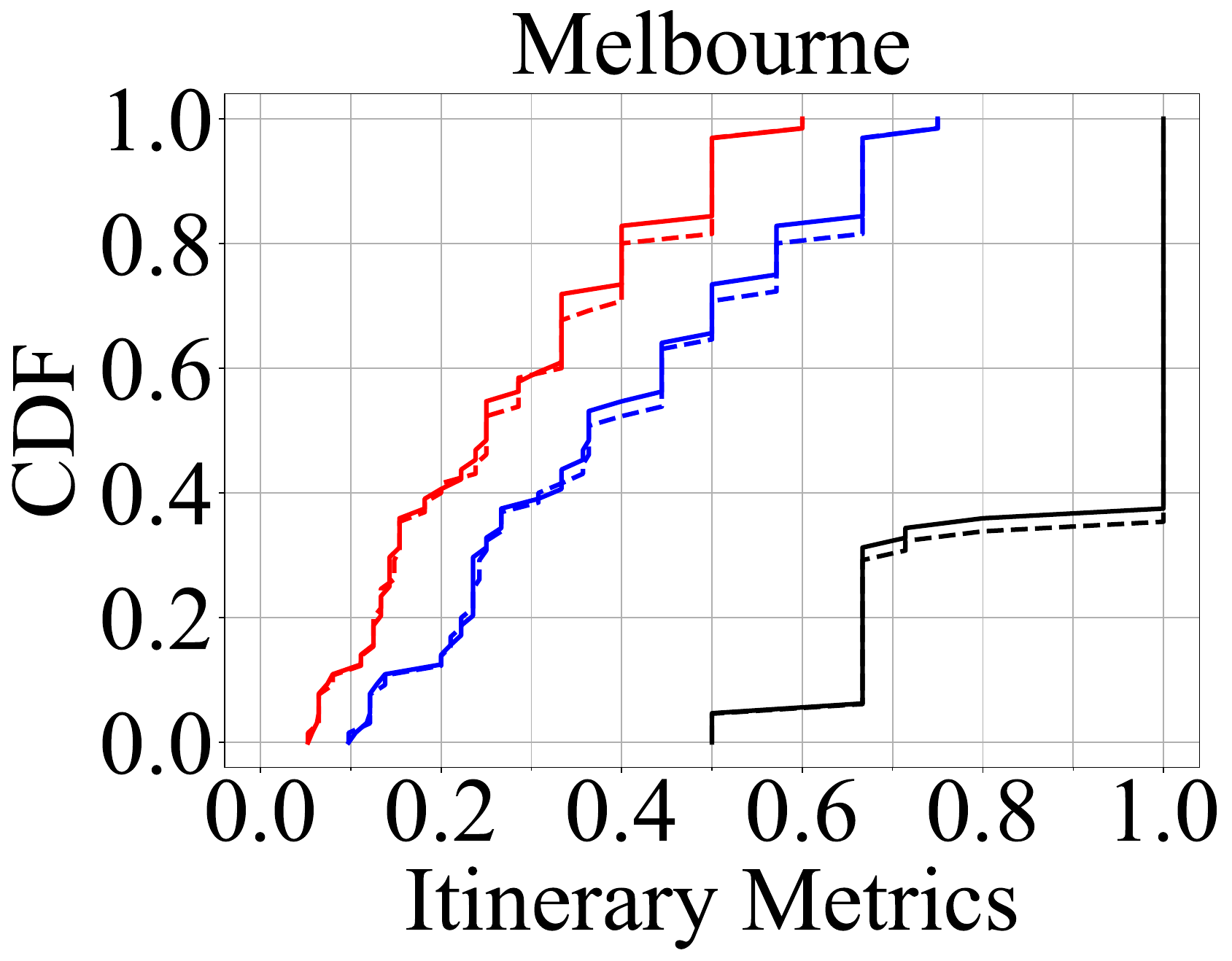} &
        \includegraphics[width=.18\textwidth]{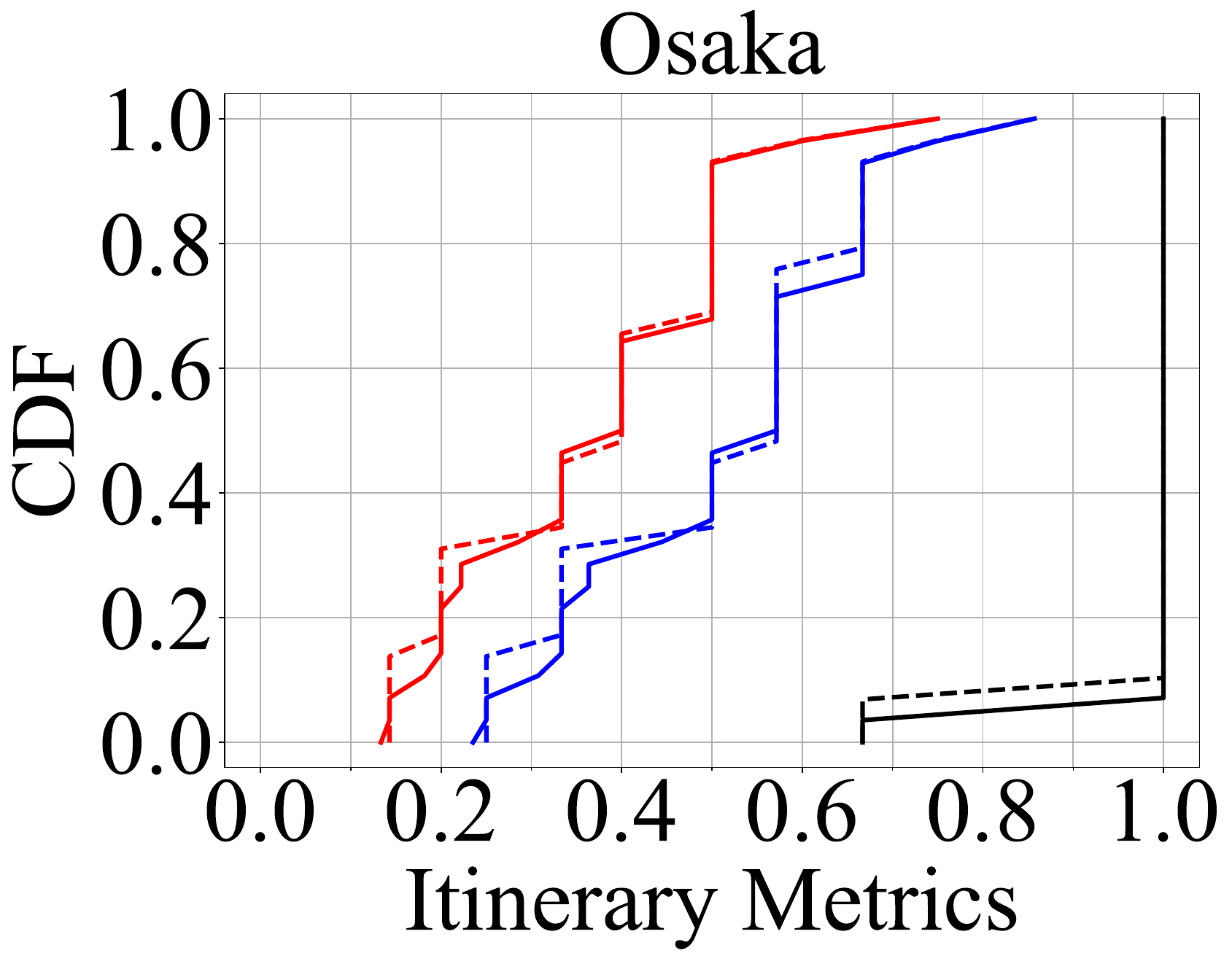} &
        \includegraphics[width=.18\textwidth]{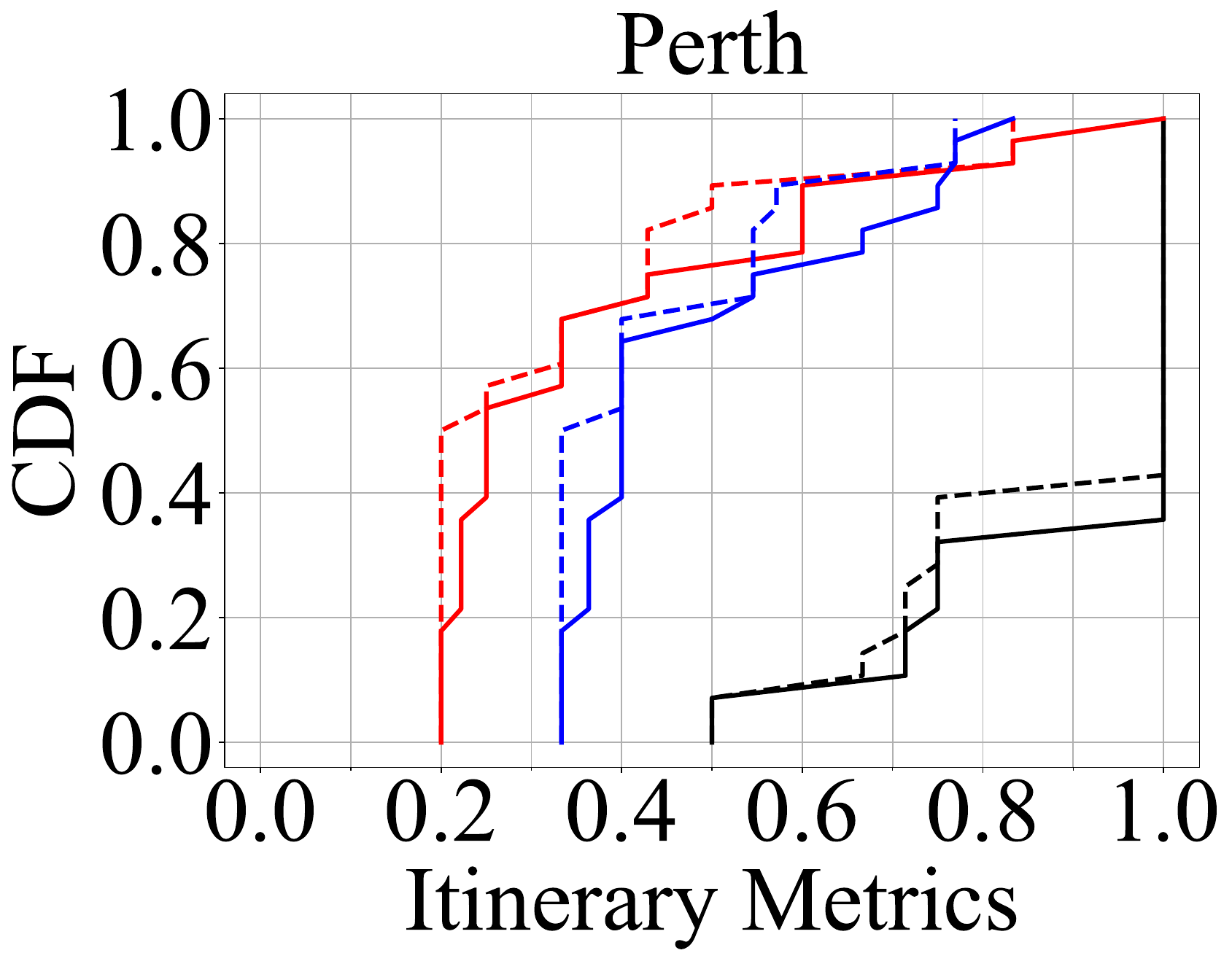} &
        \includegraphics[width=.18\textwidth]{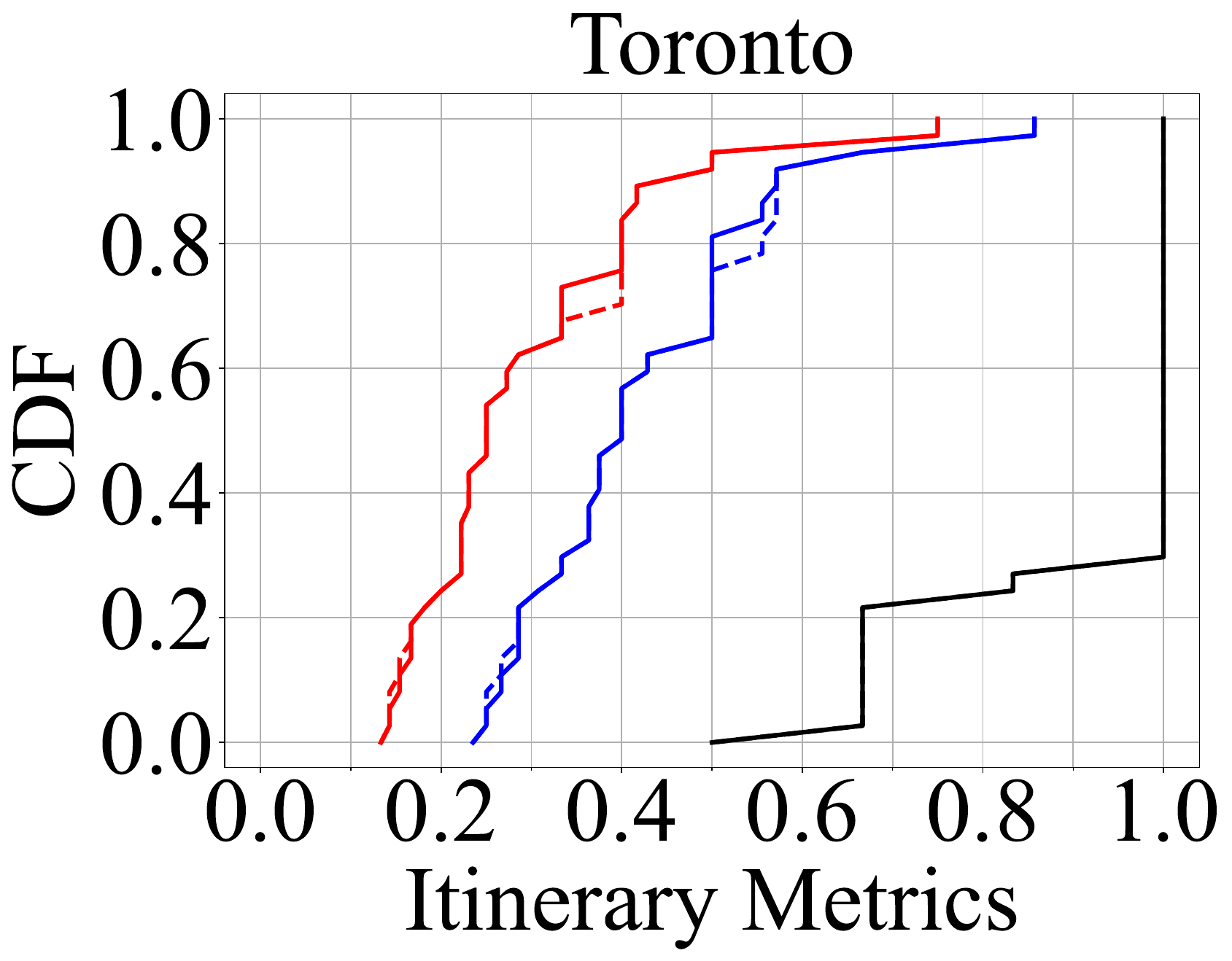} \\
    \end{tabular}
\end{figure*}

\begin{figure*}[!ht]
    \begin{tabular}{@{}ccccc@{}}
        \includegraphics[width=.18\textwidth]{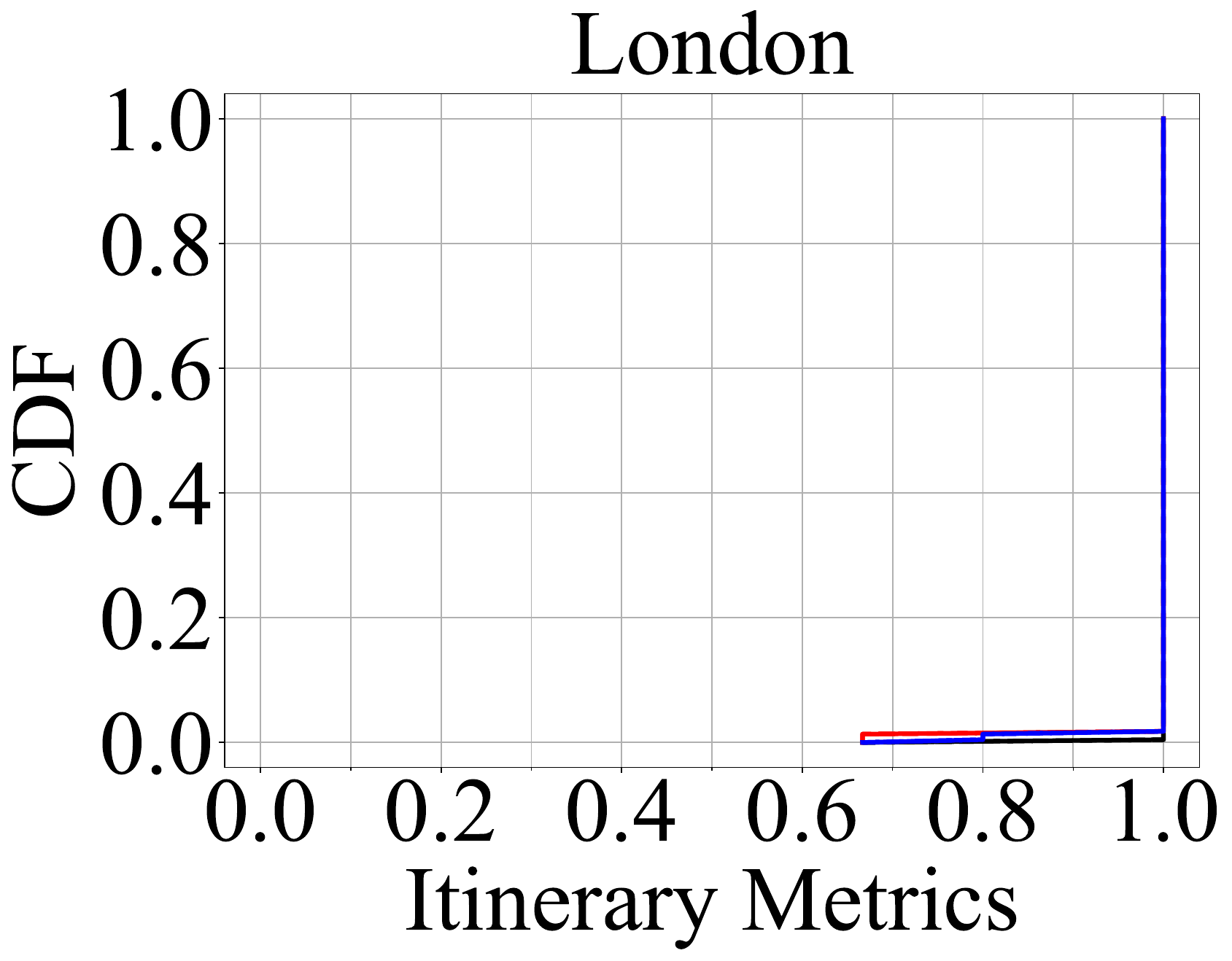} &
        \includegraphics[width=.18\textwidth]{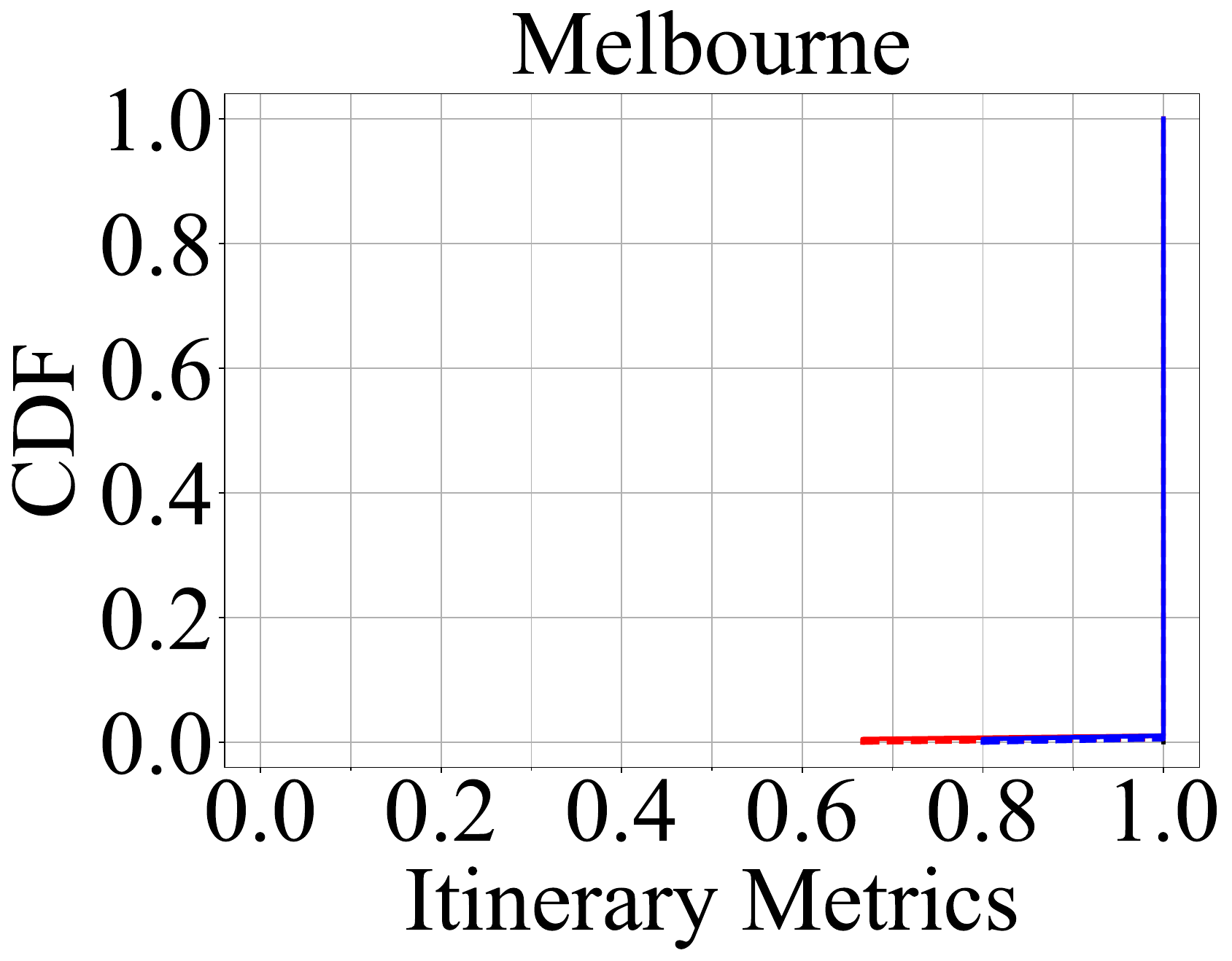} &
        \includegraphics[width=.18\textwidth]{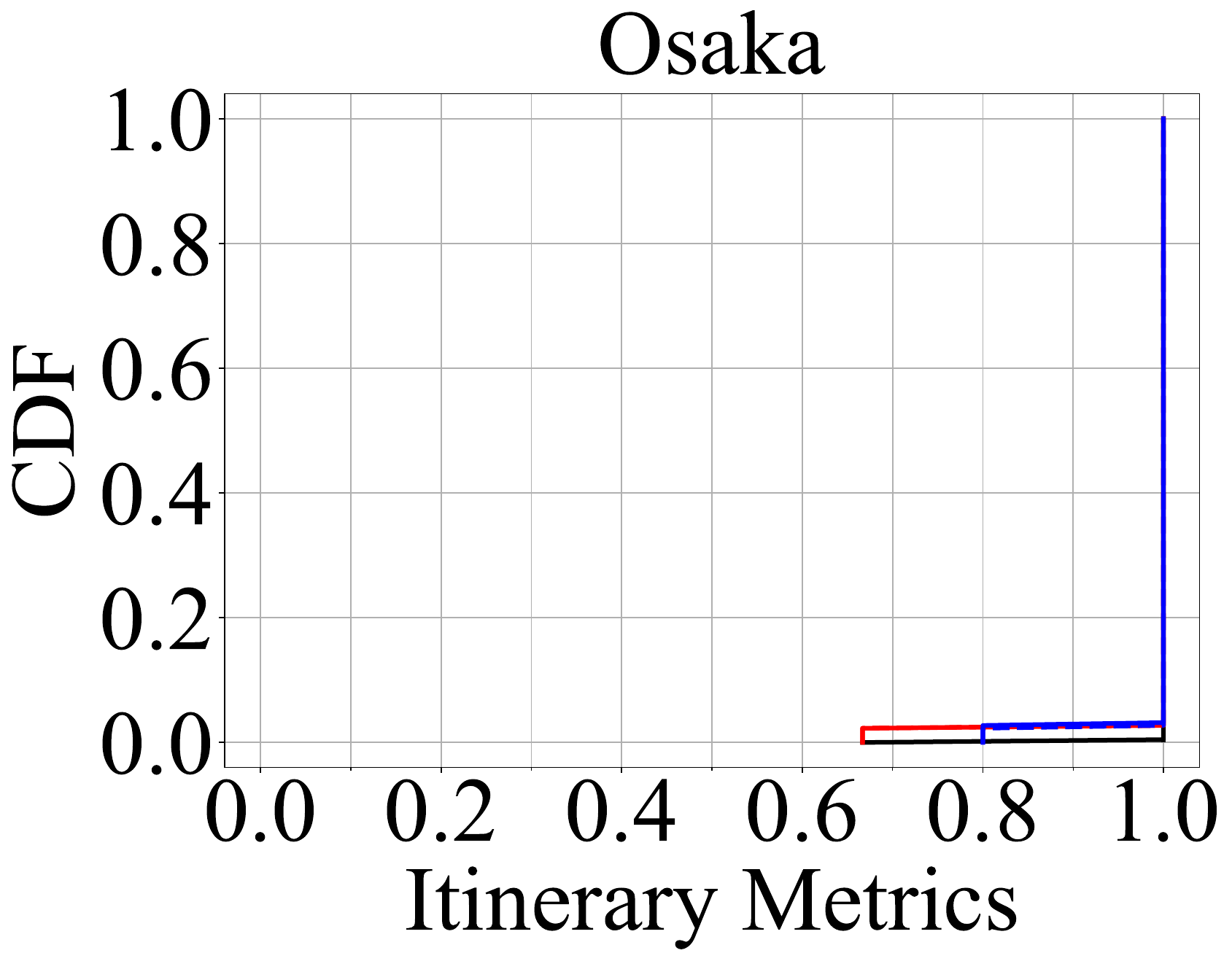} &
        \includegraphics[width=.18\textwidth]{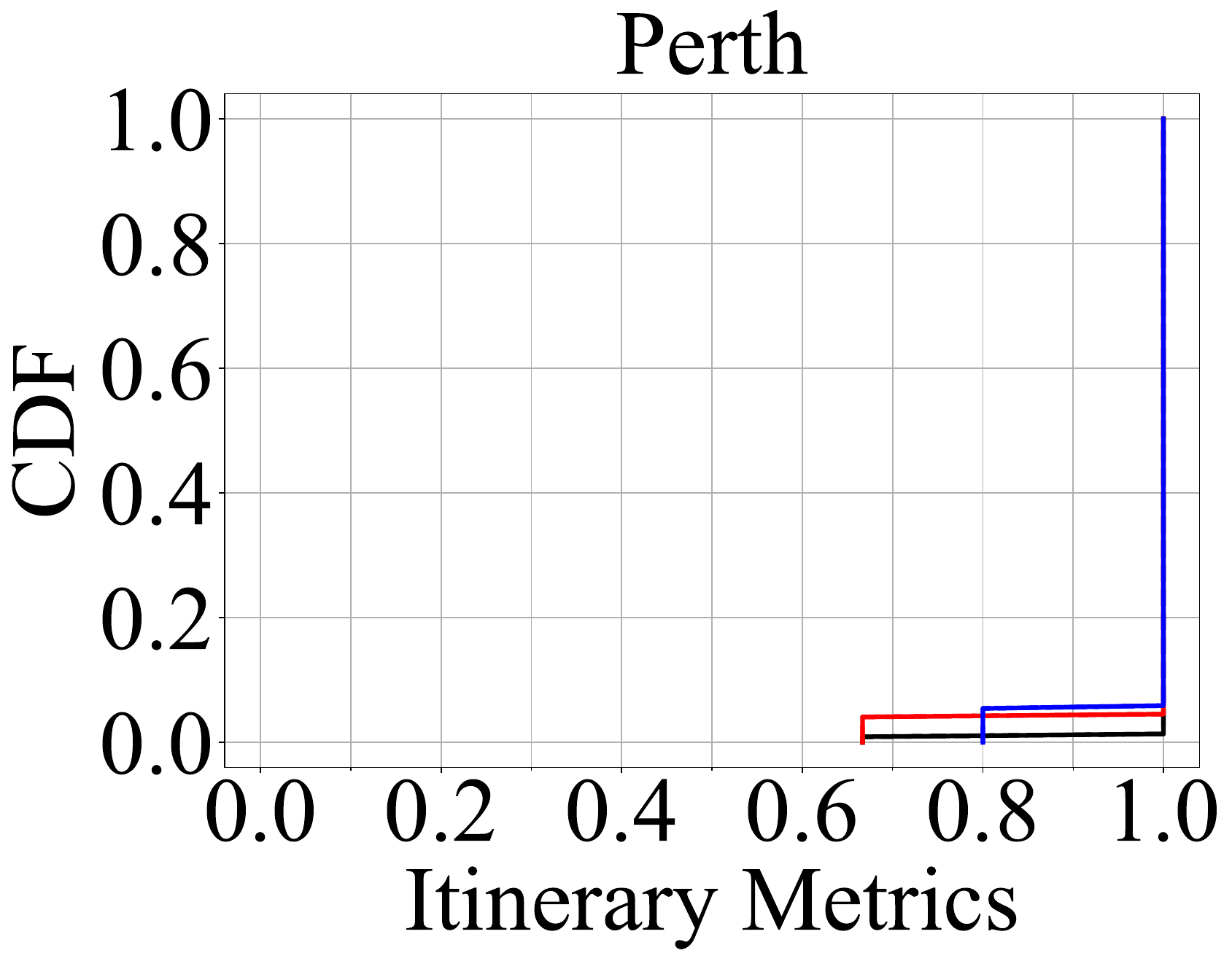} &
        \includegraphics[width=.18\textwidth]{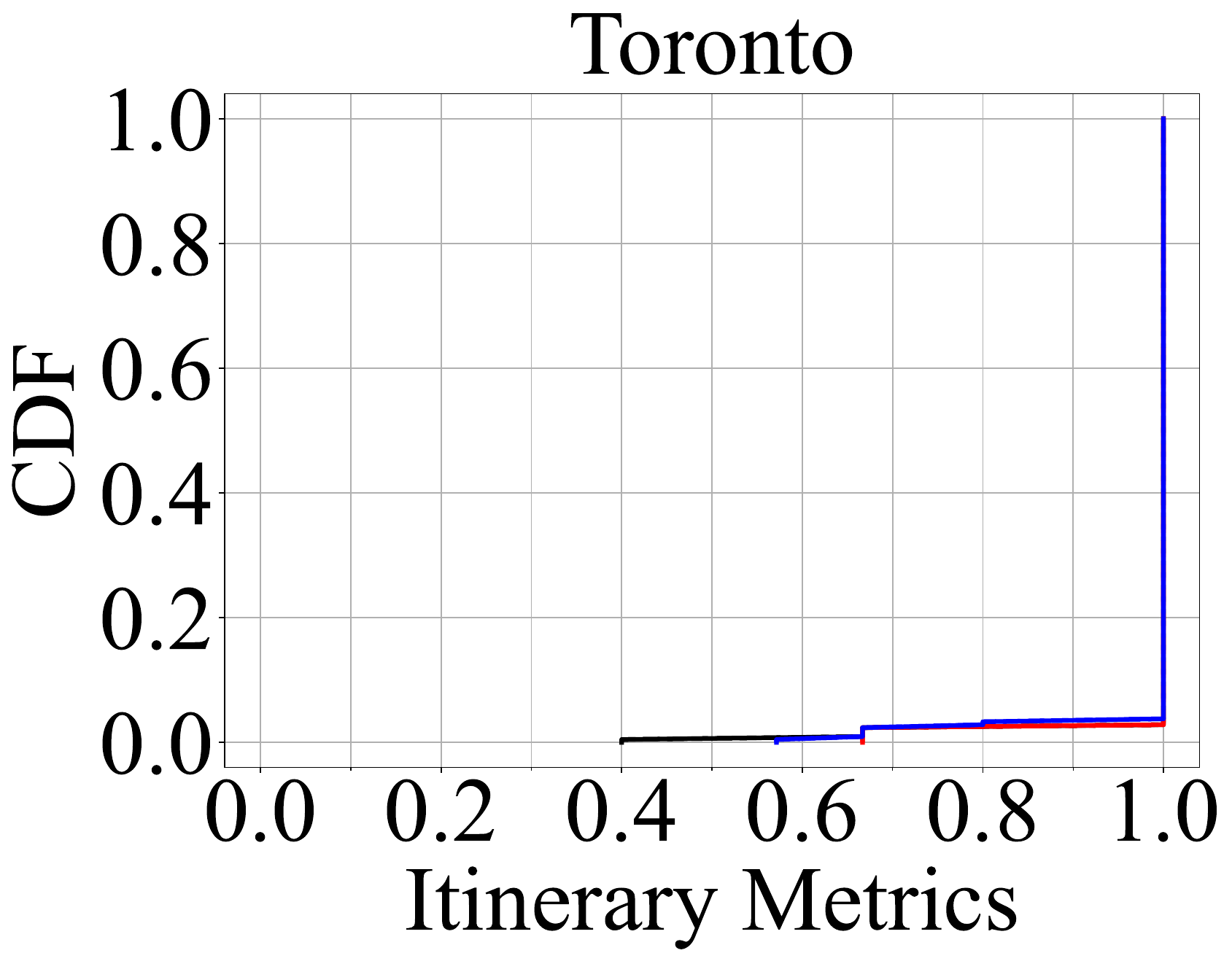} \\
    \end{tabular}
\end{figure*}

\begin{figure*}[!ht]
    \begin{tabular}{@{}ccccc@{}}
        \includegraphics[width=.18\textwidth]{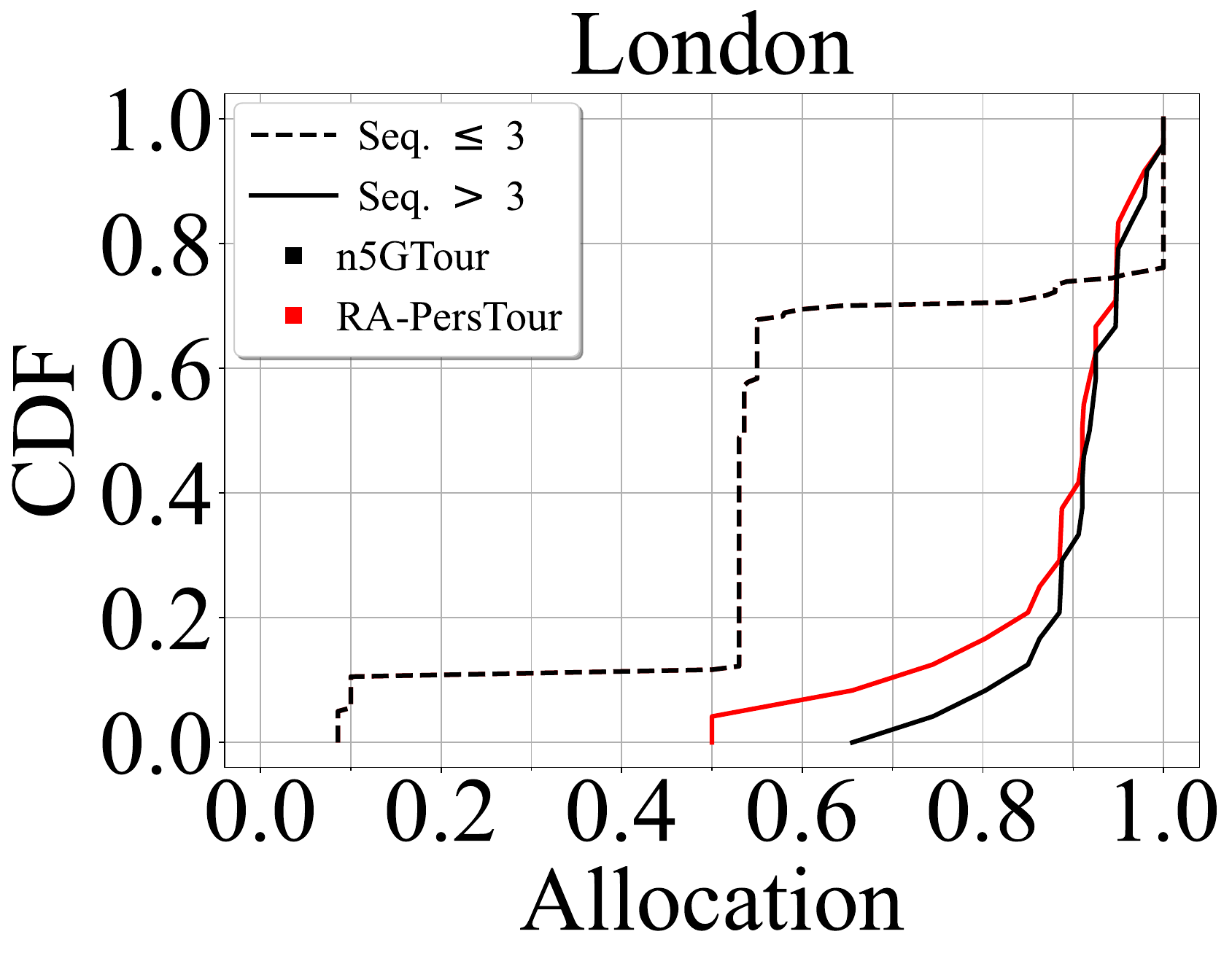} &
        \includegraphics[width=.18\textwidth]{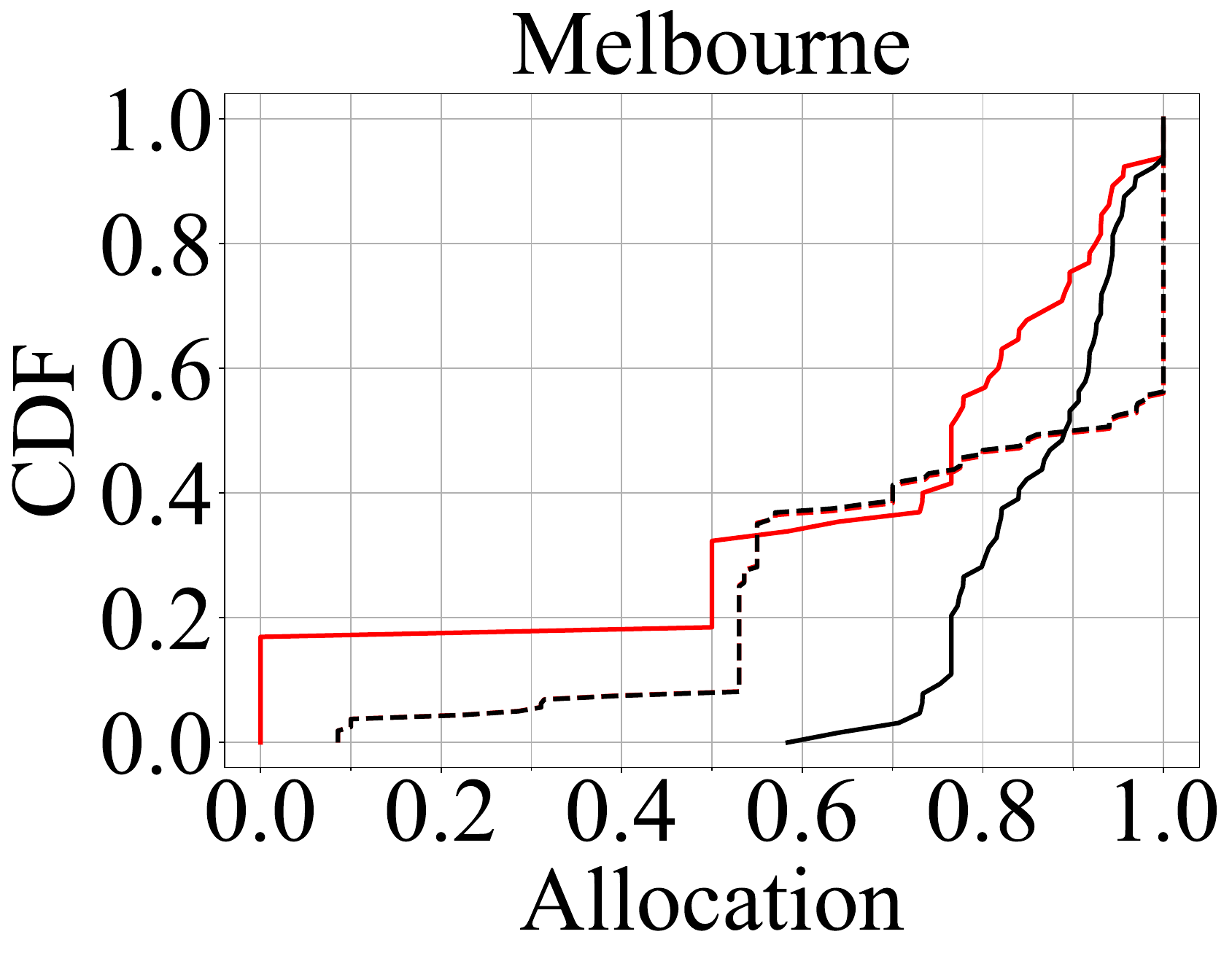} &
        \includegraphics[width=.18\textwidth]{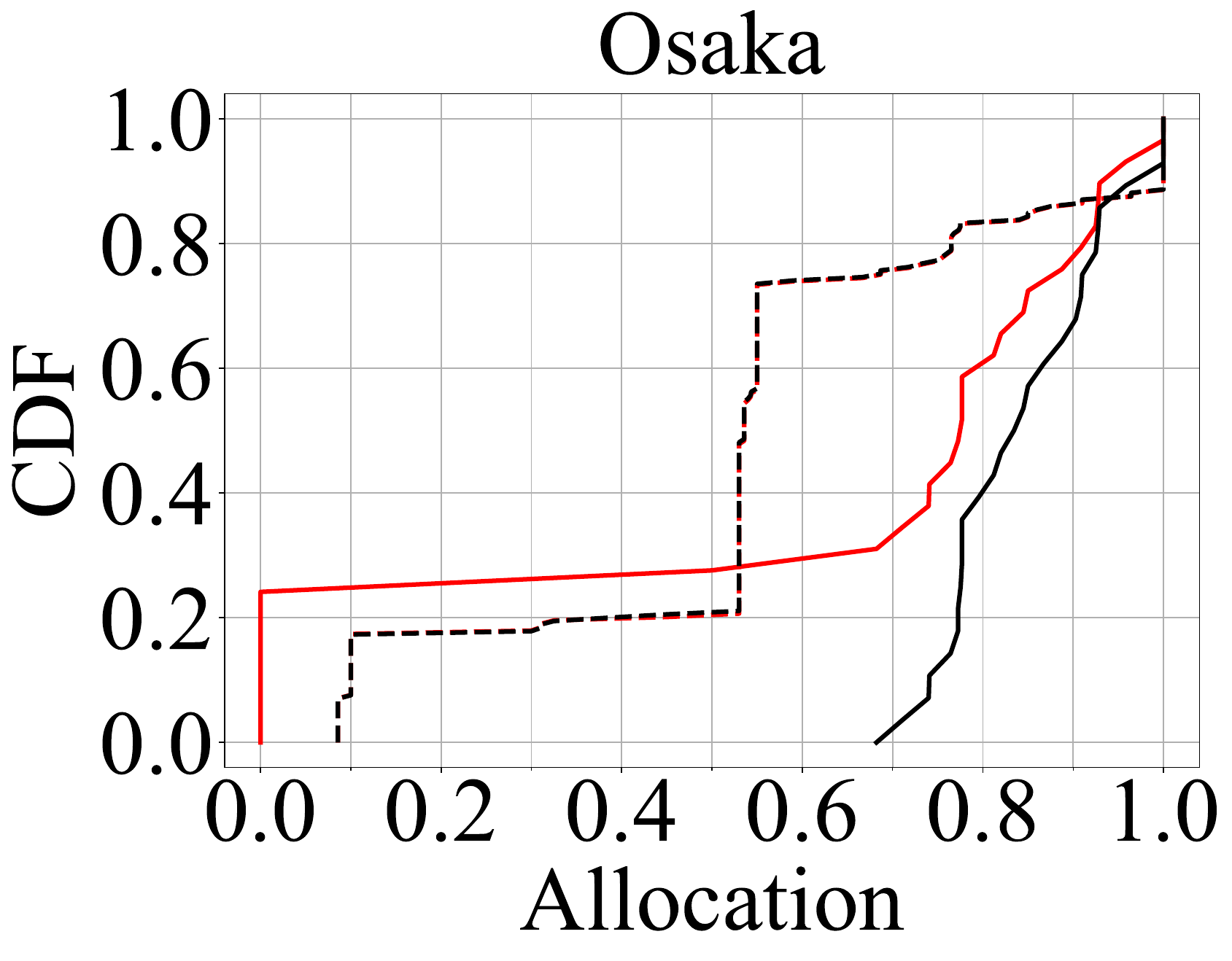} &
        \includegraphics[width=.18\textwidth]{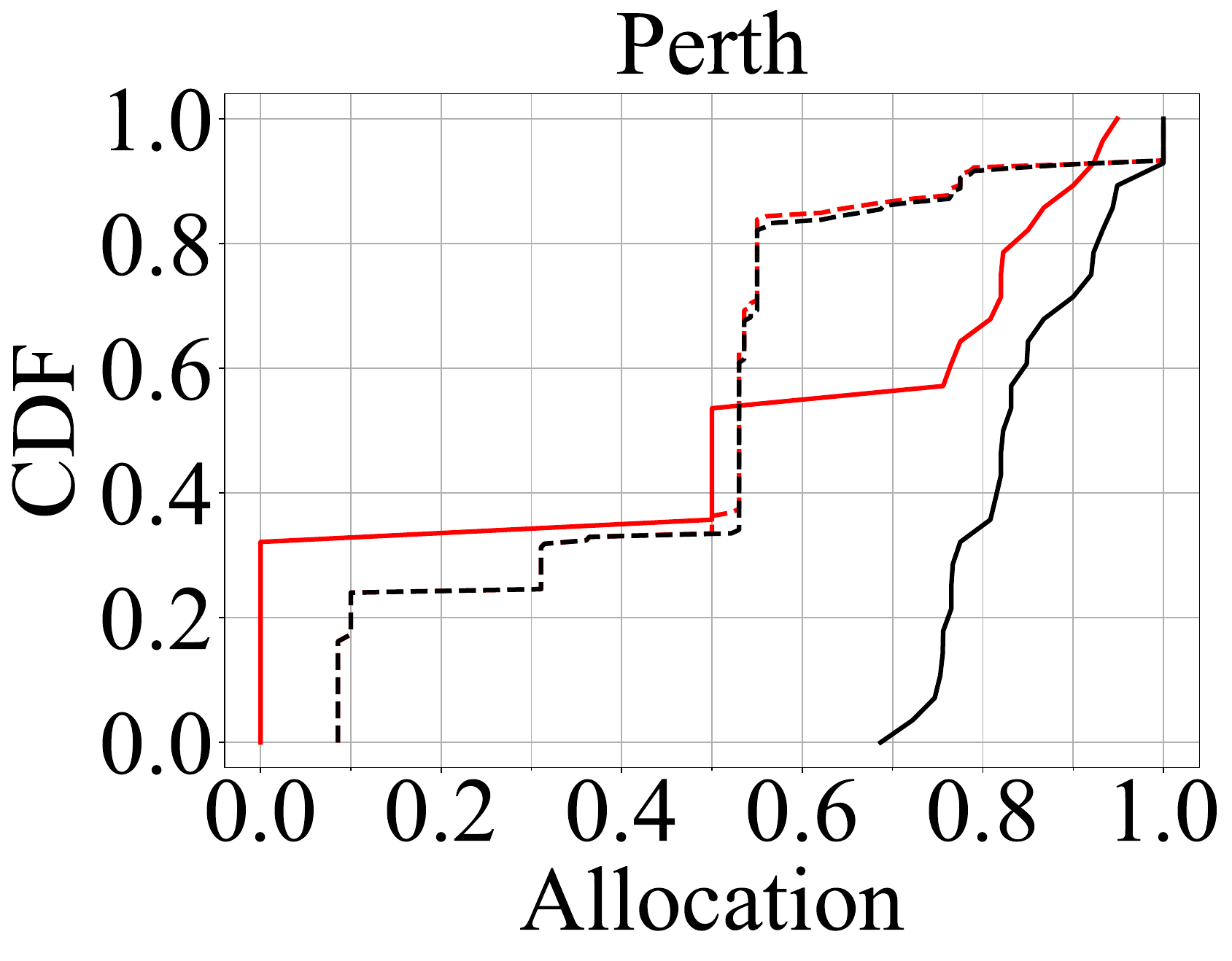} &
        \includegraphics[width=.18\textwidth]{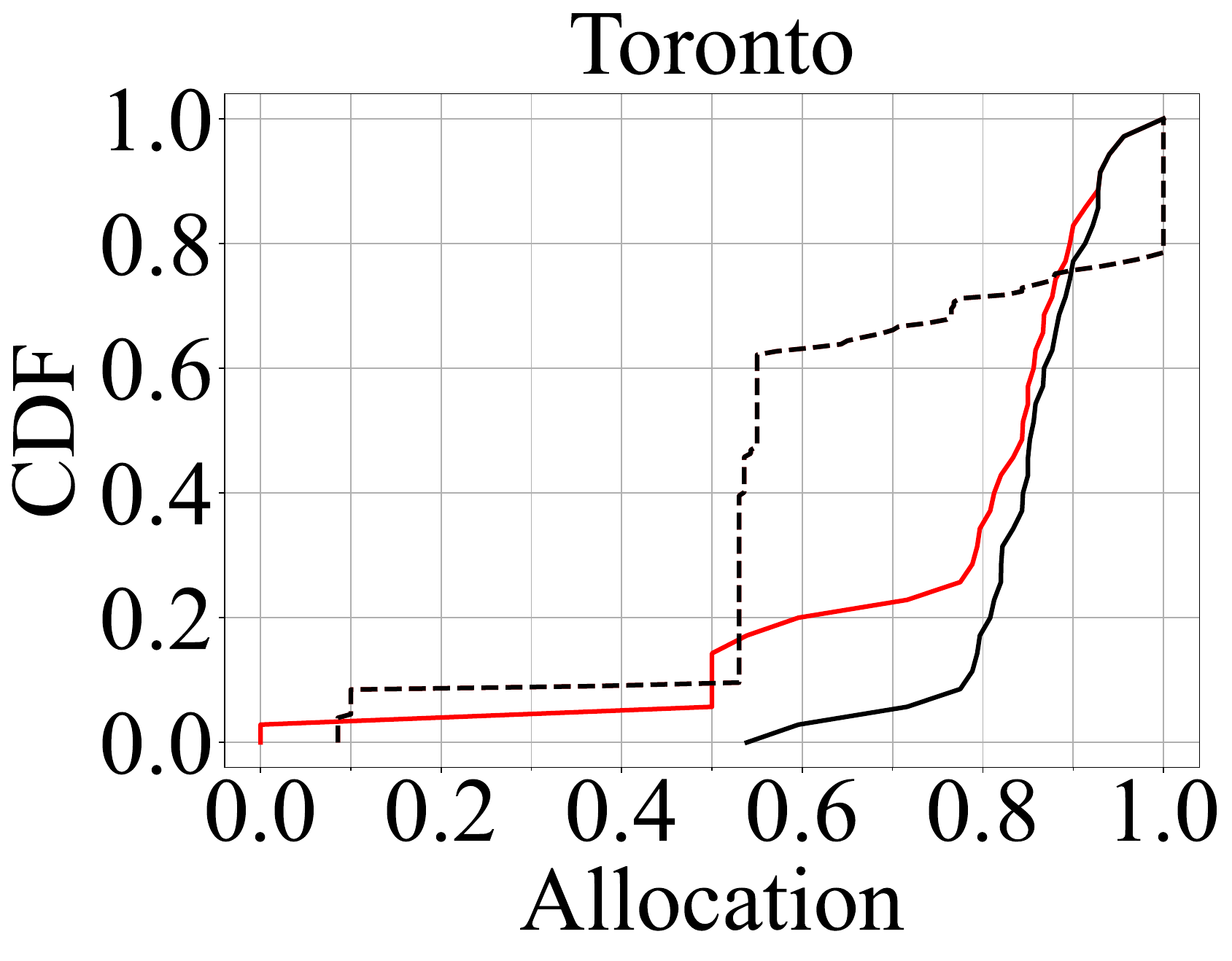} \\
    \end{tabular}
\end{figure*}

\begin{figure*}[!ht]
    \begin{tabular}{@{}ccccc@{}}
        \includegraphics[width=.18\textwidth]{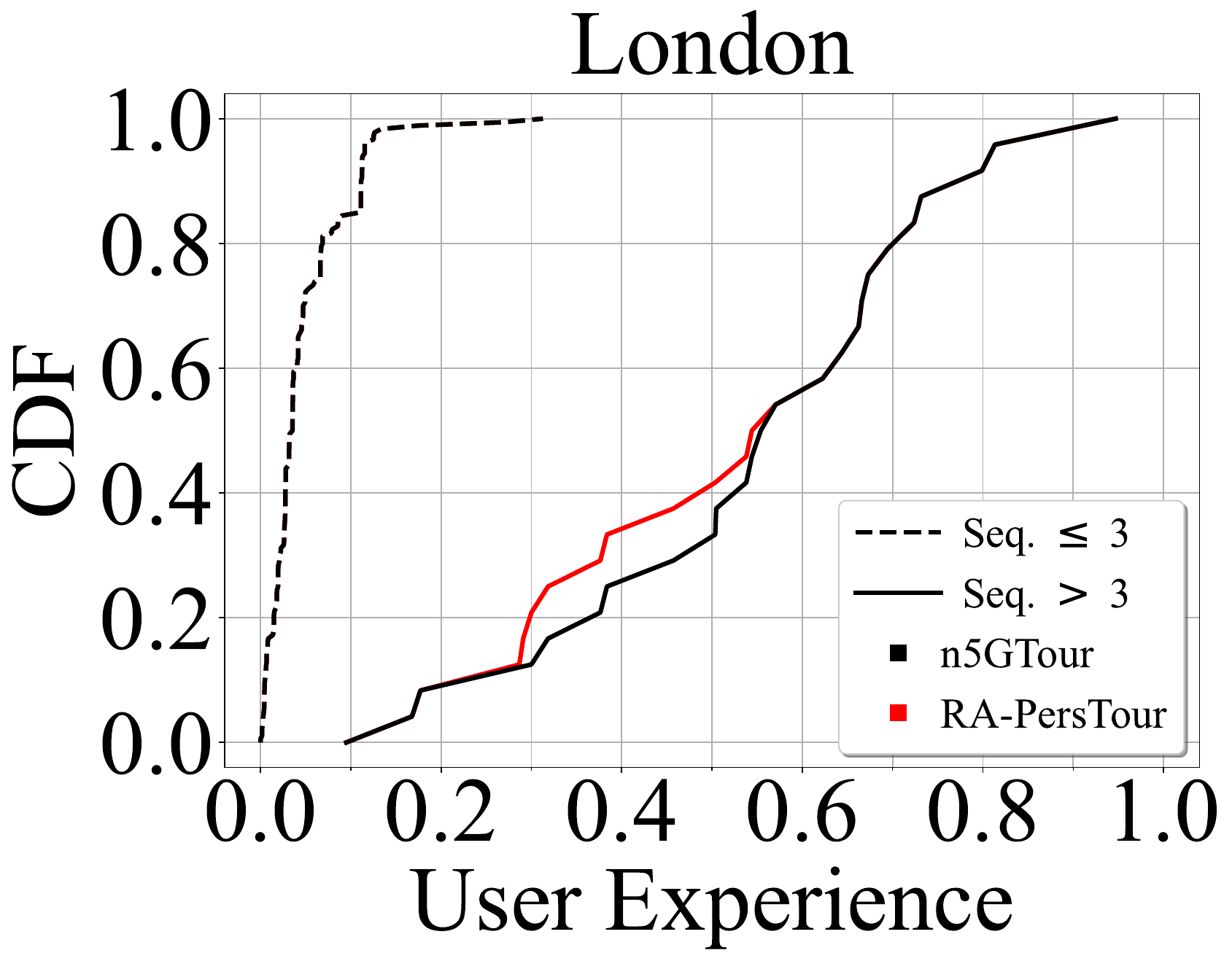} &
        \includegraphics[width=.18\textwidth]{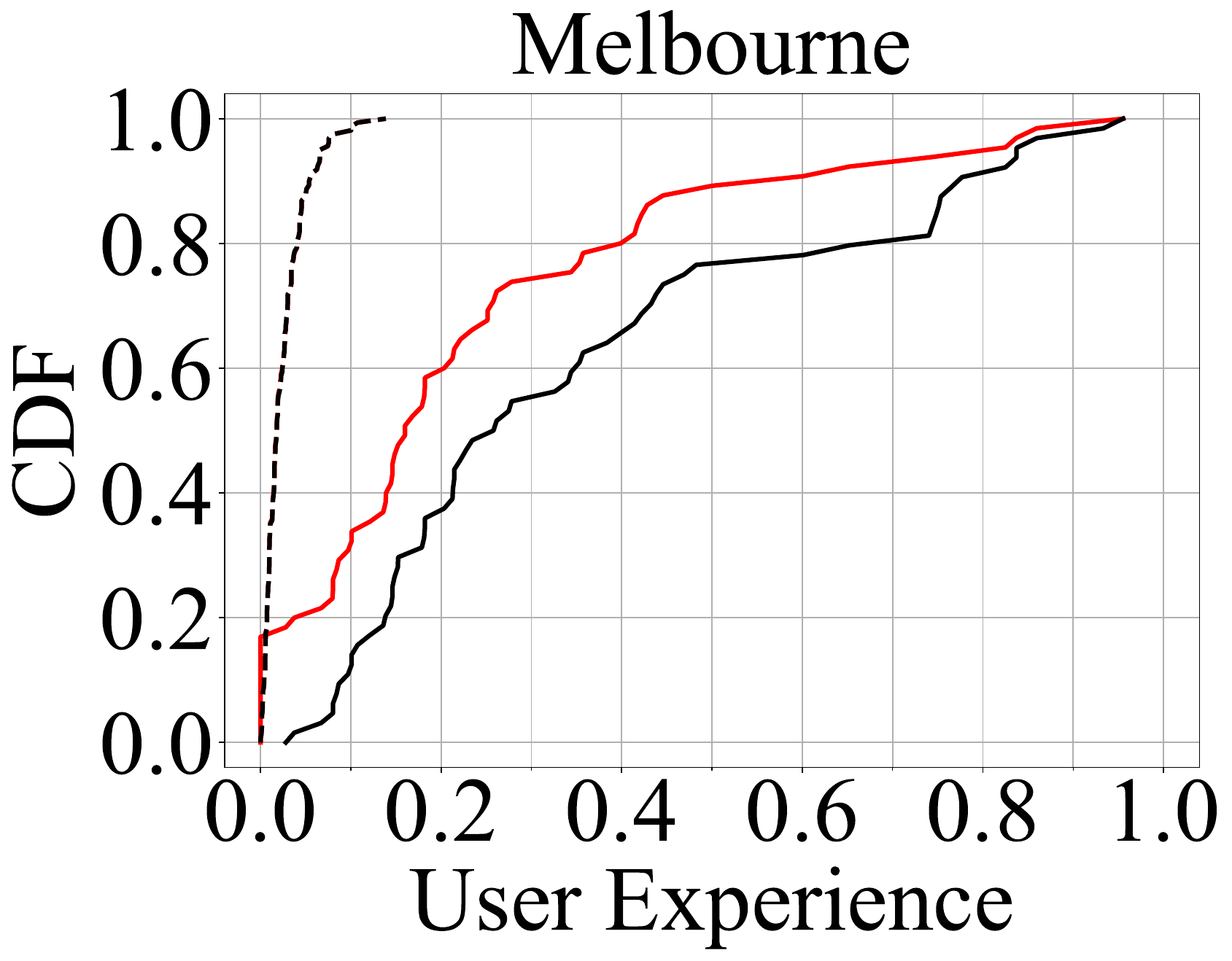} &
        \includegraphics[width=.18\textwidth]{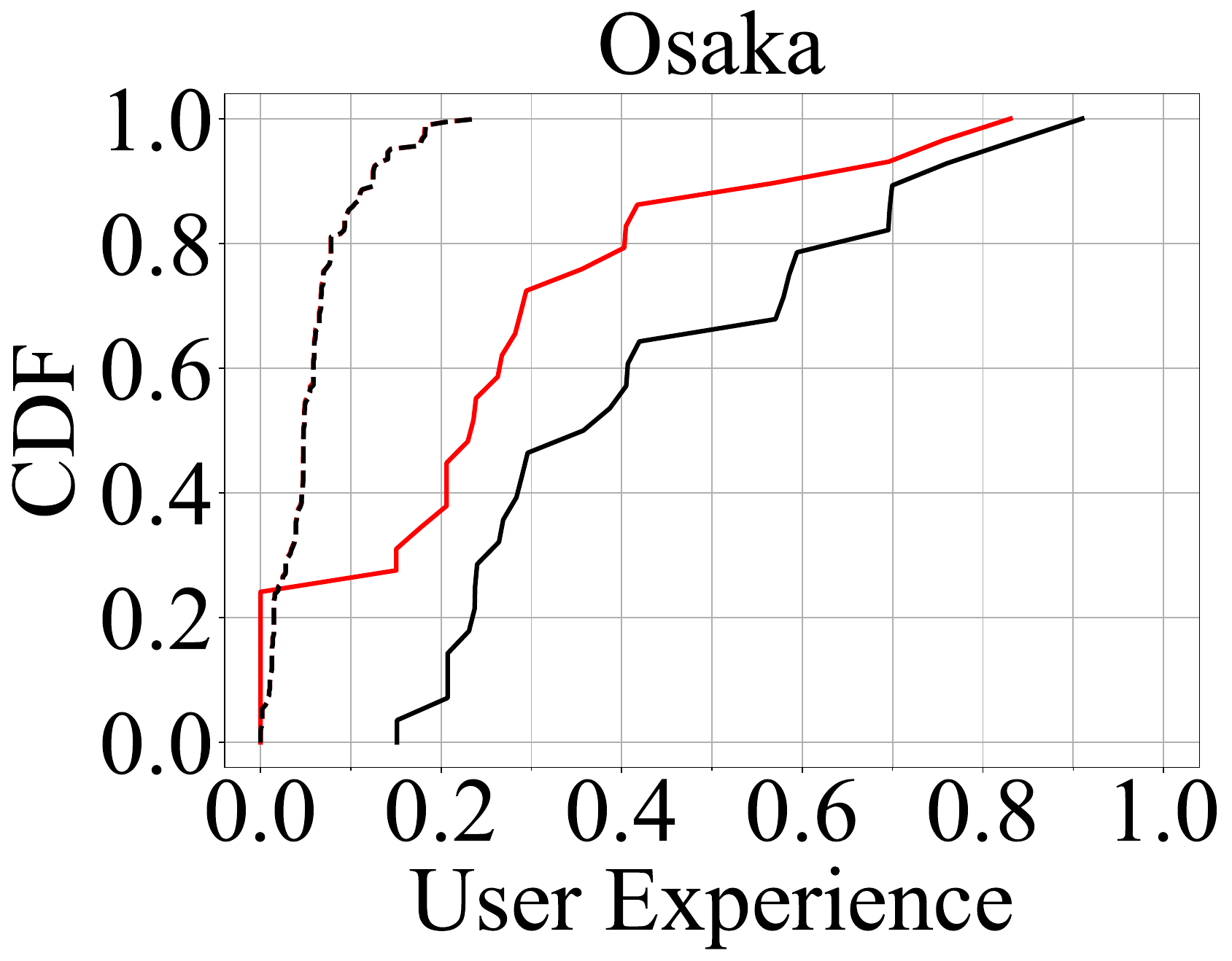} &
        \includegraphics[width=.18\textwidth]{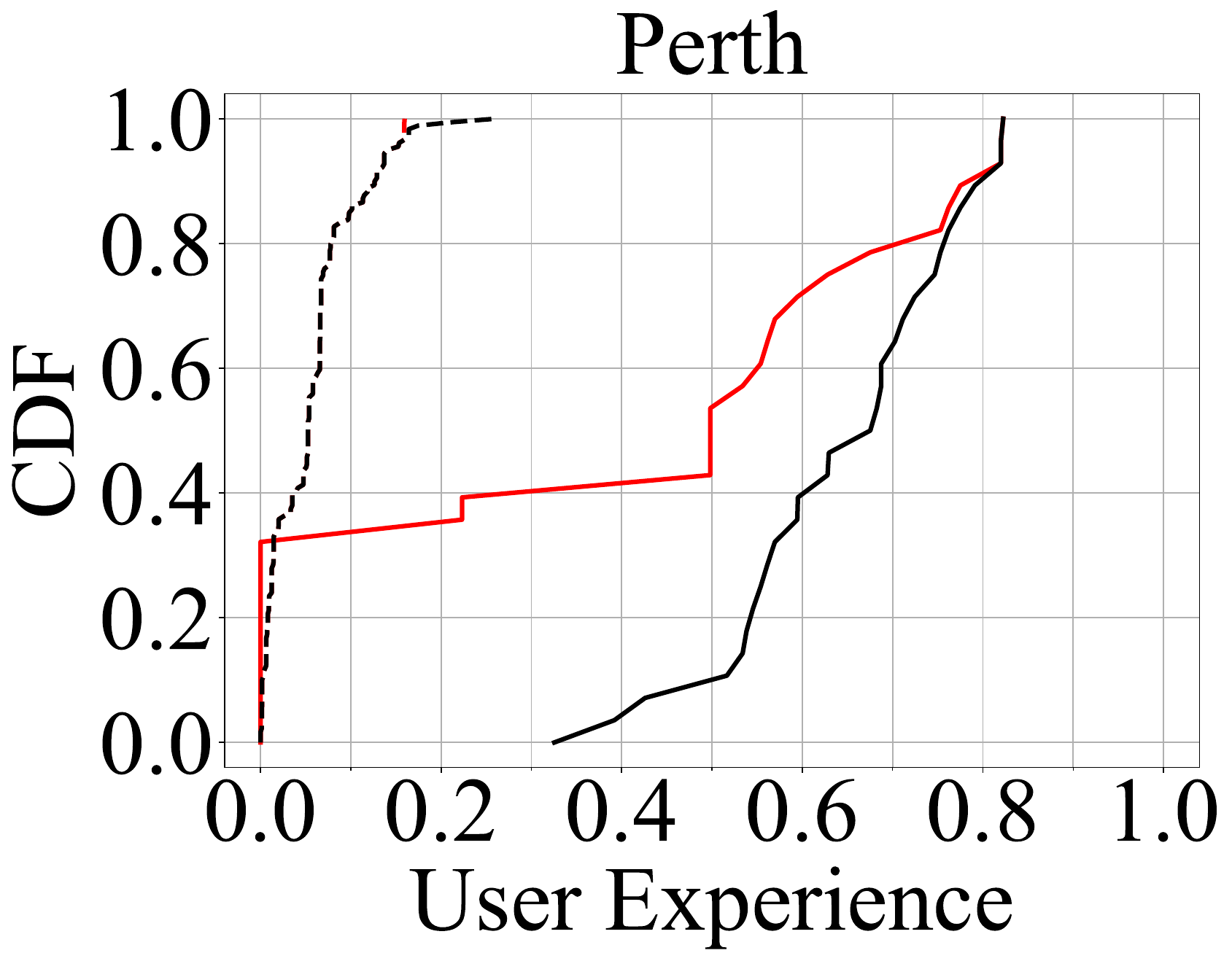} &
        \includegraphics[width=.18\textwidth]{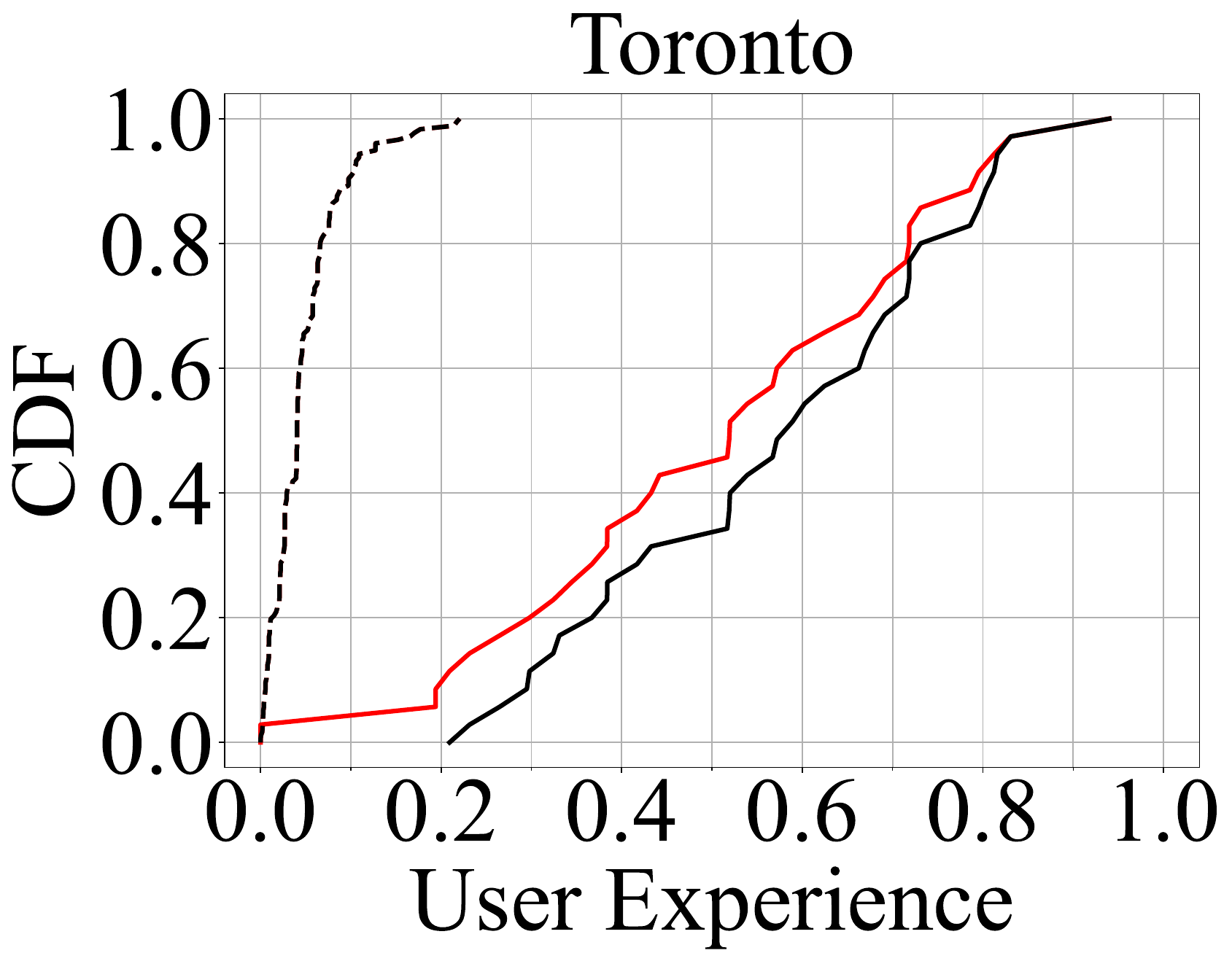} \\
    \end{tabular}
    \caption{CDF with the distribution of results for high overload scenario, relative to the itinerary metrics (first two rows), and the allocation and user experience metrics (last two rows). In itinerary metrics, the first line represents results for sequence lengths greater than 3. The second line represents results for sequences smaller or equal to 3.}
    \label{fig:CDF-metrics}
\end{figure*}

We run all experiments in a virtual machine (VM) that runs Debian 11 GNU/Linux and is configured with 4 vCPUs, 32~GB RAM, and 136~GB of virtual disk. The VM is hosted on a 2 Intel Xeon Silver 4114 @2.20GHz server. The first stage of +Tour is implemented using C++. The second stage uses Python 3.8.2, docplex 2.23.222, and IBM ILOG CPLEX 20.1.0 (as the solver). The RA-PersTour algorithm is also implemented using Python 3.8.2, docplex 2.23.222, and IBM CPLEX 20.1.0.

{Table~\ref{tab:gain}} shows the gain obtained by +Tour {(+T)} over RA-PersTour {(RA)} when comparing both algorithms in terms of \textbf{AE} and \textbf{UE} for all scenarios {(London - LO, Melbourne - ME, Osaka - OS, Perth - PE and Toronto - TO)}. %\textcolor{red}{Figure~\ref{fig:gain-All} presents the results considering all itineraries.} 
We observe that +Tour outperforms RA-PersTour in all scenarios {when considering all sequence lengths}, especially the high network overload, showing gains up to 11\% for \textbf{AE} and 40\% for \textbf{UE}. Considering only itineraries with more than 3 POIs, %\textcolor{red}{as illustrated in Figure~\ref{fig:gain-G3},} 
the gains for \textbf{AE} and \textbf{UE} are even greater, reaching, respectively, 74\% and 65\%. The CDF for each metric, including recall, precision, and F-score, is presented in Figure~\ref{fig:CDF-metrics}.

Figure \ref{fig:CDF-metrics} shows a CDF for all evaluated metrics to better illustrate the results' distribution. As can be seen in the \textbf{AE} and \textbf{UE} corresponding rows (3 and 4), our solution never presents results inferior to RA-PersTour, at most equaling the results for sequences less than or equal to 3, when there is no flexibility in choosing an itinerary from the Pareto front. For \textbf{Recall}, \textbf{Precision}, and \textbf{F-score} (rows 1 and 2), we can see that +Tour has similar accuracy to RA-PersTour with lower accuracy at certain points. This loss is justified by the gain observed in \textbf{AE} and \textbf{UE} for sequences greater than 3.

\begin{figure}[!ht]
    \begin{tabular}{@{}ccc@{}}
        \includegraphics[width=.15\textwidth]{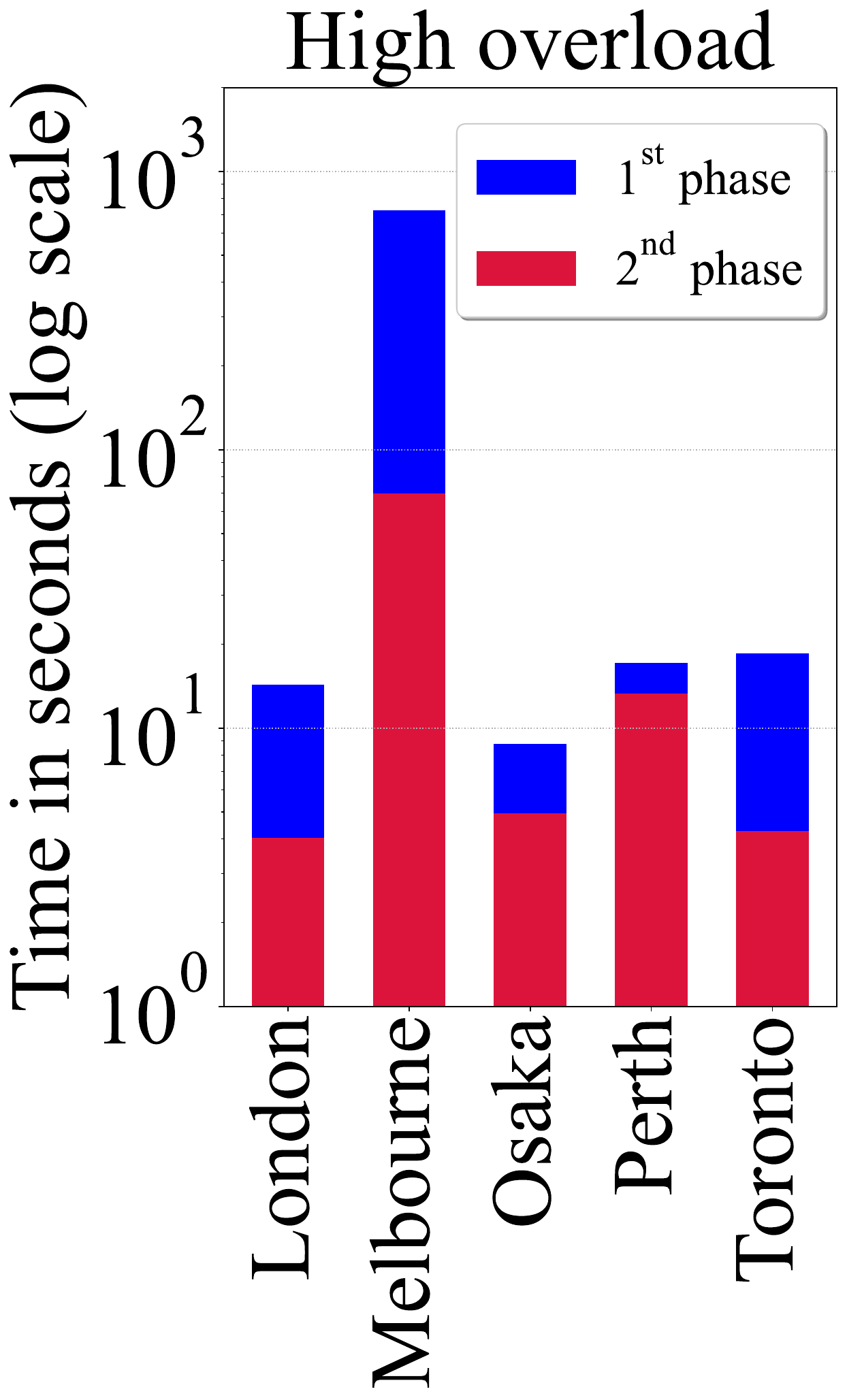} &
        \includegraphics[width=.15\textwidth]{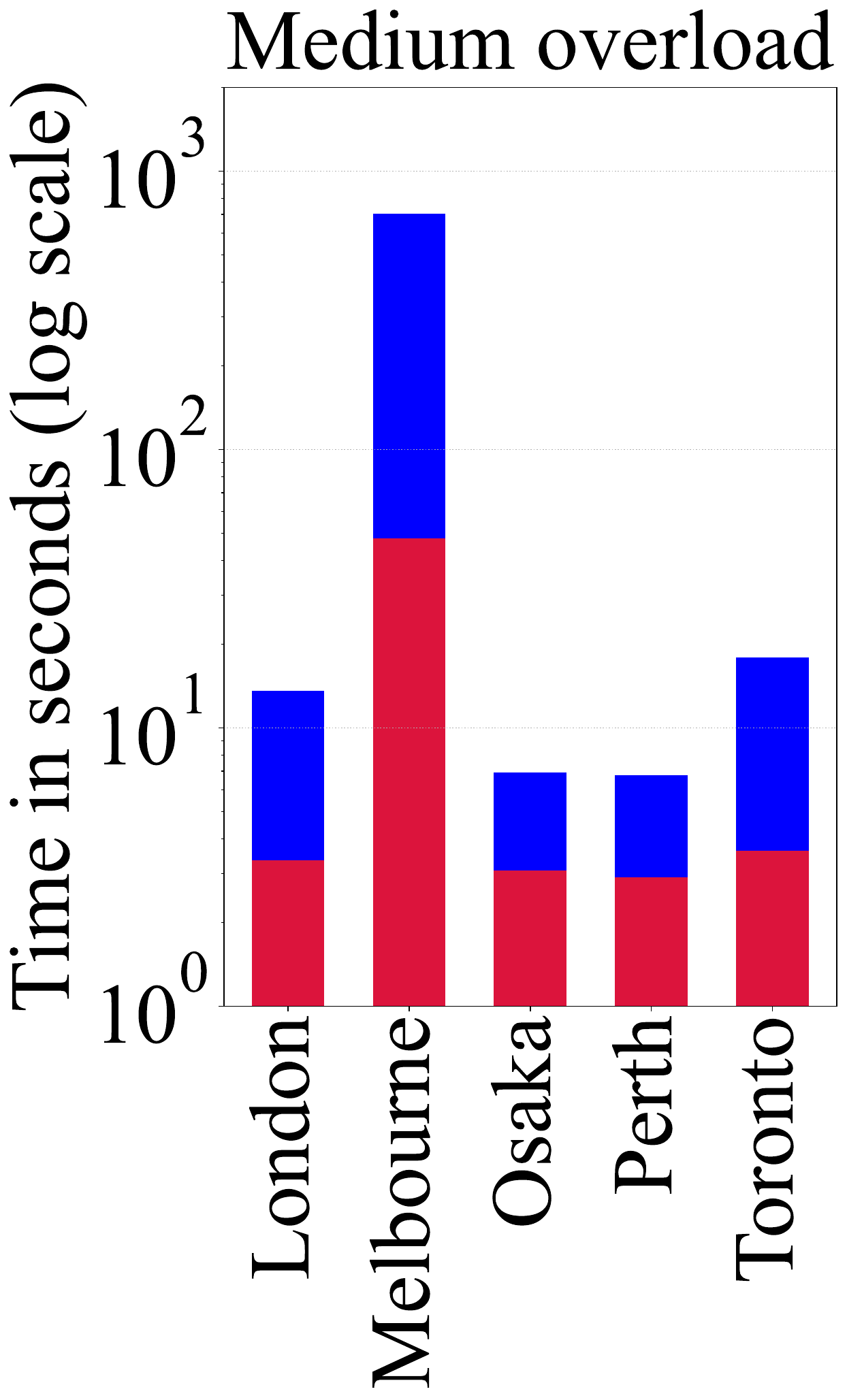} &
        \includegraphics[width=.15\textwidth]{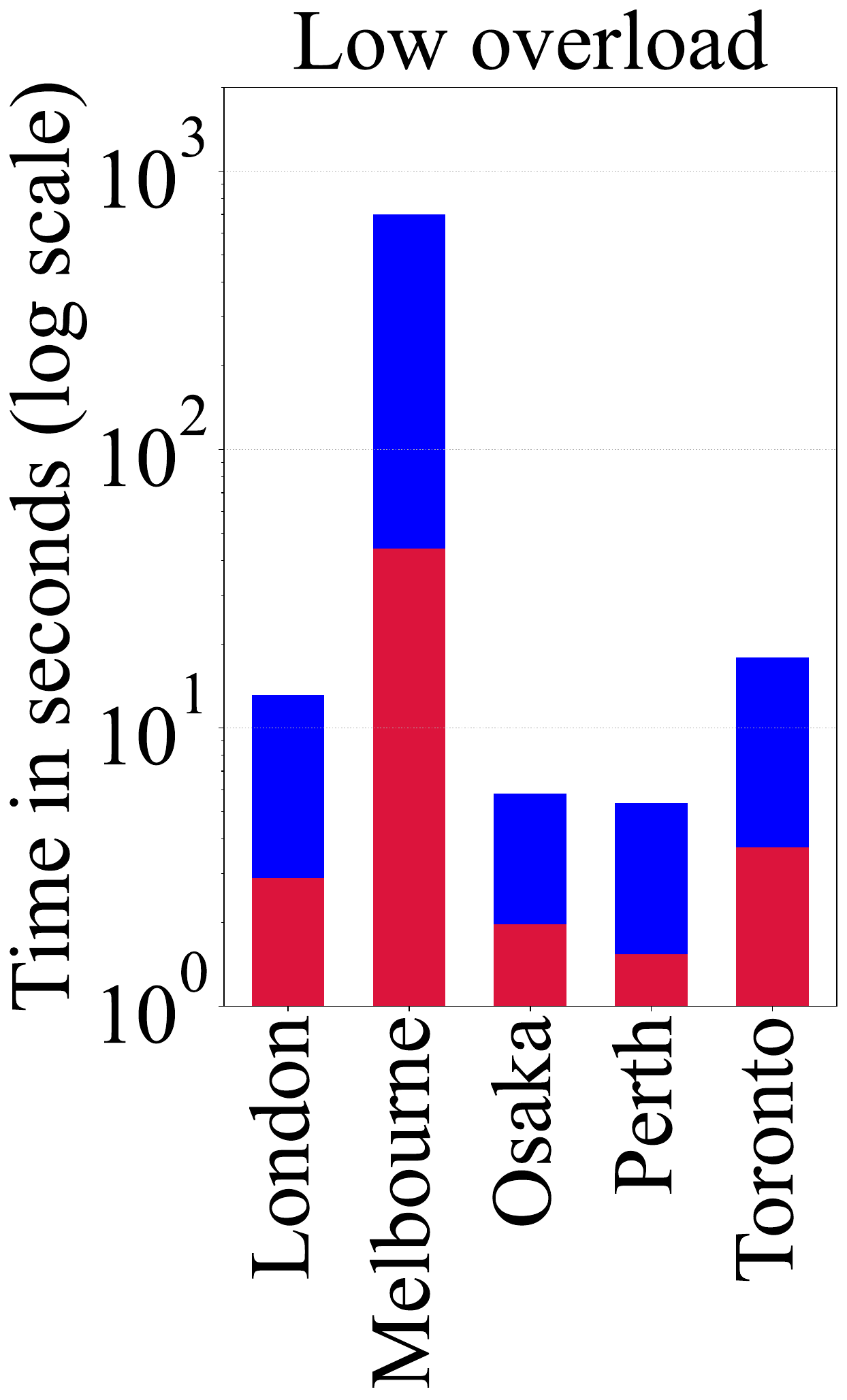} \\
    \end{tabular}
    \caption{Time of solving the two phases of +Tour.}
    \label{fig:time}
\end{figure}

Figure \ref{fig:time} shows the resolution time for +Tour's first and second phases. For all cities, except Melbourne, the total resolution time was less than 20 seconds. In Melbourne, the number of POIs is much higher than in all other cities. Therefore, it takes longer in phase 1, where the Pareto front of candidate itineraries is generated. In fact, for Melbourne, the first phase took approximately 656 seconds, while phase 2 reached 44, 47, and 69 seconds in the high, medium, and low scenarios, respectively. Indeed, our approach to solving the first stage of the problem using a dynamic programming version of the ESPPRC has proven to be very efficient, taking less than 11 minutes for Melbourne and less than 15 seconds for the other cities. This is an important result since the databases of other cities may demand scalability similar to the one observed in Melbourne.

In general, based on the experiments carried out, our solution presents superior results regarding the allocation of resources, which improves the user experience when using the applications offered in the POIs. We observed gains of up to 74\% and 65\% in terms of resource allocation and user experience, respectively, while the accuracy of the itineraries remained similar to RA-PersTour. Finally, +Tour makes itinerary recommendations, considering realistically sized instances in a reasonable amount of time.

%% file: Sections/08-Conclusions-And-Future-Work.tex
\section{Conclusions and Future Work}\label{sec:conclusions}

%In this work, we have accomplished four tasks. First, we formulated an optimization problem that jointly approaches the design of personalized tour itinerary recommendations and resource allocation at the network edge. To unveil the completeness of real-world visiting patterns, our problem formulation allows itineraries with one or more POIs. Second, we proposed +Tour, a low-complexity algorithm that efficiently solves the problem. Third, we presented an exploratory real-world dataset analysis to understand preferences and users’ visiting patterns. Based on the findings, We proposed a methodology to identify user interest in applications. Fourth, using the collected dataset, we evaluate the effectiveness of +Tour in different scenarios and compare it with a modified state-of-the-art PTIR algorithm.
{In this work, we propose +Tour, a two-stage solution to the resource-constrained itinerary recommendation problem. Our approach simultaneously maximizes profit and resource allocation while minimizing travel costs. We provide an extensive data characterization based on real-world POI visitation history from 13 cities spread across 4 continents. This dataset includes 20461 valid tour sequences generated by 8407 users who visited 401 POIs divided across 20 unique categories, which allows us to evaluate +Tour in realistic size scenarios. We also compare our solution to the state-of-the-art RA-PersTour, demonstrating that our approach achieves up to 74.2\% improvement in Allocation Efficiency and 65.1\% in User Experience while maintaining competitive performance in traditional metrics such as Recall, Precision, and F-Score. +Tour induces cooperative behavior in the tourists, which may benefit the users and the touristic service provider. In practice, we are aware that the tourists may not follow the recommendations, degrading the estimated performance. However, this issue negatively affects any recommendation system.} 

{In future work, we intend to use graph neural networks (GNN) to solve the problem of jointly designing personalized tour itinerary recommendations and resource allocation.
Among the methods for modeling preferences based on historical data, the most promising ones are machine learning techniques, especially those that consider the problem's structural characteristics. Most data in recommendation systems naturally have a graph structure. GNN takes advantage of this to learn the representation of a given POI based on all POIs visited previously and subsequently in a tour. For this reason, applying GNN has proven beneficial in recommendation problems to allow the extraction of user trends based on their preferences.}

{We also intend to broaden the scope to include other types of immersive applications with stringent requirements anticipated with the advancements of 5G and beyond, such as the metaverse. Additionally, we plan to evaluate the impact of non-cooperative users, i.e., tourists who do not follow the recommendations, in combination with other real-world aspects such as the need for waiting in lines at busy POIs and the availability of different types of transportation with different costs. A promising approach to be explored in the context of non-cooperative users is the Probabilistic Orienteering Problem~\cite{angelelli-probabilistic:17}, a variant of the OP Problem where tourists may visit a POI according to a certain probability. Finally, once developing a GNN version of the MEC-PTIR problem as well as the version that accounts for non-cooperative users, we intend to expand our evaluation using more baseline algorithms.} 